\documentclass[12pt,preprint]{aastex}
\shorttitle{COOL WDs IN THE SDSS}
\shortauthors{KILIC ET AL.}
\begin{document}
\title{A Detailed Model Atmosphere Analysis of Cool White Dwarfs in the Sloan Digital Sky Survey\altaffilmark{1}}

\author{Mukremin Kilic\altaffilmark{2,10},
S. K. Leggett\altaffilmark{3},
P.-E. Tremblay\altaffilmark{4},
Ted von Hippel\altaffilmark{5},
P. Bergeron\altaffilmark{4},
Hugh C. Harris\altaffilmark{6},
Jeffrey A. Munn\altaffilmark{6},
Kurtis A. Williams\altaffilmark{7},
Evalyn Gates \altaffilmark{8},
and J. Farihi\altaffilmark{9}}

\altaffiltext{1}{Based on observations obtained at the Hobby-Eberly Telescope (HET), Gemini Observatory, and the
Infrared Telescope Facility (IRTF).}
\altaffiltext{2}{Smithsonian Astrophysical Observatory, 60 Garden Street, Cambridge, MA 02138, USA; mkilic@cfa.harvard.edu}
\altaffiltext{3}{Gemini Observatory, 670 N. A'ohoku Place, Hilo, HI 96720, USA}
\altaffiltext{4}{D\'epartement de Physique, Universit\'e de Montr\'eal, C.P. 6128, Succursale Centre-Ville, Montr\'eal, Qu\'ebec H3C 3J7, Canada}
\altaffiltext{5}{Physics Department, Siena College, 515 Loudon Road, Loudonville, New York 12211, USA}
\altaffiltext{6}{U.S. Naval Observatory, Flagstaff Station, 10391 W. Naval Observatory Road, Flagstaff, AZ 86001, USA}
\altaffiltext{7}{Department of Astronomy, 1 University Station C1400, Austin, TX 78712, USA}
\altaffiltext{8}{Kavli Institute for Cosmological Physics and Department of Astronomy \& Astrophysics, The University of Chicago, 5640 South Ellis Avenue, Chicago, IL 60637, USA}
\altaffiltext{9}{Department of Physics \& Astronomy, University of Leicester, Leicester LE1 7RH, UK}
\altaffiltext{10}{\em Spitzer Fellow}

\begin{abstract}

We present optical spectroscopy and near-infrared photometry of
126 cool white dwarfs in the Sloan Digital Sky Survey (SDSS).
Our sample includes high proper motion targets selected using the SDSS and
USNO-B astrometry and a dozen previously known ultracool white dwarf candidates.
Our optical spectroscopic observations demonstrate that
a clean selection of large samples of cool white dwarfs in the SDSS (and the SkyMapper, Pan-STARRS, and
the Large Synoptic Survey Telescope datasets) is possible
using a reduced proper motion diagram and a tangential velocity cut-off (depending on the proper motion accuracy) of 30 km s$^{-1}$.
Our near-infrared observations reveal eight new stars with significant absorption.
We use the optical and near-infrared photometry to perform a detailed
model atmosphere analysis. More than 80\% of the stars in our sample are consistent with either
pure hydrogen or pure helium atmospheres. However,
the eight stars with significant infrared absorption and the
majority of the previously known ultracool white
dwarf candidates are best explained with mixed hydrogen and helium atmosphere models.
The age distribution of our sample is consistent with a Galactic disk age of 8 Gyr.
A few ultracool white dwarfs may be as old as 12-13 Gyr, but our models have problems matching
the spectral energy distributions of these objects. There are only two halo white dwarf
candidates in our sample. However, trigonometric parallax observations are required for
accurate mass and age determinations and to confirm their membership in the halo.

\end{abstract}


\section{Introduction}

White dwarf (WD) cosmochronology offers an independent age dating method for the Galactic disk and halo.
WDs are initially hot and consequently cool rapidly, though the cooling rate slows as their temperature
drops, allowing the oldest WDs to remain visible. Because the cooling rate slows, any census finds
more and more WDs at lower and lower temperatures until, quite abruptly, we find no more of them.
The position of this cut-off in temperature (and luminosity) is directly related to the age of the system.
Starting with \citet{winget87} and \citet{liebert88} numerous studies targeted the oldest WDs in the Galactic
disk in order to derive an accurate age. The current best-estimate for the age of the Galactic disk based on WDs
comes from \citet{leggett98}, who performed a detailed model atmosphere analysis of the optical and infrared
spectral energy distributions (SEDs) of 43 WDs to derive an age of 8 $\pm$ 1.5 Gyr.

Within the last decade, the Sloan Digital Sky Survey (SDSS) emerged as the main resource for WD studies. The SDSS
provides an unprecedented opportunity to increase the cool WD sample size from a few dozen to hundreds or thousands, and
also extend the age dating method to the thick disk and halo. \citet{kleinman04} and \citet{eisenstein06} have already
discovered over 9000 WDs in the SDSS Data Release 4 area. However, the majority of these objects are relatively warm.
WDs cooler than about 7000 K are buried in the stellar
color-color loci \citep{kilic04} and they cannot be identified based on photometry alone. Proper motion offers an efficient
means to delineate cooler WDs from the much larger number of main sequence stars. Reduced proper motion,
defined as $H = m + 5 \log\mu + 5 = M + 5 \log V_{\rm tan} - 3.379$, has long been used as a proxy for absolute magnitude
for samples with similar kinematics.
\citet{munn04} used the SDSS and USNO-B \citep[5 epochs,][]{monet03} astrometry to derive proper motions with an accuracy of 3.5 mas yr$^{-1}$.
Follow-up spectroscopy of high proper motion targets at the MMT, HET, and the McDonald 2.7m telescope showed that
WDs occupy a locus in the reduced proper motion diagram cleanly separated from most subdwarfs and that samples
of WDs can be defined using this diagram with contamination by subdwarfs and quasars of only a few percent \citep{kilic06}.
\citet{harris06} used this result to create statistically complete catalogs of WDs and 
published a substantially improved disk luminosity function including 6000 WD candidates.
This luminosity function is consistent with the \citet{leggett98} result, but the lack
of infrared photometry prevented \citet{harris06} from a definite conclusion about the implied age.
Collision induced opacity due to molecular hydrogen becomes important in the high density
atmospheres of cool WDs that contain even small traces of hydrogen.
This opacity dominates in the near-infrared, and without near-infrared photometry
the surface compositions and temperatures of WDs cooler than about 5000 K are uncertain.

In order to improve the cool end of the luminosity function presented by \citet{harris06}
we obtained optical spectroscopy and/or near-infrared photometry of 156 of the coolest WDs. Our near-infrared photometry sample
includes 126 stars.
Our target selection criteria and observations are discussed in Section 2, while an analysis of the observational
data and results from this analysis are presented in Section 3.
Various implications of these results are then discussed in Sections 4 and 5.

\section{Target Selection and Observations}

Based on a reduced proper motion diagram for the SDSS Data Release 3 footprint, \citet{harris06} derive cool WD samples 
by taking all stars below and blueward of the WD model curves for $V_{\rm tan}=20, 30,$ and 40 km s$^{-1}$. 
They use the \citet{bergeron95} models to fit all five SDSS magnitudes to determine temperature, distance, bolometric
magnitude, and tangential velocity for each star. The choice of hydrogen or helium atmosphere models has little effect on
the estimated $M_{\rm bol}$ for relatively warm WDs. However, the colors are significantly different for pure hydrogen and pure helium atmosphere
models for stars fainter than $M_{\rm bol} \approx 14.6$ ($T_{\rm eff} \approx$ 5300 K). Lacking infrared data, \citet{harris06}
make a weighted H/He assignment for each star based on the fraction of each type from the studies in the literature.
Adopting a higher or lower fraction of hydrogen-dominated stars changes the faint end of the luminosity function significantly
(see their Figure 7). Therefore, the exact luminosity of the cut-off in the luminosity function, the shape of the drop, and
the extent of the faint tail are uncertain based on the SDSS data only.
We target all WDs with $M_{\rm bol}>14.6$ from the $V_{\rm tan} \geq 20$ km s$^{-1}$ sample of \citet{harris06}
for follow-up optical spectroscopy and near-infrared photometric observations. In addition, we target
all 13 ultracool WD candidates identified by \citet{gates04} and \citet[][see also Vidrih et al. 2007]{harris08}.
Our goal is to create a clean sample of WDs that can be used to improve the faint end of the luminosity function.

\subsection{Optical Spectroscopy}

We obtained low resolution spectroscopy of 98 targets using the 9.2m HET in the queue mode and the Marcario
Low Resolution Spectrograph \citep[LRS,][]{hill98}. Our observations were performed between October 2004 and July 2006. 
We used a 1.5$\arcsec$ slit, Grism 2, and the GG385 blocking filter (with 50\% transmission
cut-on at 385 nm) to obtain spectra with a wavelength coverage of
4280 $-$ 7340 \AA\ and a spectral resolution of 6 \AA. 
A spectrophotometric standard star was observed each night for flux calibration (as part of the queue), and
Ne--Cd wavelength calibration lamp exposures were obtained after each science exposure.
The data were reduced using standard IRAF routines.

\subsection{Near-Infrared Photometry}

We obtained $JHK$ photometry of 126 WDs using the Near Infra-Red Imager and Spectrometer (NIRI) on Gemini-North,
the 0.8--5.4 $\mu$m medium-resolution spectrograph and imager (SpeX) on the IRTF, and the Wide-Field Camera (WFCAM)
on the United Kingdom Infra-Red Telescope.
The Gemini observations were obtained as part of the queue programs
GN-2005B-Q-33, GN-2006A-Q-69, and GN-2008A-Q-78. The IRTF observatons were obtained on several observing runs between 2004 December and 2006 April.
The UKIRT observations (of J0146+1404 and J2239+0018) were performed on UT 2007 September 20 under the service program U/SERV/1762.
The field of view of the NIRI observations is either $51\arcsec \times 51\arcsec$ or
$120\arcsec \times 120\arcsec$ and that of the SpeX is $60\arcsec \times 60\arcsec$.
We used a five or nine position dither pattern with 25--60s exposures.
We used the Gemini and NIRI packages in IRAF to reduce the data and the UKIRT faint standards \citep{leggett06} to calibrate the photometry.
The WFCAM images are processed through the UKIRT Infrared Deep Sky Survey \citep[UKIDSS,][]{lawrence07} pipeline.
We use the merged source catalogs from the WFCAM science archive for our service program.
The derived magnitudes are in the Mauna Kea photometric system \citep{tokunaga05}.
If available, we use observations of several standard stars in a single night to estimate the errors in nightly zero points.
The typical errors in our photometry (including the nightly zero point errors of 0.01-0.02 mag) are about 0.04 mag.
 
\subsubsection{NIRI Non-linearity Issues}

NIRI suffers from non-linearities that are correctable.
The detector response is a function of the counts in the pixel, the exposure time, the read mode,
the bias level, and the vertical position on the detector. Using NIRI spectroscopic flats, 
A. Stephens developed and kindly made available to us a Python script that calculates and applies
a per-pixel linearity correction for NIRI data. We use the
2008 April version of this script to correct our data.
Without this correction, the photometry is 10-15\% fainter than expected.

Eighteen of our targets are detected in the UKIDSS, at least
in one band. In order to make sure that our photometric reductions are reliable, we compare 
our NIRI photometry with the UKIDSS photometry. Our results are consistent
with the UKIDSS results within the errors. However, NIRI photometry is on average fainter by
0.016, 0.022, and 0.024 mag in $J, H,$ and $K$ filters, respectively. We correct
for this systematic effect by making our photometry brighter by the above amounts.
The corrected photometry is well within the errors in our original measurements.

\section{Results}

The optical spectra for the newly observed WDs at the HET are shown in Figures 1-5.
Our targets include 29 DAs that show H$\alpha$ absorption
\citep[Fig. 1, see the spectral classification system of][]{sion83}, 35 DCs with featureless
spectra (Fig. 2), 4 DQs with molecular carbon bands (+2 DQs with SDSS spectra, Fig. 3), 2 DZAs with
magnesium, sodium, and hydrogen lines (Fig. 4), and 2 unresolved DA + dM (M dwarf) pairs (Fig. 5). 
Most spectra show sky subtraction problems at 5577, 5890/5896, and 6300 \AA.

The spectra presented in Figure 2 seem featureless. However, weak H$\alpha$ absorption may be hidden
in the noise, and higher resolution and higher signal-to-noise ratio observations may reveal weak H$\alpha$
for some of the targets in this figure.

Two of the stars in our sample, J1247+0646 and J1442+4013, display shifted Swan bands. Hence,
they belong to the peculiar DQ population \citep[DQpec,][]{schmidt95}.
Excluding the WD + dM pairs, J1247+0646 is the reddest WD in our sample with a $g-i$ color of 1.64 mag.
The strong molecular absorption features in the blue causes this star to appear relatively red
compared to all other DQ, DC, and DZ WDs.

The optical spectra of the stars presented in Figure 5 are dominated by M dwarfs,
but the flux excess in the blue and the presence of strong H$\beta$ and H$\alpha$
absorption lines indicate DA WD companions. In addition, the $u-$ and $g-$band photometry for these stars is
significantly brighter than expected from single M dwarf stars. These two stars
are best explained as DA WD + dM pairs.

The coordinates, optical and near-infrared photometry, and spectroscopic
classifications for our sample are given in Table 1.
Positions are those from the SDSS astrometric pipeline \citep{pier03}. The photometric
calibration is based on the SDSS standard star system \citep{smith02} tied to the survey data
with the Photometric Telescope \citep{hogg01}. The SDSS photometry is in the AB system. We use
the corrections given in \citet{eisenstein06} for the $u-, i-$, and $z-$band photometry, and
Table 1 includes these corrections.
Even though this table includes 156 stars, optical spectroscopy \citep[this paper,][and the literature]{kilic06} and
near-infrared photometry are available only for 140 and 126 stars, respectively. $J-$, $H-$, and $K-$band
photometry is missing for 30, 32, and 38 stars in Table 1, respectively.
The 30 stars with optical-only data (without $J-$band photometry) are
not included in our model atmosphere analysis (Section 3.3). Out of the 140 stars with optical spectroscopy,
43 (31\%) are DA, 84 (60\%) are DC, 6 (4\%) are DQ, 5 (4\%) are DZ, and 2 (1\%) are WD + dM pairs.

\subsection{Reduced Proper Motion Diagram}

Figure 6 shows the reduced proper motion diagram for the SDSS DR3 area including spectroscopic classifications from
the SDSS and our observations. The SDSS observes mostly hotter white dwarfs with $g-i<0.3$ mag, whereas we
focus on WDs with $g-i\geq0.6$ mag. Both of these selection effects are evident in this diagram.
Compared to a few percent contamination rate from subdwarfs found by \citet{kilic06},
our sample has a higher rate of contamination from subdwarfs. Out of the 98 newly observed targets,
26 are subdwarfs. Table 2 presents the coordinates, optical photometry, and proper motions for these stars.
Figure 7 presents their spectra. They all show MgH, Na I, and
several other metal lines\footnote{The absorption feature at 6800 \AA\ is the atmospheric B-band}.
With a tangential velocity cut-off of 20 km s$^{-1}$, many of our targets fall in the overlap
region between the WD sequence and subdwarfs in the reduced proper motion diagram. Hence, the higher
subdwarf contamination rate in this study is not surprising.

The cool WD selection works well for $V_{\rm tan}\geq 30$ km s$^{-1}$. Only
one of the 75 cool WD candidates (J09194811+4356216, Fig. 7) in the \citet{harris06}
$V_{\rm tan}\geq 30$ km s$^{-1}$ and $M_{\rm bol}>14.6$ mag sample is
actually a subdwarf, which corresponds to a contamination rate of 1.3\%.
On the other hand, 4 out of 86 targets (4.7\%) with $V_{\rm tan}\geq 25$ km s$^{-1}$ and
30 out of 122 targets (24.6\%) with $V_{\rm tan}\geq 20$ km s$^{-1}$ are subdwarfs.
Hence, increasing the lower tangential velocity limit
from 20 to 30 km s$^{-1}$ gets rid of most of the contamination from subdwarfs.
The $V_{\rm tan}=30$ km s$^{-1}$ curve is the best choice for creating a clean
sample of WDs without losing too many of them in the overlap region with subdwarfs. 
This selection can be used to create clean samples of cool WDs from the
SDSS DR7 and SEGUE data and (depending on the astrometric accuracy) future surveys that
use the SDSS filters (e.g. the SkyMapper, Pan-STARRS, and the Large
Synoptic Survey Telescope).

\subsection{Color-Color Diagrams}

Figure 8 shows the optical color-color diagrams for the cool WDs in \citet{kilic06}, \citet{gates04},
\citet{harris08}, and this paper. A comparison of the observed colors with the predicted colors
of pure hydrogen and pure helium atmosphere WD models (see \S 3.3) show that cool WDs have redder $u-g$
colors than predicted. Our models do not include the Ly $\alpha$ far red-wing opacity \citep{kowalski06b}. Hence, the
observed mismatch in the $u-g$ colors is not surprising. On the other hand, the observed range of $g-r, r-i$, and $i-z$ colors
agree well with the pure hydrogen or helium atmosphere model predictions except for the WDs with significant $i-$ or $z-$band
absorption. The $g-r$ vs. $r-i$ diagram provides an efficient way to identify WDs that show strong absorption in the $i-$band,
and the $r-i$ vs. $i-z$ diagram provides a reliable selection of WD + dM pairs. 

The $u-g$ vs. $g-r$ diagram reveals 6 DQs in the upper right corner of the figure. Strong molecular absorption in the $g-$ band causes
the $g-r$ colors to be redder than normal. One of these DQs, J1247+0646, is about 0.7 mag redder than the other
five in our sample, and it is also redder than all other published DQ WDs\footnote{There are two redder DQ WDs,
both with $g-i=2.3$ mag, discovered in the DR7 and SEGUE observations.} including the extreme DQ
GSC2U J131147.2+292348 \citep{carollo03}. J1247+0646 is clearly a very cool, peculiar DQ WD.

Figure 9 displays optical and near-infrared color-color diagrams for the \citet{bergeron01} sample and our Gemini, IRTF, and UKIRT
sample of WDs. The \citet{bergeron01} sample includes stars with $T_{\rm eff}\leq 12,000$ K.
LHS 1126 \citep[Luyten Half Second catalog,][]{luyten79} is the only star in that sample that shows significant absorption
in the near-infrared. In addition to the previously known ultracool WD candidates, these diagrams reveal
eight more WDs that show significant absorption in the $H+K-$bands or only
in the $K-$band. There is a large scatter in the observed infrared colors of these stars. However, the previously known
and the newly found stars with flux deficits reveal, for the first time,
a pattern in the $r-J$ vs. $J-H$ diagram. 
The observed sequence for IR-faint WDs is significantly different than the pure
hydrogen model sequence indicating that these WDs most likely do not have pure hydrogen atmospheres.
The dotted, long-dashed, and dashed-dotted lines in Figure 9 show the predicted infrared colors for mixed atmosphere WDs with
H/He = 10, 1, and 0.01, respectively. The similarities between the colors for these models and the IR-faint stars
suggest that these stars have mixed H/He atmospheres.

Excluding the 14 stars from the \citet{gates04} and \citet{harris08} studies, 8 out of 112
(or 7\% of) WDs in our sample show significant absorption in the
infrared. Understanding the nature of these stars is important for WD cosmochronology.
Prior to our study, near-infrared photometry of only a few ultracool WDs has been done
\citep[e.g. LHS 1402, LHS 3250, and SDSS J1337+0001,][]{bergeron02,bergeron05}. Our Gemini
photometry for the ultracool WD candidates from \citet{gates04} and \citet{harris08} shows that six of these stars
have similar $r-J$ and $J-H$ colors
with LHS 3250 and J1337+0001. A few other stars from the \citet{harris08} sample also have similar $r-J$ colors.
However, only $J-$band photometry is available for these stars and therefore they are not
shown in Figure 9.

\subsection{Model Atmosphere Analysis}

We use new grids of pure hydrogen and mixed H/He composition models with $\log g=8$ and $T_{\rm eff}= 2000-7000$ K,
in steps of 250 K. In addition, we use a new pure helium atmosphere model grid with $T_{\rm eff}= 3500-7000$ K.
The H-rich models are very similar to those of Bergeron et al. (2001)
but with updated collision induced absorption (CIA) opacities \citep[see the discussion in][]{tremblay07}.
We use a pure helium model grid with the non-ideal equation of state of \citet{bergeron95}.
The previous models calculated with this theory had a programming error.
We use a corrected grid for this work.
The number of free electrons in cool He-rich models is about an order of
magnitude higher than the model grid used by \citet{bergeron97,bergeron01}.
However, the differences between the old and the new models are small, with shifts
in $T_{\rm eff}$ of $\approx$ 200 K for the coolest stars. The quality of the fits
is also similar in both cases.

Our mixed atmosphere models are based on those of \citet{bergeron02}
but with three significant changes. We now include the HeH$^{+}$
molecule in the equation of state \citep{harris04}, which is the
most important update here. We also include the most recent calculations
for the He$^{-}$ opacity \citep{john94}.
\citet{bergeron02} included the \citet{hummer88}
non-ideal effects only in the relative state populations, which is a
very good approximation for most DA and DB white dwarfs. However, for
ultracool white dwarfs, a full implementation of the non-ideal
effects is necessary, including corrections to the pressure (or density)
at each depth. For simplicity, we have neglected {\it all} non-ideal
effects, since these effects are poorly understood and the Hummer \&
Mihalas results are at odds with more recent calculations
\citep{kowalski06a}. In Figure 10 we present an
illustrative sequence of models at constant $T_{\rm eff}$ and $\log g$,
which shows that the maximum CIA absorption is now predicted at H/He
$\sim$10$^{-2}$ instead of 10$^{-5}$ as found by \citet{bergeron02}.
Thus we expect differences in the H/He ratios of $\sim$ 0-3 dex. 

Our fitting technique is described at length in
\citet{bergeron01}. Briefly, we convert the magnitudes into
observed fluxes using the method of \citet{holberg06}
and the appropriate filters. Then we fit the
resulting SEDs with those predicted from model atmosphere
calculations using a nonlinear least-squares method. Only $T_{\rm
eff}$ and the solid angle $\pi (R/D)^2$, where $R$ is the radius of
the star and $D$ its distance from Earth, are considered free
parameters. Since no parallax measurements are available, we assume
a surface gravity of $\log g=8$ \citep{bergeron01}. 

Our models do not include the
opacity due to the red wing of Ly $\alpha$ \citep{kowalski06b}. Hence, we omit the
$u-$band photometry from all fits. We also omit the $g-$band photometry
for stars cooler than 4600 K since the missing Ly $\alpha$ opacity has a larger
impact at lower temperatures \citep[see][]{lodieu09}.
The reason for neglecting the B and V (or $u$ and $g$) filter photometry
at lower temperatures is discussed at length in \citet[][section 5.2.2]{bergeron97}.
The atmospheric parameters for the cool WD BPM 4729 obtained with models including this opacity 
\citep[$T_{\rm eff}=5820$ K and $\log g=8.30$,][]{kowalski06b} and with models excluding
this opacity and the UV filters \citep[$T_{\rm eff}=5730 \pm 110$ K and $\log g=8.21 \pm 0.09$,][]{bergeron01} are
essentially the same. This is because
the Ly $\alpha$ opacity affects a wavelength region where there is very little
flux, hence the atmospheric structure is not affected much by a
change of the opacity in the UV (although the predicted colors in the
UV are).

We use the SEDs together with
the optical spectra at H$\alpha$ to constrain the surface
composition. If optical spectroscopy is unavailable, we
choose the best photometric fit. We use the mixed H/He atmosphere 
models only if there is a strong infrared absorption.
The majority of the objects in our sample are consistent with either pure hydrogen
or pure helium atmospheres, though some are best
explained with mixed H/He atmosphere models. 
There are almost no cases of mild absorption where mixed H/He models would
be needed. Obviously, the addition of helium into a hydrogen-rich atmosphere
provides another free parameter in our model fits and it can improve the
fits slightly, but these determinations are not statistically significant.

We use the \citet{dufour05} results for the DQ stars, but
we use regular He-rich or H-rich models for the DZA and DAZ WDs.
For the DQpec stars, we use the procedure of
\citet{bergeron97,bergeron01} in which we use the normalized observed spectra to recalibrate the
synthetic fluxes and take the molecular bands into account in first order.
Even with this procedure the fits are poor and we removed both
the $u-$ and $g-$band photometry from the fits. Table 3 presents the
best-fit model temperatures, distances, ages, and compositions
for our sample of WDs\footnote{One of the targets, J0804+2239, is a DZ WD with strong IR absorption.
This star is not included in Table 3. Its intriguing SED will be discussed in a future publication.}
assuming $\log g=8$. 

The errors in temperature are largely dominated by the observational
uncertainties with a minor role from the unknown surface gravity. Other quantities
(distances and ages) depend, of course, on the surface gravity assumption.
The thickness of the hydrogen envelope and the core composition also play a role
in age estimates.
In addition, a WD may change its surface composition several times during its
lifetime as a result of competing mechanisms (e.g. gravitational settling,
accretion, dredge-up, and convective mixing). Hence, it is difficult
to estimate the true age of a star at a given time. However, \citet{tremblay08}
demonstrate that the majority of DA stars have relatively massive hydrogen envelopes
($M_{\rm H}/M_{\star} \sim 10^{-6}$ or more) that prevent them from conversion to helium dominated
atmospheres. WDs with thin hydrogen layers likely end up with pure-He or mixed H/He atmospheres due to convective mixing.
All of these issues have been thoroughly discussed in \citet[][section 6.4]{bergeron97},
\citet[][section 2.3 and 5.5]{bergeron01}, and \citet{fontaine01}. 

\subsubsection{Pure Hydrogen and Pure Helium Atmosphere WDs}

Figure 11 shows a representative sample of fits for DA stars with $T_{\rm eff}=5000-6000$ K.
J0003$-$0111, J0250$-$0910, and J1116+0925 all show H$\alpha$ absorption and their optical and near-infrared SEDs
are best matched by pure hydrogen atmosphere models. The spectroscopic observations of H$\alpha$ are not used directly
in the fitting procedure, but they serve as an internal check of our photometric solutions.
The theoretical line profiles are calculated using the parameters obtained from the SED fits. This figure shows that the predicted
line profiles are in good agreement with the pure hydrogen atmosphere model solutions derived from the photometric observations.
On the other hand, for three of the stars in this figure (J0330+0037, J1212+4345, and J2350$-$0848) pure helium atmosphere models
fit the optical and infrared photometry better than the pure hydrogen atmosphere models.
However, the difference betwen pure hydrogen and pure helium model SEDs is relatively small at these temperatures
and the observed photometry is also consistent with the pure hydrogen model predictions
within the errors. The residuals observed in the pure hydrogen model fits
are most likely due to small photometric errors or calibration problems.
These small differences show the limitations of the entire photometric approach,
and consequently the difficulty in assigning H- versus He-composition when
H$\alpha$ is invisible. In any case,
the excellent match between the H$\alpha$ line profiles and observations rule out significant amounts of helium
in the atmosphere. Hence, we assume a pure hydrogen composition for all stars that show H$\alpha$ absorption.

Figure 12 shows sample fits for 11 helium-rich DC stars. The spectroscopic fits are not shown here since all objects are featureless
near the H$\alpha$ region. Several of the stars in this figure are warm enough to show H$\alpha$ if they were hydrogen-rich. 
With the resolution and signal-to-noise ratio of our observations, we are able to detect H$\alpha$ for stars
hotter than 5000 K.
The lack
of H$\alpha$ absorption reveals a helium-rich composition, and the pure helium models provide excellent fits to the SEDs of these objects.
There are 42 stars with $T_{\rm eff} \geq 4530$ K that are best explained as pure helium atmosphere objects, but there are none
below this temperature. Given the observed infrared colors of cool WDs, perhaps this is not surprising.
The $r-J$ vs. $J-H$ color-color diagram (Fig. 9) shows that the coolest WDs show absorption in the infrared, indicating that they
have hydrogen in their atmospheres.

Figure 13 displays the model fits to the SEDs of the eight coolest DC WDs in our sample excluding the
ultracool ($T_{\rm eff}<4000$ K) WDs. These SEDs are best explained with pure hydrogen atmosphere models with $T_{\rm eff}=4150-4420$ K.
Omitting the $u-$ and $g-$band photometry from the fits (due to the missing Ly$\alpha$ opacity),
our models are able to explain the overall SEDs of these WDs fairly well.

\subsubsection{Mixed H/He Atmosphere WDs}

\citet{bergeron01} do not find a large population of cool WDs with mixed hydrogen and helium atmospheres. Such stars would show up as outliers
in the optical and infrared color-color diagrams due to the H$_2$-He CIA, which is predicted to produce strong flux deficits in the infrared.
Our sample has half a dozen new WDs with significant absorption in the infrared (Fig. 9). Pure hydrogen and pure helium models fail to
reproduce the SEDs for these stars.
Figure 14 presents mixed H/He atmosphere model fits to eight DC WDs in our sample. The mixed H/He atmosphere models
with $\log$ (H/He) $=-5.9$ to $-3.4$ fit the observed SEDs relatively well for these stars.
Six of these targets have temperatures in the range 4500-5000 K
where there are many helium-rich DC WDs. An important implication of these temperature assignments is that
not all WDs that show infrared flux deficits are ultracool (have temperatures below 4000K).
Therefore, the classification of ultracool WDs based on photometry alone (without a detailed model atmosphere analysis) can be misleading.
A more appropriate term for WDs that show flux deficits in the infrared may be ``IR-faint''.

The best-fit model for one of the stars presented in Fig. 14, J2242+0048, implies a temperature of 3480 K.
This star is also observed in the UKIRT Infrared Deep Sky Survey. Based on the $rizYJH$ photometry,
\citet{lodieu09} find a temperature of 3820 K with a composition of equal amounts of hydrogen
and helium. Our Gemini photometry is consistent
with the UKIDSS data within the errors, but we obtain a lower temperature and a helium-dominated atmosphere using $grizJHK$ data.
There are generally two solutions for IR-faint objects.
This is because the CIA opacity peaks around $\log$ (H/He) = $-2$ and there are usually two good solutions above
and below this peak with slightly different temperatures. One of the solutions is usually better than the other one.
The cooler solution with a temperature of 3480 K explains the overall SED significantly
better than the warmer solution \citep[of][]{lodieu09} over the 0.4-2.2 $\mu$m range. Clearly, J2242+0048 is one of the coolest WDs known.

\subsubsection{Peculiar DQ WDs}

Our optical and infrared photometry sample includes two normal and two peculiar DQs presented in Figure 3.
\citet{dufour05} presented a detailed model atmosphere analysis of the DQ WDs in the SDSS Data Release 1 area
based on photometric and spectroscopic observations.
The two normal DQs in our sample are included in their study. We adopt their best-fit solutions for these two
stars. J0320$-$0716 and J2053$-$0702 have best-fit
temperatures of 6390 and 6570 K and $\log$ (C/He) $=-4.88$ and $-$5.25, respectively \citep{dufour05}.

There is considerable interest in understanding the nature of the peculiar DQ WDs. The model fits to the optical
and near-infrared SEDs of peculiar DQs indicate mixed H/He/C atmospheres \citep{bergeron94}.
Figure 15 presents our model fits to the optical and near-infrared SEDs of the two DQpec WDs in our sample.
J1442+4013 is a $T_{\rm eff}=5740$ K WD with $\log$ (H/He) = $-2.7$ and J1247+0646 is a
$T_{\rm eff}=5120$ K WD with $\log$ (H/He) = $-0.7$.
\citet{hall08} argue that the molecular bands observed in peculiar DQs are most likely pressure shifted
C$_2$ bands in a helium-rich atmosphere. However, the coolest known DQpec J1247+0646 and
the coolest known DQ WDs GSC2U J131147.2+292348 and SDSS J080843.15+464028.7 have essentially the same temperature
\citep[5120-5140 K,][]{carollo03,dufour05}. Normal DQ stars (including the coolest DQs) with helium-dominated,
high-pressure atmospheres do not show pressure shifts.
The only difference between these two classes of DQ stars
(other than the shifted bands) seems to be infrared absorption, most likely from collision induced absorption 
due to molecular hydrogen. The peculiar DQ WDs should have atmospheric pressures that are much
lower than the normal DQs as they have obviously larger opacities,
contradicting the scenario proposed
by \citet{hall08}. The observed mid-infrared flux deficits for the peculiar DQ
LHS 1126 demonstrate that the mixed H/He models have problems in the mid-infrared \citep{kilic06b,kilic08},
but mixed H/He atmospheres remain the best explanation for the observed SEDs for these WDs.
Clearly, these objects deserve further investigation.

\subsubsection{Ultracool WDs}

WDs cooler than about 4000 K can be classified as ultracool. Starting with \citet{hambly99} and
\citet{harris99}, the SDSS and various other proper motion surveys have discovered ultracool WDs.
LHS 1402 \citep{oppenheimer01}, LHS 3250 \citep{harris99}, and SDSS J1337+00 \citep{harris01} are the best studied ultracool
WDs with significant absorption in the optical and infrared. \citet{bergeron02} and \citet{bergeron05} performed detailed model
atmosphere analysis of these three stars
using $BVRI$ and $JH(K)$ photometry. While none of their fits reproduce the SEDs perfectly, they rule out pure hydrogen composition
based on the non-detection of the CIA feature near 0.8 $\mu$m. Instead, the SEDs are better fit with helium-dominated
atmosphere models with small amounts of hydrogen. They find best-fit $T_{\rm eff}=$ 3240-3480 K and $\log$ (H/He) = $-3.8$ to $-4.7$
for these stars.

We include all 13 ultracool WD candidates discovered by \citet{gates04} and
\citet[][see also Vidrih et al. 2007]{harris08} in our sample. We now have a chance
to increase the sample of well studied
ultracool WDs significantly. The optical and infrared color-color diagrams show that several of these ultracool
WDs have colors similar to the previously studied WDs LHS 3250 and SDSS J1337+00. Their $J-H$ and $H-K$ colors are also similar to
the other eight IR-faint stars in our sample. Our model fits to these eight stars imply mixed H/He
compositions. Therefore, based on the near-infrared colors, the newly observed ultracool WDs are likely to have mixed compositions as well.

Figure 16 presents our model fits to the SEDs of the five WDs from the \citet{gates04} sample assuming pure hydrogen and mixed H/He atmosphere
compositions. The SED for J0854+3503 is different than the other four stars, and our best-fit model
implies a temperature above 4000 K. Admittedly, the best-fit model is not a very
good fit to the data and there is clearly something at odds with this object.
The best fit solution has $T_{\rm eff}=$ 4070 K and $\log$ (He/H) $=-0.95$, implying
that J0854+3503 may not be an ultracool WD afterall. The unusual SED may be due to a binary WD system
and parallax observations will be helpful in understanding the nature of this system.

The SEDs for the remaining four stars in the \citet{gates04} sample, J0947+4459, J1001+3903, J1220+0914, and
J1403+4533 are similar to the SEDs for LHS 3250 and SDSS 1337+00.
The best-fit pure hydrogen models predict strong absorption bumps in the near-infrared including a strong absorption feature at 0.8 $\mu$m, which
is not observed in any of these targets. Pure helium models also fail to reproduce the SEDs as these stars all show significant absorption
in the optical and infrared. The solid lines in Figure 16 show the fits using mixed H/He atmosphere models.
The best-fit model temperatures are 2670-3410 K with $\log$ (H/He) = $-2.7$ to $-5.1$. These are similar to the best-fit
model solutions for LHS 1402, LHS 3250, and SDSS J1337+00. 

Figure 17 presents our model fits to the SEDs of two ultracool WD candidates from \citet{harris08} and one from
\citet{vidrih07}. The observed optical and near-infrared
SEDs of J0310$-$0110 and J1452+4522 are best explained with pure helium atmosphere models with $T_{\rm eff}=$ 4970 and 5780 K and that of J2239+0018B is best
explained as a pure hydrogen atmosphere WD with $T_{\rm eff}= 4440$ K. J0310$-$0110 and J1452+4522 were classified as ultracool based on the SDSS photometry
and spectroscopy. However, our near-infrared observations show that these two stars do not display infrared flux deficits, and they are relatively warm WDs.
Our model fit for J1452+4522 is unusual in the sense that the observations behave like
there is an additional opacity source in the UV, although we are not aware of similar objects.
Using mixed H/He models do not improve the fits. The best-fit mixed atmosphere model
has $T_{\rm eff}=4730$ K and $\log$ (H/He) = $-4.8$ (assuming $\log g=8$), but this model significantly
over(under)-predicts the $J(K)-$band photometry.
J2239+0018B is a common proper motion companion to the ultracool WD candidate J2239+0018A. \citet{vidrih07} identified J2239+0018B as an ultracool WD based
on $K-$band photometry from the UKIDSS survey. Our observations in the $JH$ bands, UKIDSS photometry in the $K-$band, and a detailed model atmosphere
analysis demonstrate that J2239+0018B is not an ultracool WD.

\citet{harris08} report narrow H$\alpha$ (and possibly H$\beta$, see Figure 17) emission in the SDSS spectrum of J0310$-$0110 indicating the presence of a substellar
companion. Our model fit for this star is consistent with a cool He-atmosphere WD with no sign of a companion.
Based on our age and distance estimates of 6.1 Gyr and 117 pc, J0310$-$0110 has $M_{\rm K}=$ 13.2 mag. 
The 3$\sigma$ limit of the $K-$band photometry limits possible companions to fainter than about $M_{\rm K}=$ 16.3 mag.
At 5 Gyr of age, a 0.05 $M_\odot$ companion would have an absolute magnitude of $M_{\rm K}=$ 16.4 mag \citep{baraffe03}.
Thus, if J0310$-$0110 has a companion, it must be less massive than about 0.05 $M_\odot$.

Figure 18 presents our model fits to the SEDs of the remaining five stars from the \citet{harris08} sample. The observed SEDs for these stars
are best fit with mixed H/He atmosphere models with $T_{\rm eff}=$ 2290-4160 K and $\log$ (H/He) = $-2.3$ to $-7.8$.
While these fits are better than the fits using pure hydrogen or pure helium models, they are not perfect.
Like LHS 1402, LHS 3250, and SDSS J1337+00, the peaks of the energy distributions of the ultracool WDs in our sample (Figure 16 and 18) are predicted
too sharp compared to the observations. The best-fit model temperature for J1238+3502 is unusually low (2290 K) and
the required helium abundance is relatively high. However, only $J-$band observations are available in the infrared,
and our model fit parameters may be improved with additional near-infrared observations (although such observations
are difficult to obtain since J1238+3502 is relatively faint in the near-infrared).
The poor model-fit for J1251+4403 is similar to that of J1403+4533; the observations peak at a bluer wavelength compared
to the best-fit model. The $u-$ and $g-$band photometry suggests very cool ($\sim$ 2000 K) WDs, but the $riz$ and infrared photometry
require a hotter WD. This is probably an indication that the CIA opacities are wrong at such low temperatures.
Despite the fact that our current model atmospheres do not find perfect fits to the observed photometry, we
can rule out extreme hydrogen-rich compositions for these stars based on the current CIA opacity calculations.

\section{Discussion}

\subsection{Chemical Abundance Patterns}

Our detailed model atmosphere analysis of 126 cool WDs with optical and near-infrared photometry shows that 61 stars (48\%) have pure hydrogen,
44 stars (35\%) have pure helium (including DQ WDs with helium-dominated atmospheres), and 21 stars have mixed H/He atmospheres.
The latter include 10 ultracool WD candidates and 2 peculiar DQ WDs. Based on a detailed model atmosphere analysis of 150 WDs with $T_{\rm eff}\leq$
12,000 K, \citet{bergeron01} find the frequency of pure hydrogen and pure helium atmosphere WDs to be 64\% and 33\%, respectively.

Figure 19 presents the surface composition measurements for our sample and the \citet{bergeron01} sample of WDs.
\citet{bergeron01} find helium-rich atmosphere WDs down to about 4500 K and hydrogen-rich WDs down to 4000 K. 
The coolest and oldest WDs are likely to accrete from the interstellar medium in their $\sim$10 Gyr lifetimes. The lack of
pure helium WDs below 4500 K supports this scenario. \citet{bergeron01} also find a non-DA
gap (or a deficiency in number) between about 5000 K and 6000 K. They find non-DA stars below and above this temperature range,
but they find only three non-DA stars in the gap.
In addition, they do not find a large population of mixed H/He atmosphere WDs. In contrast, our sample is restricted to stars
cooler than about 6600 K, and the fraction of mixed H/He atmosphere WDs is larger.
Our sample fills in the non-DA gap
somewhat. However, there is still a gap between 5600 K and 6200 K in both our and the \citet{bergeron01} sample
(though a selection bias is evident in our sample, which has only a few stars warmer than 6000 K).
None of our WD targets identified as pure helium atmosphere objects with $T_{\rm eff}=$ 5400-5600 K have optical spectroscopy available.
Given the slight differences between the pure hydrogen and pure helium model SEDs for this temperature range, our pure helium
classification based on the SED fits should be taken with caution. Therefore, the non-DA gap may extend down to 5400 K.
Perhaps another important piece of evidence for the existence of this non-DA gap comes from the work by \citet{kilic06} who identified
5 DCs in the non-DA gap. So far four of these DCs are observed in the near-infrared \citep[in this work
and also in][]{kilic09}, and it turns out that all four stars
have mixed H/He atmospheres. The SEDs of three of these stars, J0157+1335, J1648+3939, and J1722+5752, are shown in Figure 14.
In addition, J1203+0426 is identified as a mixed H/He atmosphere WD by \citet{kilic09}. Our model fits with mixed H/He atmospheres
place these stars outside of the non-DA gap.

Our model fits imply that a significant fraction of WDs in the temperature range 4500-5000 K
are He-rich. Since H$\alpha$ is invisible at these temperatures, the choice of composition depends on the
quality of the fits to the SEDs. The best H-rich model fit is sometimes not too different from
the He-rich model fit. It is possible that
small shifts in the $ugriz$ and $JHK$ calibration may explain the overabundance of He-rich objects in
this temperature range. However, a similar number distribution is also evident in the \citet{bergeron01}
sample, which relies on $BVRI$ and $JHK$ photometry. The non-inclusion of the Ly $\alpha$ opacity in our models
or problems with the CIA calculations may cause incorrect assignment of the atmospheric
types to WDs in this temperature range and further work is required to understand if the observed
overabundance of He-rich atmosphere WDs at this temperature range is real.

Overall, our analysis reveals a complex spectral evolutionary history for cool WDs \citep[see also][]{bergeron01}.
The model atmosphere analysis by \citet{kowalski06b} presents a completely different picture, in which
WDs below 6000 K are hydrogen-rich. They come to this conclusion by excluding the DQ and DZ WDs from their sample and
also using a different set of pure helium atmosphere models that have colors essentially the same as blackbodies.
Since the cool WD SEDs are not blackbodies, they assign
hydrogen-rich composition for most cool WDs and they propose a simple evolutionary scenario in which
WDs accrete hydrogen from the interstellar medium and turn into hydrogen-rich WDs even if they start with
pure helium atmospheres. Our pure helium atmosphere models have colors slightly different than simple blackbodies
(see Figure 8) and more similar to the observed colors of cool DC WDs. 
Resolving the discrepancy between our interpretation and the \citet{kowalski06b} results
requires a thorough study of the differences between these models. 
DZ WDs are the only cool ($T_{\rm eff}<5000$ K) WDs with atomic absorption lines.
A detailed model atmosphere
analysis of cool DZs \citep[e.g.][]{dufour07} would be a crucial test for identifying problems
with both sets of models.

\subsection{Stellar Ages}

Our sample is complete for the faint end of the luminosity function ($M_{\rm bol}\geq 14.6$ mag) from
\citet[][although this luminosity function is limited to $V_{\rm tan}\geq 30$ km s$^{-1}$ objects and is not complete itself]{harris06}
and it can
be used to constrain the age of the Galactic disk.
WD cooling rates depend on the radius (mass) of the star. Without a parallax measurement, we cannot reliably determine
the radii and masses of our targets, and hence ages. However, the average mass for the 48 WDs cooler than 6000 K
in the \citet{bergeron01} sample of WDs with parallax measurements is 0.61 $M_\odot$. Therefore,
our assumption of $\log g=8$ is
reasonable. Based on our model fits, the implied cooling ages and distances for our sample are given in Table 3. 
The predicted distances range from 15 pc to 130 pc.
Given the uncertainties in the distance estimates, several targets may be part of the 20 pc sample.
The closest star in our sample is J2325+1403, predicted to be
at 15.4 pc. \citet{lepine09} present trigonometric parallax observations of J2325+1403,
which indicate that it is at 22 pc. Using this distance measurement in our model fits, J2325+1403
has $T_{\rm eff}= 5070$ K and $\log g= 7.39$. Not surprisingly, its mass (0.28 $\pm$ 0.03 $M_\odot$)
has to be significantly lower than the average ($\log g=8$ or $M=0.58 M_\odot$) to match the distance.
Alternatively, J2325+1403 could be an unresolved double
degenerate binary since it is overluminous.

The estimated WD cooling ages for our sample range from 1.8 Gyr to 12.1 Gyr.
Excluding the ultracool WDs \citep[due to poor model fits and relatively uncertain ages,][]{bergeron02} and
J2242+0048 (with strong infrared absorption),
the coolest WDs in our sample have temperatures around 4200 K (see Fig. 13).
This is comparable to the coolest WDs in the \citet{bergeron01} sample.
Figure 20 shows the age distribution of our sample of WDs (excluding the ultracool WD candidates)
compared to that of \citet{bergeron01} sample, which also includes the majority of the objects from
the \citet{leggett98} WD luminosity function. Compared to the \citet{bergeron01} sample, we
have a significantly larger number of stars older than 5 Gyr. Both samples show a significant drop
in the number of stars at 8 Gyr.
Even though the individual ages for our targets
cannot be trusted due to the unknown masses, the average mass for our sample should be about 0.6 $M_\odot$ and
the {\it average age} for the oldest stars in our sample should be reliable.
Adding 1.4 Gyr for the main-sequence lifetime of the 2 $M_\odot$ solar-metallicity progenitor stars \citep{marigo08} brings the total age
to about 9.4 Gyr, entirely consistent with the oldest WDs in Table 2 of \citet{leggett98} and the Galactic
disk age of 8 $\pm$ 1.5 Gyr.

There are four common-proper motion WD pairs in our sample. These systems provide a consistency
check in our distance and age measurements. Two of these sytems, J0947+4459 and J2239+0018, include ultracool WDs and are discussed in section 4.3.
We derive distances of 52 and 54 pc and cooling ages of 2.6 and 6.3 Gyr for the two WDs in the J0747+2438 system, respectively.
The difference in cooling ages can be explained by
a small mass difference (of order 0.3 $M_\odot$) between the two stars.
We derive distances of 51 and 82 pc and cooling ages of 2.6 and 7.8 Gyr for the two WDs in the J2321+0102 system.
However, these differences can also be explained by a small mass difference between the two stars in the system.
Parallax measurements will be useful to constrain the individual masses in these systems.

\subsection{Disk or Halo Membership}

\citet{bergeron05} demonstrate the importance of determining total stellar ages in order to associate any WD with the
thick disk or halo. 
Given the total estimated ages of $<$ 10 Gyr, the majority of the objects in our sample most likely belong to the disk population. 
Figure 21 shows the Galactic UV velocities of our targets (based on the model fits assuming $\log g=8$) compared to the velocity ellipses of the disks and
the halo \citep{chiba00}. All but two of the objects in our sample, including the ultracool WDs, have velocities consistent with the thin disk or thick disk objects. 
\citet{reid05} argue that the thick disk is likely to contribute about 20\% of the solar neighborhood WDs, and these WDs should dominate at faint
absolute magnitudes. About 27\% of our targets have UV velocities inconsistent with the thin disk population at 2$\sigma$, and these objects
may belong to the thick disk. However, this fraction is very uncertain due to the lack of parallax measurements.

J0301$-$0044 and J1255+4655 have UV velocites that are inconsistent with the disk population at more than 2$\sigma$, and they are
halo WD candidates. Both have $T_{\rm eff}\approx4500$ K with WD cooling ages of 7.1-7.3 Gyr and total ages
$\approx$ 8 Gyr for average mass WDs. These ages are relatively young for halo objects. However, a slightly lower mass around
0.5 $M_\odot$ would make the total ages for J0301$-$0044 and J1255+4655 consistent with the halo population.

\subsection{Ultracool WDs}

Our model fits to the SEDs of ultracool WDs are not perfect, but the best-fit models imply mixed H/He atmosphere composition.
The near-infrared colors of ultracool WDs are similar to the relatively warmer WDs that also show flux deficits in the near-infrared.
In addition, the observed WD sequence in $r-J$ vs. $J-H$ color-color diagram suggests that
the ultracool WD sequence is an extension of the warmer WDs with near-infrared flux deficits. Since our mixed H/He atmosphere models
fit the SEDs of warmer WDs with flux deficits fairly well, there is a high probability that the ultracool WDs also have mixed H/He atmospheres.
Based on our model fits with $\log g=8$, we estimate that the 10 ultracool WD candidates presented in Figure 16 and 18
are 24-66 pc away and they have
WD cooling ages in the range 8.0-12.1 Gyr. These would correspond to total ages (including main-sequence lifetimes) of $\approx$ 9-13 Gyr.
The implied tangential velocities are in the range 16-92 km s$^{-1}$, indicating that these ultracool WDs
most likely belong to the thick disk population.

Trigonometric parallax measurements are available for only one ultracool WD so far. The distance measurement of 30.3 $\pm$ 0.5 pc for LHS 3250 \citep{harris99}
implies that it is brighter than expected for a 0.6 $M_\odot$ WD, unless it is warmer by about 700 K.
Given the similarities between the SEDs of LHS 3250 and the ultracool WDs in our sample, it is likely that these WDs are also brighter,
more distant, and younger than our model fits with $\log g=8$ imply. In fact, our preliminary parallax observations at the MDM 2.4m
telescope for a few of these ultracool WDs show that the distances are underestimated in our model fits. Hence, they are likely
to be less massive than 0.6 $M_\odot$ (perhaps a binary origin) or hotter than currently predicted from the model fits.
Further improvements in the CIA opacity and model atmosphere calculations may help in matching the observed SEDs and 
luminosities. Until then, the nature of these stars remain uncertain.

Two of the ultracool WDs, J0947+4459 and J2239+0018, have common proper motion companions.
The model fits to the J2239+0018A/B system are rather good and the ages (8.9 and 7.8 Gyr) and distances (60 and 79 pc) agree reasonably well.
This is encouraging because one of the components is a mixed H/He atmosphere WD and the other is a very cool H-rich WD.
For an average mass WD, J2239+0018B has a total age (WD cooling + main-sequence lifetime) of about 9.2 Gyr.

The optical spectrum of the companion for J0947+4459 displays weak H$\alpha$ absorption.
Figure 22 shows our pure hydrogen atmosphere model fits to the J0947+4500 SED.
Omitting the $u-$ and $g-$band photometry, the best-fit temperature is 5095 $\pm$ 130 K.
This corresponds to a distance of 61 pc and a WD cooling age of 5.2 Gyr.
The observed H$\alpha$ absorption is consistent with the model prediction.
The ultracool WD J0947+4459 has an estimated distance of 39 pc and an age of 9.5 Gyr.
If the ultracool WD is slightly lower-mass or hotter, it  would be more distant (at 61 pc),
brighter, and its implied WD cooling age would be similar to its companion. 
Of course, all of this can be resolved
when accurate parallax measurements are available.

\section{Conclusions}

We present follow-up optical spectroscopy and near-infrared photometry of the cool and ultracool WDs in the SDSS DR3 that are
identified by \citet{kilic06}, \citet{gates04}, and \citet{harris06,harris08}. We demonstrate that a clean selection of WDs is possible
using a reduced proper motion diagram and a tangential velocity cut-off of 30 km s$^{-1}$. This can be used to select large
samples of cool WDs from the SDSS DR7 and SEGUE data or any other survey that uses the SDSS filter set.

Our near-infrared observations reveal eight new stars with significant absorption. All of these stars are best explained with mixed H/He atmosphere
models. The infrared colors of ultracool WDs are similar to these eight stars as well. Hence, there is indirect evidence that they
also have mixed H/He atmospheres. Our model fits to the ultracool WD SEDs show that they may be as cool as $\approx$ 2300 K and as old
as 13 Gyr (including main-sequence age of 1 Gyr for a 2$M_\odot$ thick disk star). Their implied tangential velocities and Galactic
space velocities are consistent with the thick disk objects. 
Further progress in understanding the ultracool white dwarfs and estimating
reliable temperatures, masses, and ages can be achieved with the help of trigonometric parallax observations. Such observations
for the ultracool WDs identified by \citet{gates04} and \citet{harris08} are currently underway at the MDM 2.4m and the USNO 1.55m telescopes.

Only two objects in our sample have space velocities consistent with the halo population. Trigonometric parallax observations for these
targets will be required to confirm their halo membership. The absence of more halo WD candidates in our sample is not surprising as our survey is
limited to objects brighter than about $g=19.5$ mag due to the brightness limit of the Palomar Observatory Sky Survey plates. A deeper,
wide-field astrometric survey is currently being conducted at the Bok 90" and USNO 1.3m telescopes \citep{liebert07,munn10}.
Initial follow-up observations of candidates from this survey are already finding the elusive old halo WD candidates in
the solar neighborhood \citep{kilic10}. This survey, along with the Pan-Starrs and LSST surveys, will be an
invaluable resource for halo WD studies. 

\acknowledgements
We thank J. Liebert for many useful discussions, Andy Stephens for the derivation
of the NIRI nonlinearity correction, and an anonymous referee for useful suggestions.
Support for this work was provided by NASA through the Spitzer Space Telescope Fellowship Program,
under an award from Caltech. This material is also based on work supported by the National Science
Foundation under grants AST-0607480 and AST-0602288 and by the NSERC Canada and by the Fund FQRNT (Qu\'ebec).
SKL's research is supported by Gemini Observatory.
P. Bergeron is a Cottrell Scholar of Research Corporation for Science Advancement.

The Hobby-Eberly Telescope (HET) is a joint project of the University of Texas at Austin, The Pennsylvania State
University, Stanford University, Ludwig-Maximilians-Universitat Munchen, and Georg-August-Universitat Gottingen. The
HET is named in honor of its principal benefactors, William P. Hobby and Robert E. Eberly.
The Gemini observatory is operated by the 
Association of Universities for Research in Astronomy, Inc., under a cooperative agreement 
with the NSF on behalf of the Gemini partnership: the National Science Foundation (United 
States), the Science and Technology Facilities Council (United Kingdom), the 
National Research Council (Canada), CONICYT (Chile), the Australian Research Council (Australia), 
Minist\'{e}rio da Ci\^{e}ncia e Tecnologia (Brazil) 
and Ministerio de Ciencia, Tecnolog\'{i}a e Innovaci\'{o}n Productiva (Argentina).
The IRTF is operated by the University of Hawaii under Cooperative Agreement no. NCC 5-538 with the
National Aeronautics and Space Administration, Office of Space Science, Planetary Astronomy Program.

\begin{deluxetable}{lcccccccccc}
\tabletypesize{\scriptsize}
\tablecolumns{11}
\tablewidth{0pt}
\tablecaption{Optical and Near-Infrared Photometry of Cool White Dwarfs}
\tablehead{
\colhead{Name (SDSS J)}&
\colhead{$u$}&
\colhead{$g$}&
\colhead{$r$}&
\colhead{$i$}&
\colhead{$z$}&
\colhead{$J$}&
\colhead{$H$}&
\colhead{$K$}&
\colhead{Type}&
\colhead{Source}
}
\startdata
00:03:16.69$-$01:11:17.9 & 20.53 & 19.32 & 18.79 & 18.56 & 18.52 & 17.63 $\pm$ 0.04 & 17.31 $\pm$ 0.04 & 17.35 $\pm$ 0.05 & DA & 1\\
00:33:00.80+14:51:09.8 & 19.27 & 18.56 & 18.13 & 17.99 & 18.01 & 17.39 $\pm$ 0.05 & 17.45 $\pm$ 0.08 & \nodata & \nodata & \nodata\\
00:45:21.88+14:20:45.3 & 20.64 & 19.20 & 18.45 & 18.20 & 18.10 & 17.24 $\pm$ 0.04 & 16.99 $\pm$ 0.04 & 16.89 $\pm$ 0.04 & DZA & 1\\
01:02:59.98+14:01:08.1 & 21.21 & 19.48 & 18.69 & 18.41 & 18.30 & 17.39 $\pm$ 0.04 & 17.05 $\pm$ 0.04 & 17.05 $\pm$ 0.06 & DC & 1\\
01:46:29.01+14:04:38.2 & 21.21 & 19.99 & 19.37 & 19.24 & 19.71 & 19.56 $\pm$ 0.05 & 20.07 $\pm$ 0.12 & \nodata          & DC & 3\\
01:57:43.25+13:35:58.2 & 20.56 & 19.35 & 18.65 & 18.47 & 18.41 & 17.75 $\pm$ 0.04 & 17.53 $\pm$ 0.04 & 17.49 $\pm$ 0.06 & DC & 1\\
02:12:06.36$-$00:40:05.8 & 19.80 & 18.96 & 18.59 & 18.41 & 18.39 & 17.60 $\pm$ 0.04 & 17.27 $\pm$ 0.04 & 17.42 $\pm$ 0.05 & \nodata & \nodata\\
02:50:05.81$-$09:10:02.8 & 20.02 & 18.96 & 18.45 & 18.27 & 18.21 & 17.41 $\pm$ 0.04 & 17.14 $\pm$ 0.04 & 17.03 $\pm$ 0.06 & DA & 1\\
02:56:41.62$-$07:00:33.8 & 20.74 & 19.00 & 18.13 & 17.79 & 17.69 & 16.71 $\pm$ 0.05 & 16.62 $\pm$ 0.05 & 16.48 $\pm$ 0.06 & DC & 1\\
03:01:44.09$-$00:44:39.5 & 22.23 & 20.43 & 19.38 & 18.99 & 18.92 & 17.96 $\pm$ 0.04 & 17.73 $\pm$ 0.04 & 17.68 $\pm$ 0.08 & DC & 2\\
03:07:13.91$-$07:15:06.2 & 18.65 & 17.65 & 17.18 & 17.01 & 16.98 & 16.20 $\pm$ 0.04 & 15.95 $\pm$ 0.04 & 15.87 $\pm$ 0.05 & DA & 4\\
03:09:24.87+00:25:25.3 & 19.15 & 18.19 & 17.72 & 17.53 & 17.50 & 16.64 $\pm$ 0.04 & 16.54 $\pm$ 0.04 & 16.87 $\pm$ 0.04 & DC & 1\\
03:10:49.53$-$01:10:35.3 & 22.49 & 20.95 & 20.20 & 19.89 & 19.97 & 18.94 $\pm$ 0.02 & 18.73 $\pm$ 0.02 & 18.58 $\pm$ 0.02 & DC & 3\\
03:14:49.81$-$01:05:19.3 & 19.53 & 18.59 & 18.14 & 17.98 & 17.97 & 17.08 $\pm$ 0.04 & 16.82 $\pm$ 0.04 & 16.85 $\pm$ 0.06 & DA & 1\\
03:20:54.11$-$07:16:25.4 & 19.93 & 19.75 & 19.27 & 19.21 & 19.19 & 18.80 $\pm$ 0.04 & 18.69 $\pm$ 0.05 & 18.62 $\pm$ 0.06 & DQ & 3\\
03:30:54.88+00:37:16.5 & 20.72 & 19.79 & 19.32 & 19.16 & 19.10 & 18.35 $\pm$ 0.04 & 18.16 $\pm$ 0.05 & 18.09 $\pm$ 0.08 & DA & 1\\
04:06:47.32$-$06:44:36.9 & 18.83 & 18.02 & 17.58 & 17.48 & 17.40 & 16.59 $\pm$ 0.05 & 16.44 $\pm$ 0.04 & 16.27 $\pm$ 0.05 & DA & 1\\
07:47:21.56+24:38:47.7 & 21.09 & 19.27 & 18.54 & 18.26 & 18.17 & 17.16 $\pm$ 0.04 & 16.99 $\pm$ 0.04 & 16.85 $\pm$ 0.04 & DC & 2\\
07:47:23.50+24:38:23.7 & 19.34 & 18.36 & 17.92 & 17.75 & 17.72 & 16.78 $\pm$ 0.04 & 16.58 $\pm$ 0.04 & 16.53 $\pm$ 0.05 & DA & 2\\
07:53:13.28+42:30:01.6 & 19.97 & 18.09 & 17.19 & 16.87 & 16.75 & 15.69 $\pm$ 0.04 & 15.49 $\pm$ 0.04 & 15.47 $\pm$ 0.04 & DC & 1\\
08:01:32.83+39:49:25.9 & 21.88 & 20.17 & 19.37 & 19.09 & 18.96 & \nodata & \nodata & \nodata & DC & 2\\
08:04:40.63+22:39:48.7 & 19.73 & 18.30 & 17.59 & 17.39 & 17.33 & 16.71 $\pm$ 0.04 & 16.92 $\pm$ 0.04 & 17.29 $\pm$ 0.06 & DZ & 2\\
08:17:45.33+24:51:05.5 & 20.59 & 19.44 & 18.91 & 18.74 & 18.62 & 17.93 $\pm$ 0.04 & 17.76 $\pm$ 0.04 & 17.71 $\pm$ 0.09 & \nodata & \nodata\\
08:17:51.52+28:22:03.1 & 21.59 & 19.49 & 18.61 & 18.30 & 18.22 & 17.33 $\pm$ 0.04 & 17.01 $\pm$ 0.04 & 16.91 $\pm$ 0.09 & DC & 2\\
08:19:24.32+31:59:56.8 & 21.65 & 19.75 & 18.90 & 18.59 & 18.43 & 17.47 $\pm$ 0.04 & 17.32 $\pm$ 0.04 & 17.21 $\pm$ 0.09 & DC & 2\\
08:21:08.18+37:27:38.3 & 20.68 & 19.14 & 18.43 & 18.15 & 18.04 & 17.25 $\pm$ 0.04 & 17.00 $\pm$ 0.04 & 16.85 $\pm$ 0.05 & DA & 2\\
08:25:00.61+28:41:49.9 & 20.14 & 18.98 & 18.44 & 18.29 & 18.18 & 17.38 $\pm$ 0.04 & 17.12 $\pm$ 0.04 & 17.03 $\pm$ 0.04 & \nodata & \nodata\\
08:25:19.70+50:49:20.1 & 21.09 & 19.34 & 18.43 & 18.09 & 18.00 & 17.08 $\pm$ 0.04 & 16.83 $\pm$ 0.04 & 16.74 $\pm$ 0.04 & DC & 1\\
08:28:42.31+35:27:29.5 & 21.43 & 19.85 & 19.05 & 18.73 & 18.71 & \nodata & \nodata & \nodata & DC & 2\\
08:36:41.56+45:56:58.7 & 21.64 & 20.01 & 19.14 & 18.87 & 18.73 & 17.90 $\pm$ 0.04 & 17.70 $\pm$ 0.04 & 17.59 $\pm$ 0.05 & DC & 1\\
08:38:31.82+28:04:59.7 & 20.55 & 19.14 & 18.49 & 18.29 & 18.21 & 17.33 $\pm$ 0.04 & 17.07 $\pm$ 0.04 & 17.04 $\pm$ 0.05 & DC & 2\\
08:53:45.93+41:18:50.1 & 20.38 & 19.33 & 18.75 & 18.59 & 18.60 & 17.83 $\pm$ 0.04 & 17.70 $\pm$ 0.05 & 17.67 $\pm$ 0.06 & \nodata & \nodata\\
08:54:54.45+29:26:41.8 & 20.80 & 19.33 & 18.63 & 18.48 & 18.36 & \nodata & \nodata & \nodata & DA & 2\\
08:54:43.33+35:03:52.7 & 23.57 & 20.53 & 19.39 & 19.09 & 18.95 & 18.44 $\pm$ 0.04 & 18.23 $\pm$ 0.04 & 17.98 $\pm$ 0.04 & DC & 5\\
08:55:49.89+37:00:16.7 & 19.01 & 17.97 & 17.47 & 17.29 & 17.27 & 16.37 $\pm$ 0.05 & 16.17 $\pm$ 0.04 & 16.08 $\pm$ 0.05 & \nodata & \nodata\\
09:02:44.02+56:30:32.7 & 20.84 & 19.56 & 18.88 & 18.70 & 18.52 & \nodata & \nodata & \nodata & DA & 2\\
09:03:04.38+08:38:15.6 & 20.06 & 18.85 & 18.27 & 18.08 & 17.99 & \nodata & \nodata & \nodata & DA & 2\\
09:04:06.89+34:03:53.1 & 20.01 & 18.73 & 18.10 & 17.87 & 17.85 & 17.23 $\pm$ 0.04 & 16.93 $\pm$ 0.05 & 16.85 $\pm$ 0.06 & DC & 2\\
09:09:14.56+47:00:17.5 & 20.64 & 19.29 & 18.74 & 18.50 & 18.42 & 18.11 $\pm$ 0.04 & 18.62 $\pm$ 0.07 & 19.10 $\pm$ 0.10 & DC & 2\\
09:33:45.58+37:43:49.8 & 20.71 & 19.17 & 18.59 & 18.42 & 18.36 & 17.63 $\pm$ 0.04 & 17.37 $\pm$ 0.04 & 17.26 $\pm$ 0.04 & DZA & 2\\
09:42:44.96+44:37:43.1 & 21.37 & 19.47 & 18.58 & 18.22 & 18.05 & 17.15 $\pm$ 0.04 & 16.97 $\pm$ 0.04 & 16.86 $\pm$ 0.04 & DC & 1\\
09:43:16.62+51:34:40.9 & 20.52 & 18.89 & 18.13 & 17.84 & 17.75 & 16.72 $\pm$ 0.05 & 16.61 $\pm$ 0.06 & 16.53 $\pm$ 0.06 & DC & 2\\
09:47:23.00+44:59:48.7 & 20.67 & 19.44 & 18.84 & 18.95 & 19.45 & 19.69 $\pm$ 0.07 & 20.34 $\pm$ 0.06 & 20.96 $\pm$ 0.12 & DC & 5\\
09:47:24.47+45:00:01.9 & 21.27 & 19.52 & 18.77 & 18.53 & 18.32 & 17.43 $\pm$ 0.04 & 17.24 $\pm$ 0.04 & 17.11 $\pm$ 0.04 & DA & 5\\
10:00:29.47+42:36:31.2 & 21.70 & 19.58 & 18.79 & 18.46 & 18.39 & 17.47 $\pm$ 0.04 & 17.23 $\pm$ 0.04 & 17.13 $\pm$ 0.05 & DC & 2\\
10:01:03.42+39:03:40.5 & 21.36 & 20.05 & 19.60 & 20.02 & 20.61 & 20.65 $\pm$ 0.06 & 21.05 $\pm$ 0.07 & \nodata & DC & 5\\
10:01:19.48+46:56:50.6 & 21.34 & 19.27 & 18.22 & 17.90 & 17.82 & 16.79 $\pm$ 0.04 & 16.69 $\pm$ 0.06 & 16.99 $\pm$ 0.06 & DC & 1\\
10:02:04.11+43:26:45.7 & 20.40 & 18.56 & 17.74 & 17.45 & 17.30 & \nodata & \nodata & \nodata & DC & 2\\
10:02:25.85+61:08:58.1 & 21.70 & 19.40 & 18.42 & 18.06 & 17.87 & 16.85 $\pm$ 0.06 & 16.69 $\pm$ 0.07 & 16.72 $\pm$ 0.05 & DC & 1\\
10:19:59.36+52:14:08.4 & 19.83 & 18.51 & 17.95 & 17.73 & 17.67 & \nodata & \nodata & \nodata & DA & 2\\
10:42:44.79+49:32:47.0 & 20.62 & 19.17 & 18.59 & 18.40 & 18.31 & 17.51 $\pm$ 0.04 & 17.32 $\pm$ 0.04 & 17.32 $\pm$ 0.05 & DC & 2\\
10:49:39.97+45:43:57.5 & 21.97 & 20.04 & 19.18 & 18.87 & 18.74 & \nodata & \nodata & \nodata & DC & 2\\
10:58:35.49+08:18:28.6 & 20.17 & 19.45 & 18.67 & 18.15 & 17.81 & \nodata & \nodata & \nodata & WD+dM & 2\\
11:02:13.70+67:07:52.6 & 21.37 & 19.62 & 18.95 & 18.68 & 18.65 & 17.76 $\pm$ 0.04 & 17.49 $\pm$ 0.04 & 17.41 $\pm$ 0.05 & DC & 1\\
11:02:29.26+40:30:04.8 & 20.44 & 18.90 & 18.24 & 18.00 & 17.87 & 17.09 $\pm$ 0.04 & 16.78 $\pm$ 0.04 & 16.67 $\pm$ 0.04 & DA & 2\\
11:04:00.98+04:36:16.8 & 22.06 & 19.90 & 18.96 & 18.54 & 18.46 & 17.42 $\pm$ 0.05 & 17.31 $\pm$ 0.06 & 17.26 $\pm$ 0.06 & DC & 2\\
11:07:31.38+48:55:23.0 & 21.50 & 19.49 & 18.54 & 18.23 & 18.11 & 17.05 $\pm$ 0.05 & 16.95 $\pm$ 0.07 & 16.86 $\pm$ 0.07 & DC & 1\\
11:16:08.81+09:25:32.6 & 20.26 & 19.03 & 18.47 & 18.25 & 18.22 & 17.31 $\pm$ 0.04 & 17.09 $\pm$ 0.04 & 16.98 $\pm$ 0.05 & DA & 2\\
11:17:08.63+50:10:33.9 & 21.17 & 19.34 & 18.57 & 18.30 & 18.16 & 17.24 $\pm$ 0.04 & 17.07 $\pm$ 0.04 & 16.97 $\pm$ 0.05 & DC & 2\\
11:19:40.62$-$01:07:55.1 & 21.99 & 19.95 & 19.06 & 18.80 & 18.65 & 17.76 $\pm$ 0.04 & 17.55 $\pm$ 0.04 & 17.41 $\pm$ 0.05 & DC & 1\\
11:30:50.44+10:02:59.3 & 20.41 & 18.89 & 18.23 & 17.98 & 17.89 & 17.10 $\pm$ 0.04 & 16.85 $\pm$ 0.05 & 16.77 $\pm$ 0.06 & DA & 2\\
11:47:09.81+43:03:06.4 & 20.60 & 19.09 & 18.33 & 18.04 & 17.95 & 17.04 $\pm$ 0.04 & 16.84 $\pm$ 0.04 & 16.72 $\pm$ 0.04 & DC & 2\\
11:51:16.42+12:53:45.6 & 20.72 & 19.38 & 18.80 & 18.57 & 18.48 & 17.64 $\pm$ 0.04 & 17.41 $\pm$ 0.04 & 17.45 $\pm$ 0.05 & DC & 6\\
11:58:14.52+00:04:58.3 & 20.86 & 18.89 & 17.85 & 17.54 & 17.34 & 16.36 $\pm$ 0.04 & 16.31 $\pm$ 0.05 & 16.18 $\pm$ 0.05 & DC & 2\\
12:02:00.48$-$03:13:47.4 & 22.34 & 19.97 & 19.08 & 18.77 & 18.68 & 17.61 $\pm$ 0.05 & 17.55 $\pm$ 0.08 & 17.51 $\pm$ 0.09 & DC & 1\\
12:04:39.54+62:22:16.4 & 20.91 & 19.25 & 18.43 & 18.14 & 18.06 & 17.07 $\pm$ 0.04 & 16.86 $\pm$ 0.04 & 16.80 $\pm$ 0.04 & DC & 1\\
12:08:15.60+08:45:43.1 & 20.33 & 18.75 & 18.04 & 17.79 & 17.69 & \nodata & \nodata & \nodata & DA & 2\\
12:11:18.81+07:24:47.5 & 18.51 & 17.15 & 16.53 & 16.33 & 16.24 & \nodata & \nodata & \nodata & DA & 2\\
12:12:02.47+43:45:09.8 & 20.52 & 19.36 & 18.81 & 18.56 & 18.52 & 17.69 $\pm$ 0.04 & 17.45 $\pm$ 0.04 & 17.28 $\pm$ 0.05 & DA & 2\\
12:12:07.01+04:40:12.0 & 22.07 & 20.04 & 19.09 & 18.79 & 18.66 & 17.67 $\pm$ 0.04 & 17.50 $\pm$ 0.04 & 17.50 $\pm$ 0.05 & DC & 2\\
12:14:51.48$-$01:42:11.3 & 20.93 & 19.49 & 18.85 & 18.61 & 18.50 & \nodata & \nodata & \nodata & DA & 2\\
12:20:48.66+09:14:12.3 & 22.34 & 20.35 & 19.33 & 19.47 & 19.91 & 19.97 $\pm$ 0.07 & 20.68 $\pm$ 0.07 & \nodata & DC & 5\\
12:20:52.87+45:19:41.7 & 20.59 & 19.45 & 18.94 & 18.74 & 18.63 & 17.83 $\pm$ 0.04 & 17.57 $\pm$ 0.04 & 17.52 $\pm$ 0.05 & \nodata & \nodata\\
12:34:08.12+01:09:47.4 & 21.23 & 19.82 & 19.26 & 19.03 & 19.01 & 18.19 $\pm$ 0.04 & 17.91 $\pm$ 0.04 & 17.79 $\pm$ 0.05 & DA & 1\\
12:34:44.88+66:05:08.9 & 20.81 & 19.31 & 18.73 & 18.55 & 18.38 & 17.56 $\pm$ 0.04 & 17.38 $\pm$ 0.04 & 17.24 $\pm$ 0.05 & DA & 2\\
12:37:43.16+60:23:20.6 & 19.82 & 18.52 & 17.87 & 17.70 & 17.63 & \nodata & \nodata & \nodata & DA & 2\\
12:37:52.12+41:56:25.8 & 17.80 & 17.77 & 17.12 & 16.85 & 16.97 & \nodata & \nodata & \nodata & DQ & 2\\
12:38:12.85+35:02:49.1 & 24.73 & 21.76 & 20.31 & 19.87 & 20.31 & 21.19 $\pm$ 0.06  & \nodata & \nodata & DC & 3\\
12:47:39.05+06:46:04.6 & 20.89 & 20.04 & 18.67 & 18.40 & 18.30 & 17.55 $\pm$ 0.04 & 17.54 $\pm$ 0.04 & 17.35 $\pm$ 0.05 & DQpec & 2\\
12:51:06.12+44:03:03.1 & 21.44 & 20.17 & 20.39 & 20.69 & 20.86 & 21.78 $\pm$ 0.08 & \nodata & \nodata & DC & 3\\
12:55:08.13+46:55:18.5 & 21.01 & 19.19 & 18.38 & 18.06 & 17.95 & 16.87 $\pm$ 0.05 & 16.78 $\pm$ 0.06 & 16.68 $\pm$ 0.06 & DC & 2\\
12:59:25.91+04:42:09.6 & 21.27 & 19.43 & 18.61 & 18.32 & 18.24 & 17.39 $\pm$ 0.06 & 17.18 $\pm$ 0.06 & 16.90 $\pm$ 0.11 & DC & 2\\
13:13:13.12+02:26:45.8 & 20.98 & 18.93 & 17.83 & 17.48 & 17.30 & 16.25 $\pm$ 0.04 & 16.22 $\pm$ 0.04 & 16.13 $\pm$ 0.06 & DC & 1\\
13:17:37.46+06:21:21.1 & 19.94 & 18.60 & 17.96 & 17.75 & 17.69 & 16.86 $\pm$ 0.04 & 16.74 $\pm$ 0.05 & 16.69 $\pm$ 0.07 & DA & 2\\
13:22:54.60$-$00:50:42.8 & 20.65 & 18.91 & 18.13 & 17.82 & 17.73 & 16.81 $\pm$ 0.05 & 16.62 $\pm$ 0.08 & \nodata & DC & 1\\
13:24:15.17+41:49:05.8 & 20.78 & 19.38 & 18.85 & 18.62 & 18.55 & \nodata & \nodata & \nodata & DA & 2\\
13:24:51.92+46:19:53.4 & 20.44 & 19.04 & 18.47 & 18.27 & 18.23 & 17.32 $\pm$ 0.04 & 17.05 $\pm$ 0.04 & 16.95 $\pm$ 0.05 & \nodata & \nodata\\
13:40:19.66+40:33:03.1 & 21.06 & 19.43 & 18.65 & 18.36 & 18.26 & 17.31 $\pm$ 0.04 & 17.01 $\pm$ 0.04 & 16.84 $\pm$ 0.04 & DC & 2\\
13:41:18.68+02:27:37.0 & 18.35 & 17.97 & 17.30 & 17.19 & 17.23 & \nodata & \nodata & \nodata & DQ & 2\\
13:45:32.92+42:00:44.2 & 19.70 & 17.86 & 17.01 & 16.72 & 16.57 & \nodata & \nodata & \nodata & DC & 2\\
13:46:33.40+09:18:36.8 & 20.03 & 18.78 & 18.22 & 18.03 & 17.98 & \nodata & \nodata & \nodata & DA & 2\\
13:47:10.98+50:36:06.3 & 20.66 & 19.14 & 18.51 & 18.27 & 18.18 & 17.29 $\pm$ 0.04 & 17.07 $\pm$ 0.04 & 16.91 $\pm$ 0.05 & DA & 2\\
13:52:30.45+09:07:14.2 & 20.90 & 19.29 & 18.67 & 18.41 & 18.30 & 17.49 $\pm$ 0.04 & 17.23 $\pm$ 0.04 & 17.17 $\pm$ 0.05 & DC & 2\\
13:58:15.98+37:04:20.2 & 20.68 & 19.46 & 18.92 & 18.78 & 18.72 & 17.95 $\pm$ 0.04 & 17.66 $\pm$ 0.04 & 17.67 $\pm$ 0.05 & DA & 3\\
14:03:24.67+45:33:32.7 & 20.09 & 18.91 & 19.01 & 19.55 & 19.85 & 20.19 $\pm$ 0.07 & 20.61 $\pm$ 0.07 & 20.87 $\pm$ 0.11 & DC & 5\\
14:05:22.28+14:14:03.4 & 20.56 & 19.32 & 18.78 & 18.58 & 18.51 & 17.68 $\pm$ 0.04 & 17.46 $\pm$ 0.04 & 17.34 $\pm$ 0.06 & \nodata & \nodata\\
14:07:46.96+11:33:20.2 & 20.06 & 18.81 & 18.23 & 18.03 & 17.95 & \nodata & \nodata & \nodata & DA & 2\\
14:16:53.24+07:52:44.9 & 20.52 & 19.34 & 18.82 & 18.64 & 18.62 & 17.79 $\pm$ 0.04 & 17.61 $\pm$ 0.04 & 17.42 $\pm$ 0.05 & \nodata & \nodata\\
14:22:25.73+04:59:39.7 & 20.98 & 19.44 & 18.58 & 18.27 & 18.18 & 17.15 $\pm$ 0.05 & 17.10 $\pm$ 0.08 & 17.02 $\pm$ 0.05 & DC & 1\\
14:24:29.52+62:46:17.1 & 20.33 & 18.83 & 18.15 & 17.90 & 17.74 & \nodata & \nodata & \nodata & DA & 1\\
14:26:59.36+09:37:00.9 & 20.63 & 19.35 & 18.78 & 18.58 & 18.52 & 17.62 $\pm$ 0.04 & 17.36 $\pm$ 0.04 & 17.17 $\pm$ 0.05 & DC & 2\\
14:36:42.78+43:32:35.7 & 19.83 & 18.04 & 17.19 & 16.85 & 16.75 & 15.78 $\pm$ 0.04 & 15.62 $\pm$ 0.04 & 15.51 $\pm$ 0.04 & DC & 7\\
14:37:18.15+41:51:51.5 & 20.06 & 19.03 & 18.45 & 18.23 & 18.12 & 17.43 $\pm$ 0.04 & 17.76 $\pm$ 0.05 & 18.42 $\pm$ 0.08 & DC & 2\\
14:40:18.81+13:18:35.4 & 20.10 & 18.88 & 18.30 & 18.09 & 18.01 & \nodata & \nodata & \nodata & DC & 2\\
14:42:39.69+40:13:19.2 & 20.15 & 19.47 & 18.81 & 18.66 & 18.63 & 18.15 $\pm$ 0.04 & 18.24 $\pm$ 0.05 & 18.20 $\pm$ 0.08 & DQpec & 2\\
14:42:43.52+55:46:14.4 & 20.82 & 19.46 & 18.88 & 18.71 & 18.62 & 17.76 $\pm$ 0.04 & 17.52 $\pm$ 0.04 & 17.45 $\pm$ 0.07 & DA & 2\\
14:47:01.85+54:27:44.6 & 21.23 & 19.46 & 18.64 & 18.36 & 18.25 & 17.26 $\pm$ 0.07 & 17.20 $\pm$ 0.07 & 17.07 $\pm$ 0.06 & DC & 2\\
14:52:00.08+40:49:07.3 & 21.77 & 20.02 & 19.25 & 18.93 & 18.79 & \nodata & \nodata & \nodata & DC & 2\\
14:52:34.12$-$00:51:06.7 & 20.50 & 19.31 & 18.79 & 18.62 & 18.60 & 17.66 $\pm$ 0.04 & 17.48 $\pm$ 0.04 & 17.42 $\pm$ 0.05 & \nodata & \nodata\\
14:52:39.00+45:22:38.3 & 21.59 & 20.03 & 19.35 & 19.26 & 19.31 & 18.60 $\pm$ 0.02 & 18.43 $\pm$ 0.02 & 18.37 $\pm$ 0.02 & DC & 3\\
14:53:20.66+32:44:22.5 & 20.87 & 19.31 & 18.58 & 18.32 & 18.27 & \nodata & \nodata & \nodata & DC & 2\\
14:56:03.92+08:53:58.8 & 20.32 & 19.16 & 18.64 & 18.46 & 18.39 & 17.49 $\pm$ 0.04 & 17.35 $\pm$ 0.04 & 17.24 $\pm$ 0.04 & \nodata & \nodata\\
14:58:48.52+11:46:55.9 & 20.62 & 18.85 & 18.02 & 17.72 & 17.64 & 16.63 $\pm$ 0.04 & 16.47 $\pm$ 0.05 & 16.31 $\pm$ 0.06 & DC & 2\\
15:19:12.06+48:17:10.8 & 20.27 & 19.70 & 19.31 & 18.44 & 17.80 & \nodata & \nodata & \nodata & WD+dM & 2\\
15:26:59.16$-$00:07:31.5 & 20.66 & 19.39 & 18.86 & 18.66 & 18.60 & 17.71 $\pm$ 0.04 & 17.39 $\pm$ 0.04 & 17.37 $\pm$ 0.05 & \nodata & \nodata\\
15:28:15.02+32:54:10.3 & 20.62 & 19.28 & 18.67 & 18.44 & 18.37 & 17.53 $\pm$ 0.04 & 17.30 $\pm$ 0.04 & 17.17 $\pm$ 0.05 & DC & 2\\
15:34:18.29+07:11:48.7 & 20.50 & 19.15 & 18.51 & 18.27 & 18.21 & 17.44 $\pm$ 0.04 & 17.09 $\pm$ 0.04 & 17.06 $\pm$ 0.05 & DA & 2\\
15:34:51.02+46:49:49.5 & 20.90 & 18.76 & 17.74 & 17.36 & 17.19 & 16.17 $\pm$ 0.04 & 16.12 $\pm$ 0.04 & 16.04 $\pm$ 0.05 & DC & 6\\
15:49:00.06+31:56:56.5 & 20.63 & 19.72 & 19.94 & 20.50 & 20.85 & \nodata & \nodata & \nodata & DA & 2\\
16:00:00.78+00:19:06.9 & 20.09 & 18.95 & 18.40 & 18.24 & 18.17 & 17.31 $\pm$ 0.04 & 17.10 $\pm$ 0.04 & 16.98 $\pm$ 0.05 & \nodata & \nodata\\
16:06:19.81+25:47:02.9 & 20.99 & 19.24 & 18.45 & 18.17 & 18.07 & 17.07 $\pm$ 0.04 & 17.09 $\pm$ 0.06 & 16.84 $\pm$ 0.06 & DZA & 2\\
16:08:09.48+42:35:15.3 & 21.02 & 19.47 & 18.70 & 18.50 & 18.37 & 17.42 $\pm$ 0.05 & 17.30 $\pm$ 0.07 & 17.17 $\pm$ 0.08 & DC & 2\\
16:15:44.67+44:49:42.5 & 21.18 & 19.59 & 18.84 & 18.57 & 18.52 & 17.44 $\pm$ 0.04 & 17.24 $\pm$ 0.05 & 17.26 $\pm$ 0.07 & DC & 1\\
16:22:40.08+29:19:12.2 & 21.92 & 19.89 & 18.93 & 18.52 & 18.34 & \nodata & \nodata & \nodata & DC & 2\\
16:27:24.58+37:26:43.2 & 21.77 & 19.80 & 18.94 & 18.64 & 18.60 & \nodata & \nodata & \nodata & DC & 2\\
16:27:31.09+48:59:19.0 & 20.70 & 19.22 & 18.62 & 18.39 & 18.34 & 17.48 $\pm$ 0.04 & 17.18 $\pm$ 0.04 & 17.16 $\pm$ 0.05 & DZA & 1\\
16:32:42.23+24:26:55.2 & 21.33 & 19.60 & 18.72 & 18.49 & 18.47 & 17.67 $\pm$ 0.02 & 18.10 $\pm$ 0.02 & 18.04 $\pm$ 0.02 & DC & 3\\
16:48:47.07+39:39:17.0 & 20.13 & 18.87 & 18.31 & 18.16 & 18.11 & 17.19 $\pm$ 0.05 & 17.33 $\pm$ 0.09 & 17.58 $\pm$ 0.09 & DC & 1\\
17:04:47.70+36:08:47.4 & 20.50 & 18.72 & 17.94 & 17.66 & 17.55 & 16.62 $\pm$ 0.04 & 16.34 $\pm$ 0.04 & 16.32 $\pm$ 0.06 & DC & 1\\
17:22:51.94+28:48:46.9 & 20.61 & 19.23 & 18.68 & 18.42 & 18.31 & 17.43 $\pm$ 0.04 & 17.28 $\pm$ 0.04 & 17.31 $\pm$ 0.07 & DA & 2\\
17:22:57.78+57:52:50.7 & 20.39 & 19.28 & 18.79 & 18.56 & 18.50 & 17.74 $\pm$ 0.04 & 17.84 $\pm$ 0.05 & 18.75 $\pm$ 0.12 & DC & 1\\
20:41:28.99$-$05:20:27.7 & 20.95 & 19.27 & 18.51 & 18.24 & 18.14 & 17.25 $\pm$ 0.04 & 16.97 $\pm$ 0.04 & 16.93 $\pm$ 0.04 & DC & 1\\
20:42:59.23+00:31:56.6 & 21.67 & 19.95 & 19.05 & 18.73 & 18.61 & 17.65 $\pm$ 0.04 & 17.45 $\pm$ 0.04 & 17.36 $\pm$ 0.05 & DC & 1\\
20:45:06.97+00:37:34.4 & 20.45 & 19.77 & 19.42 & 19.26 & 19.21 & 18.43 $\pm$ 0.04 & 18.23 $\pm$ 0.05 & 18.22 $\pm$ 0.08 & DA & 1\\
20:45:57.53$-$07:10:03.5 & 21.00 & 19.33 & 18.60 & 18.34 & 18.18 & 17.32 $\pm$ 0.04 & 17.10 $\pm$ 0.04 & 17.03 $\pm$ 0.04 & DC & 1\\
20:53:16.34$-$07:02:04.2 & 19.25 & 19.19 & 18.74 & 18.62 & 18.74 & 18.27 $\pm$ 0.04 & 18.04 $\pm$ 0.04 & 18.21 $\pm$ 0.07 & DQ & 3\\
21:03:36.68$-$00:55:45.2 & 21.65 & 20.06 & 19.29 & 19.02 & 18.89 & \nodata & \nodata & \nodata & DC & 2\\
21:16:40.30$-$07:24:52.7 & 20.25 & 18.43 & 17.59 & 17.26 & 17.13 & 16.16 $\pm$ 0.04 & 16.02 $\pm$ 0.05 & 15.90 $\pm$ 0.05 & DC & 1\\
21:18:05.21$-$07:37:29.1 & 23.38 & 20.70 & 19.48 & 19.01 & 18.76 & 17.90 $\pm$ 0.04 & 17.82 $\pm$ 0.04 & 17.81 $\pm$ 0.05 & DC & 1\\
21:25:01.48$-$07:34:56.0 & 20.74 & 19.88 & 19.49 & 19.36 & 19.28 & 18.61 $\pm$ 0.04 & 18.32 $\pm$ 0.04 & 18.21 $\pm$ 0.07 & DA & 1\\
21:47:25.17+11:27:56.1 & 20.83 & 19.19 & 18.43 & 18.13 & 18.01 & 17.14 $\pm$ 0.04 & 16.84 $\pm$ 0.04 & 16.79 $\pm$ 0.04 & DA & 2\\
21:51:53.79$-$07:31:31.0 & 19.85 & 18.85 & 18.36 & 18.19 & 18.14 & 17.32 $\pm$ 0.04 & 16.96 $\pm$ 0.04 & 16.89 $\pm$ 0.05 & \nodata & \nodata\\
22:04:14.16$-$01:09:31.2 & 22.16 & 20.21 & 19.29 & 18.99 & 18.83 & 18.01 $\pm$ 0.04 & 17.68 $\pm$ 0.04 & 17.70 $\pm$ 0.05 & DC & 1\\
22:25:43.50$-$01:13:59.6 & 21.47 & 19.90 & 19.12 & 18.89 & 18.75 & \nodata & \nodata & \nodata & DA & 2\\
22:39:54.07+00:18:49.2 & 24.21 & 21.02 & 19.93 & 19.59 & 19.41 & 18.40 $\pm$ 0.02 & 18.25 $\pm$ 0.02 & 18.48 $\pm$ 0.27 & DC & 3\\
22:39:54.12+00:18:47.3 & 21.51 & 20.16 & 19.53 & 19.47 & 20.09 & 19.73 $\pm$ 0.06 & 19.99 $\pm$ 0.08 & \nodata          & DC & 3\\
22:42:06.19+00:48:22.8 & 22.11 & 19.63 & 18.65 & 18.28 & 18.16 & 18.06 $\pm$ 0.04 & 18.72 $\pm$ 0.07 & 19.16 $\pm$ 0.10 & DC & 1\\
22:54:08.64+13:23:57.2 & 21.57 & 19.51 & 18.49 & 18.14 & 18.00 & 17.04 $\pm$ 0.04 & 16.88 $\pm$ 0.04 & 16.85 $\pm$ 0.04 & DC & 1\\
23:07:22.35+14:00:46.2 & 20.07 & 19.18 & 18.73 & 18.59 & 18.59 & 17.82 $\pm$ 0.04 & 17.46 $\pm$ 0.04 & 17.48 $\pm$ 0.05 & \nodata & \nodata\\
23:21:15.32+01:02:11.3 & 20.45 & 19.35 & 18.81 & 18.63 & 18.60 & 17.78 $\pm$ 0.04 & 17.54 $\pm$ 0.04 & 17.53 $\pm$ 0.05 & DA & 8\\
23:21:15.68+01:02:23.9 & 21.64 & 19.84 & 18.93 & 18.64 & 18.49 & 17.54 $\pm$ 0.04 & 17.37 $\pm$ 0.04 & 17.32 $\pm$ 0.04 & DC & 6\\
23:25:19.89+14:03:39.7 & 18.02 & 16.46 & 15.84 & 15.55 & 15.44 & 14.53 $\pm$ 0.04 & 14.34 $\pm$ 0.04 & 14.21 $\pm$ 0.05 & DA & 9\\
23:30:55.20+00:28:52.3 & 21.85 & 19.88 & 18.95 & 18.66 & 18.53 & 17.63 $\pm$ 0.04 & 17.36 $\pm$ 0.04 & 17.32 $\pm$ 0.04 & DC & 1\\
23:42:45.75$-$10:01:21.4 & 20.45 & 18.95 & 18.21 & 17.94 & 17.89 & 17.08 $\pm$ 0.06 & 16.90 $\pm$ 0.06 & 16.79 $\pm$ 0.07 & DA & 1\\
23:44:05.54$-$14:29:23.5 & 21.28 & 19.79 & 19.19 & 18.94 & 18.89 & \nodata & \nodata & \nodata & DA & 2\\
23:50:42.52$-$08:46:18.9 & 20.22 & 19.17 & 18.61 & 18.38 & 18.31 & 17.52 $\pm$ 0.04 & 17.27 $\pm$ 0.04 & 17.19 $\pm$ 0.05 & DA & 1\\
\enddata
\tablecomments{The last column in the table refers to the source of the optical spectroscopic observations: (1) Kilic et al. (2006),
(2) This paper, (3) SDSS, (4) Oppenheimer et al. (2001), (5) Gates et al. (2004), (6) Oswalt et al. (1996), (7) Hintzen et al. (1986),
(8) Carollo et al. (2006), (9) Vennes \& Kawka (2003).}
\end{deluxetable}

\begin{deluxetable}{lcccccrrc}
\tabletypesize{\scriptsize}
\tablecolumns{9}
\tablewidth{0pt}
\tablecaption{Spectroscopically Identified Subdwarf Stars}
\tablehead{
\colhead{Name (SDSS J)}&
\colhead{$u$}&
\colhead{$g$}&
\colhead{$r$}&
\colhead{$i$}&
\colhead{$z$}&
\colhead{$\mu_{\rm RA}$}&
\colhead{$\mu_{\rm DEC}$}&
\colhead{Epochs}\\
                       & (mag) & (mag) & (mag) & (mag) & (mag) & (mas/yr) & (mas/yr) & 
}
\startdata
09:19:48.11+43:56:21.6 & 21.90 & 19.42 & 18.45 & 17.94 & 17.60 & $-$73 & $-$171 & 6\\
10:13:29.64+51:04:12.8 & 20.83 & 18.57 & 17.47 & 17.03 & 16.80 & $-$117 & $-$156 & 6\\
10:19:57.78+62:19:48.1 & 20.57 & 18.48 & 17.63 & 17.30 & 17.18 & $-$83 & $-$128 & 6\\
10:21:36.30+38:08:39.8 & 21.09 & 18.80 & 17.68 & 17.23 & 17.00 & 24 & $-$199 & 6\\
10:29:22.43+02:44:53.3 & 21.96 & 19.49 & 18.36 & 17.87 & 17.64 & $-$86 & $-$103 & 6\\
11:06:10.50+11:34:24.2 & 20.53 & 18.77 & 17.84 & 17.48 & 17.26 & $-$77 & $-$97 & 6\\
11:19:58.69+43:54:54.8 & 21.72 & 19.46 & 18.43 & 18.02 & 17.75 & $-$18 & $-$103 & 6\\
12:00:18.05+11:46:48.4 & 20.57 & 18.63 & 17.67 & 17.27 & 17.05 & $-$116 & $-$166 & 6\\
12:04:50.43+05:11:54.1 & 21.44 & 19.22 & 18.04 & 17.57 & 17.30 & $-$41 & $-$164 & 6\\
12:08:51.72+43:24:10.4 & 20.81 & 18.53 & 17.39 & 16.96 & 16.64 & $-$183 & 24 & 6\\
12:33:30.40+10:00:31.8 & 20.87 & 18.81 & 17.77 & 17.35 & 17.10 & $-$54 & $-$159 & 6\\
12:34:13.46+02:01:39.2 & 21.42 & 18.80 & 17.65 & 17.17 & 16.93 & $-$183 & $-$54 & 6\\
12:39:07.85+47:22:16.5 & 21.53 & 19.46 & 18.39 & 17.99 & 17.76 & $-$82 & $-$69 & 6\\
12:44:25.95$-$01:44:25.2 & 19.66 & 17.51 & 16.48 & 16.03 & 15.80 & $-$217 & $-$187 & 6\\
13:11:06.30+51:54:45.0 & 19.93 & 17.76 & 16.75 & 16.30 & 16.09 & $-$14 & $-$262 & 6\\
13:12:29.65+41:02:20.5 & 21.47 & 19.44 & 18.41 & 17.98 & 17.77 & $-$49 & $-$100 & 6\\
13:12:43.88+59:27:10.1 & 21.66 & 19.47 & 18.33 & 17.86 & 17.66 & 11 & $-$125 & 6\\
13:16:33.66+02:28:17.9 & 20.52 & 18.49 & 17.45 & 17.07 & 16.84 & $-$63 & $-$170 & 6\\
13:37:15.76+01:14:56.4 & 20.75 & 18.44 & 17.33 & 16.86 & 16.66 & $-$145 & $-$141 & 6\\
13:58:18.78+03:22:59.5 & 19.29 & 17.08 & 15.97 & 15.54 & 15.34 & $-$257 & $-$273 & 6\\
14:20:28.35+07:24:54.5 & 21.15 & 19.16 & 18.11 & 17.71 & 17.48 & $-$138 & $-$132 & 4\\
14:23:24.99+12:40:38.1 & 20.32 & 18.25 & 17.42 & 17.07 & 16.86 & $-$174 & 94 & 6\\
14:49:29.72+34:28:43.2 & 21.86 & 19.43 & 18.30 & 17.83 & 17.60 & $-$107 & $-$102 & 6\\
14:58:44.11+00:44:03.6 & 21.13 & 19.48 & 18.60 & 18.26 & 18.07 & 16 & 87 & 6\\
15:11:45.76+03:31:16.2 & 25.36 & 17.80 & 16.67 & 16.20 & 15.98 & $-$210 & $-$210 & 6\\
16:06:44.68+48:34:51.6 & 21.43 & 19.07 & 17.99 & 17.55 & 17.32 & $-$154 & 28 & 6
\enddata
\tablecomments{The proper motion measurements are from \citet{munn04}.
The last column indicates the number of epochs an object is detected in the USNO-B + SDSS.}
\end{deluxetable}

\begin{deluxetable}{llccccrrl}
\tabletypesize{\scriptsize}
\tablecolumns{10}
\tablewidth{0pt}
\tablecaption{Physical Parameters}
\tablehead{
\colhead{Object}&
\colhead{$T_{\rm eff}$}&
\colhead{$d$}&
\colhead{Cooling Age}&
\colhead{Comp}&
\colhead{Type}&
\colhead{$\mu_{\rm RA}$}&
\colhead{$\mu_{\rm DEC}$}&
\colhead{Notes}\\
           & ~~~~(K)            & (pc) & (Gyr) & ($\log$ H/He) &      & (mas/yr) &  (mas/yr) &      
}
\startdata
J0003$-$0111 & 5450 $\pm$ 70  & 72  & 3.4  & H    & DA      & 98     & $-$16  & a \\
J0033+1451 & 5360 $\pm$ 180 & 53  & 4.5  & $-3.4$ & \nodata & $-$209 & $-$190 &  \\
J0045+1420 & 5070 $\pm$ 90  & 54  & 5.4  & H    & DZA     & 260    & $-$53  & b,d \\
J0102+1401 & 4830 $\pm$ 50  & 54  & 6.4  & He   & DC      & 12     & 106    & \\
J0146+1404 & 3930 $\pm$ 235 & 60  & 8.5  & $-2.5$ & DC    & 252    & 38     & \\
J0157+1335 & 4820 $\pm$ 70  & 56  & 6.4  & $-5.0$ & DC      & 87     & $-$62  & e \\
J0212$-$0040 & 6010 $\pm$ 80  & 79  & 2.2  & H    & \nodata & 132    &  18    & \\
J0250$-$0910 & 5640 $\pm$ 80  & 67  & 2.8  & H    & DA      & 106    &   2    & a \\
J0256$-$0700 & 4420 $\pm$ 110 & 35  & 7.8  & H    & DC      & 373    & $-$202 & \\
J0301$-$0044 & 4530 $\pm$ 50  & 65  & 7.1  & He   & DC      & 108    & $-$532 & \\
J0307$-$0715 & 5840 $\pm$ 80  & 40  & 2.4  & H    & DA      & $-$199 & $-$452 & \\
J0309+0025 & 4920 $\pm$ 50  & 37  & 6.2  & $-4.4$ & DC      & $-$6   & $-$106 & \\
J0310$-$0110 & 4970 $\pm$ 50  & 117 & 6.1  & He   & DC      & $-$36  & $-$80  & \\
J0314$-$0105 & 5800 $\pm$ 80  & 61  & 2.5  & H    & DA      & $-$77  & $-$71  & \\
J0320$-$0716 & 6570 $\pm$ 80  & 129 & 1.8  & He   & DQ      & 120    & $-$11 & \\
J0330+0037 & 5870 $\pm$ 90  & 108 & 2.4  & H    & DA      & 77     & 34    & a,c \\
J0406$-$0644 & 5960 $\pm$ 80  & 50  & 2.3  & H    & DA      & 67     & 27    & \\
J0747+2438N & 4890 $\pm$ 50  & 52  & 6.3  & He   & DC     & 140    & $-$70 & b \\
J0747+2438S & 5740 $\pm$ 80  & 54  & 2.6  & H    & DA     & 137    & $-$69 & a \\
J0753+4230 & 4650 $\pm$ 100 & 24  & 7.1  & H    & DC      & 113    & $-$403 & \\
J0817+2451 & 5520 $\pm$ 70  & 81  & 3.9  & He   & \nodata & 81     & $-$211 & \\
J0817+2822 & 4720 $\pm$ 40  & 50  & 6.7  & He   & DC      & 63     & $-$205 & \\
J0819+3159 & 4700 $\pm$ 40  & 56  & 6.7  & He   & DC      & 211    & $-$22 & \\
J0821+3727 & 5060 $\pm$ 60  & 52  & 5.4  & H    & DA      & 165    & $-$154 & a,c \\
J0825+2841 & 5600 $\pm$ 70  & 66  & 2.9  & H    & \nodata & $-$114 & $-$110 & \\
J0825+5049 & 4690 $\pm$ 40  & 46  & 6.7  & He   & DC      & $-$331 & $-$330 & \\
J0836+4556 & 4880 $\pm$ 60  & 70  & 6.3  & He   & DC      & $-$64  & $-$169 & \\
J0838+2804 & 5170 $\pm$ 60  & 58  & 5.4  & He   & DC      & 63     & $-$200 & \\
J0853+4118 & 5590 $\pm$ 70  & 78  & 3.6  & He   & \nodata & $-$128 & 103    & \\
J0854+3503 & 4070 $\pm$ 130 & 57  & 8.8  & 1.0 & DC      & $-$133 & $-$179 & \\
J0855+3700 & 5660 $\pm$ 70  & 42  & 2.8  & H    & \nodata & 148    & $-$89  & \\
J0904+3403 & 5350 $\pm$ 40  & 53  & 4.6  & He   & DC      & $-$180 & $-$291 & \\
J0909+4700 & 4510 $\pm$ 160 & 53  & 7.1  & $-3.4$ & DC      & $-$117 & $-$179 & \\
J0933+3743 & 5430 $\pm$ 60  & 68  & 4.2  & He   & DZA     & 134    & $-$113 & \\
J0942+4437 & 4450 $\pm$ 90  & 42  & 7.8  & H    & DC      & $-$135 & $-$189 & \\
J0943+5134 & 4880 $\pm$ 60  & 43  & 6.3  & He   & DC      & 131    & $-$261 & \\
J0947+4459 & 3410 $\pm$ 90  & 39  & 9.5  & $-2.7$ & DC      & 74     & 45     & \\
J0947+4500 & 5100 $\pm$ 130 & 61  & 5.2  & H    & DA      & 74     & 45     & b \\
J1000+4236 & 4840 $\pm$ 50  & 57  & 6.4  & He   & DC      & $-$342 & $-$80  & \\
J1001+3903 & 3050 $\pm$ 150 & 45  & 10.2 & $-2.9$  & DC      & $-$301 & $-$185 & \\
J1001+4656 & 4150 $\pm$ 70  & 33  & 8.6  & H    & DC      & $-$17  & $-$339 & \\
J1002+6108 & 4420 $\pm$ 120 & 39  & 7.8  & H    & DC      & $-$448 & $-$328 & \\
J1042+4932 & 5380 $\pm$ 60  & 66  & 4.5  & He   & DC      & $-$120 & $-$85  & \\
J1102+6707 & 5080 $\pm$ 60  & 68  & 5.7  & He   & DC      & $-$380 & $-$185 & \\
J1102+4030 & 5110 $\pm$ 50  & 49  & 5.2  & H    & DA      & 193    & $-$256 & b,c \\
J1104+0436 & 4400 $\pm$ 120 & 49  & 7.9  & H    & DC      & 101    & $-$385 & \\
J1107+4855 & 4640 $\pm$ 160 & 46  & 7.1  & H    & DC      & $-$726 & $-$79  & \\
J1116+0925 & 5470 $\pm$ 90  & 63  & 3.3  & H    & DA      &  7     & $-$152 & a\\
J1117+5010 & 4900 $\pm$ 60  & 53  & 6.3  & He   & DC      & 170    & $-$120 & \\
J1119$-$0107 & 4760 $\pm$ 50  & 64  & 6.6  & He   & DC      & $-$291 & $-$28  & \\
J1130+1002 & 5230 $\pm$ 70  & 52  & 4.5  & H    & DA      & 145    & $-$311 & a,c\\
J1147+4303 & 4900 $\pm$ 30  & 48  & 6.2  & He   & DC      & 114    & $-$174 & \\
J1151+1253 & 5260 $\pm$ 70  & 69  & 5.0  & He   & DC      & 4      & $-$210 & \\
J1158+0004 & 4390 $\pm$ 100 & 30  & 7.9  & H    & DC      & $-$17  & 182    & \\
J1202$-$0313 & 4490 $\pm$ 160 & 55  & 7.6  & H    & DC      & $-$73  & 134    & \\
J1204+6222 & 4820 $\pm$ 50  & 48  & 6.4  & He   & DC      & $-$21  & $-$159 & \\
J1212+4345 & 5450 $\pm$ 70  & 73  & 3.4  & H    & DA      & $-$210 & 59     & a,c\\
J1212+0440 & 4450 $\pm$ 110 & 55  & 7.8  & H    & DC      & $-$279 & $-$72  & \\
J1220+0914 & 2950 $\pm$ 130 & 39  & 10.5 & $-5.1$ & DC      & $-$341 & $-$372 & \\
J1220+4519 & 5570 $\pm$ 80  & 81  & 3.0  & H    & \nodata & $-$203 & $-$21  & \\
J1234+0109 & 5500 $\pm$ 80  & 92  & 3.2  & H    & DA      & $-$284 & $-$55  & a\\
J1234+6605 & 5430 $\pm$ 90  & 70  & 3.5  & H    & DA      & $-$185 & $-$72  & a,c\\
J1238+3502 & 2290 $\pm$ 120 & 36  & 12.1 & $-7.8$ & DC    & $-$130 & $-$124& \\  
J1247+0646 & 5120 $\pm$ 180 & 60  & 5.6  & $-0.7$  & DQpec     & $-$382 & 71     & \\
J1251+4403 & 3200 $\pm$ 60  & 66  & 10.0 & $-2.3$ & DC    & $-$167 & 30      & \\
J1255+4655 & 4580 $\pm$ 160 & 41  & 7.3  & H    & DC      & $-$1089 & $-$114 & \\
J1259+0442 & 4840 $\pm$ 50  & 53  & 6.4  & He   & DC      & 201    & $-$59  & \\
J1313+0226 & 4360 $\pm$ 90  & 29  & 8.0  & H    & DC      & $-$744 & $-$116 & \\
J1317+0621 & 5400 $\pm$ 70  & 49  & 3.6  & H    & DA      & 248    & $-$152 & a,c,f\\
J1322$-$0050 & 4840 $\pm$ 80  & 42  & 6.4  & He   & DC      & $-$156 & 118    & \\
J1324+4619 & 5410 $\pm$ 70  & 62  & 3.6  & H    & \nodata & 67     & $-$158 & \\
J1340+4033 & 4770 $\pm$ 60  & 52  & 6.6  & He   & DC      & 76     & $-$195 & \\
J1347+5036 & 5210 $\pm$ 70  & 58  & 4.6  & H    & DA      & 166    & $-$178 & a\\
J1352+0907 & 5150 $\pm$ 50  & 61  & 5.5  & He   & DC      & 113    & $-$305 & \\
J1358+3704 & 5810 $\pm$ 100 & 89  & 2.5  & H    & DA      & 86     & $-$59  & a\\
J1403+4533 & 2670 $\pm$ 1500 & 24 & 11.1 & $-3.0$ & DC      & $-$271 & $-$84  & \\
J1405+1414 & 5360 $\pm$ 60  & 71  & 4.5  & He   & \nodata & $-$26  & $-$105 & \\
J1416+0752 & 5480 $\pm$ 60  & 76  & 4.0  & He   & \nodata & 0      & $-$92  & \\
J1422+0459 & 4430 $\pm$ 110 & 43  & 7.8  & H    & DC      & $-$277 & $-$62  & \\
J1426+0937 & 5200 $\pm$ 70  & 67  & 5.2  & He   & DC      & $-$180 & $-$8   & \\
J1436+4332 & 4750 $\pm$ 40  & 26  & 6.6  & He   & DC      & $-$314 & 505    & \\
J1437+4151 & 4480 $\pm$ 50  & 44  & 7.2  & $-4.3$ & DC      & $-$157 & $-$69  & \\
J1442+4013 & 5740 $\pm$ 430 & 81  & 3.1  & $-2.7$ & DQpec     & $-$199 & $-$89  & \\
J1442+5546 & 5500 $\pm$ 90  & 77  & 3.2  & H    & DA      & 91     & 0      & a\\
J1447+5427 & 4890 $\pm$ 80  & 55  & 6.3  & He   & DC      & $-$231 & 39     & \\
J1452$-$0051 & 5640 $\pm$ 80  & 78  & 2.8  & H    & \nodata & 92     & $-$168 & \\
J1452+4522 & 5780 $\pm$ 50  & 115 & 3.0  & He   & DC      & $-$54  & 76     & \\
J1456+0853 & 5420 $\pm$ 60  & 68  & 4.3  & He   & \nodata & 19     & $-$138 & \\
J1458+1146 & 4780 $\pm$ 50  & 39  & 6.5  & He   & DC      & $-$131 & $-$94  & \\
J1526$-$0007 & 5440 $\pm$ 50  & 74  & 3.4  & H    & \nodata & 54     & $-$132 & \\
J1528+3254 & 5210 $\pm$ 50  & 64  & 5.2  & He   & DC      & 136    & $-$103 & \\
J1534+0711 & 5320 $\pm$ 70  & 61  & 4.0  & H    & DA      & 126    & 114    & a,c\\
J1534+4649 & 4340 $\pm$ 80  & 27  & 8.1  & H    & DC      & $-$465 & 220    & \\
J1600+0019 & 5410 $\pm$ 40  & 61  & 4.3  & He   & \nodata & $-$122 & $-$29  & \\
J1606+2547 & 4880 $\pm$ 50  & 50  & 6.3  & He   & DZA     & $-$225 & $-$125 & \\
J1608+4235 & 4990 $\pm$ 70  & 60  & 6.0  & He   & DC      & $-$123 & 102    & \\
J1615+4449 & 4780 $\pm$ 80  & 56  & 6.6  & H    & DC      & 44     & $-$237 & \\
J1627+4859 & 5330 $\pm$ 70  & 64  & 4.0  & H    & DZA     & $-$91  & 77     & a\\
J1632+2426 & 4160 $\pm$ 40  & 44  & 8.0  & $-5.6$ & DC    & $-$10  & $-$340 & \\
J1648+3939 & 4850 $\pm$ 70  & 48  & 6.4  & $-4.4$ & DC      & $-$126 & 0      & \\
J1704+3608 & 4850 $\pm$ 30  & 39  & 6.4  & He   & DC      & 186    & $-$175 & \\
J1722+2848 & 5370 $\pm$ 70  & 66  & 3.8  & H    & DA      & 2      & $-$255 & a,f\\
J1722+5752 & 4620 $\pm$ 80  & 54  & 6.9  & $-4.2$ & DC      & $-$37  & 390    & \\
J2041$-$0520 & 4910 $\pm$ 30  & 52  & 6.2  & He   & DC      & $-$149 & $-$29  & \\
J2042+0031 & 4680 $\pm$ 40  & 60  & 6.8  & He   & DC      & $-$71  & $-$244 & \\
J2045+0037 & 6110 $\pm$ 100 & 119 & 2.1  & H    & DA      & 32     & $-$32  & a \\
J2045$-$0710 & 4930 $\pm$ 70  & 53  & 6.0  & H    & DC      & $-$73  & $-$134 & b,c \\
J2053$-$0702 & 6390 $\pm$ 40  & 96  & 2.1  & He   & DQ      & 21     & $-$105 & \\
J2116$-$0724 & 4600 $\pm$ 100 & 29  & 7.3  & H    & DC      & 111    & $-$223 & \\
J2118$-$0737 & 4220 $\pm$ 70  & 56  & 8.4  & H    & DC      & 115    & $-$144 & \\
J2125$-$0734 & 6120 $\pm$ 100 & 126 & 2.1  & H    & DA      & 64     & 13     & a\\
J2147+1127 & 4770 $\pm$ 80  & 46  & 6.7  & H    & DA      & 103    & $-$254 & b,c \\
J2151$-$0731 & 5620 $\pm$ 80  & 63  & 2.8  & H    & \nodata & 117    & 30     & \\
J2204$-$0109 & 4750 $\pm$ 50  & 70  & 6.6  & He   & DC      & 112    & $-$303 & \\
J2239+0018A & 3740 $\pm$ 120 & 60 & 8.9  & $-3.1$ & DC    & 7      & 98     & \\
J2239+0018B & 4440 $\pm$ 80  & 79 & 7.8  &  H   & DC      & 7      & 98     & \\ 
J2242+0048 & 3480 $\pm$ 60  & 32  & 9.4  & $-5.9$ & DC      & 132    & $-$76  & \\
J2254+1323 & 4390 $\pm$ 80  & 40  & 7.9  & H    & DC      & 329    & $-$199 & \\
J2307+1400 & 5940 $\pm$ 90  & 84  & 2.3  & H    & \nodata & $-$122 & $-$38  & \\
J2321+0102S & 5750 $\pm$ 100 & 82 & 2.6  & H    & DA      & $-$104 & $-$255 & c\\
J2321+0102N & 4430 $\pm$ 100 & 51 & 7.8  & H    & DC      & $-$106 & $-$258 & \\
J2325+1403 & 5030 $\pm$ 60  & 15  & 5.5  & H    & DA      &  336   & 115    & \\
J2330+0028 & 5130 $\pm$ 130 & 67  & 5.0  & H    & DC      & 151    & 91     & \\
J2342$-$1001 & 5160 $\pm$ 90  & 50  & 4.9  & H    & DA      & $-$28  & $-$95  & a,f\\
J2350$-$0846 & 5600 $\pm$ 110 & 70  & 2.9  & H    & DA      & 209    & $-$139 & a,c
\enddata
\tablecomments{(a) H$\alpha$ clearly visible and in good agreement with the best fit pure-H model,
(b) H$\alpha$ is barely visible and can not be confirmed,
(c) The pure-helium fit is significantly better,
(d) The object has to be H-rich for the H$\alpha$ line to be visible at this temperature,
(e) The fit is poor and unusual, and the photometry seems offset,
(f) Mild infrared absorption could be fitted with mixed models but the
H$\alpha$ line would be incompatible with predictions.
}
\end{deluxetable}

\clearpage
\begin{figure}
\includegraphics[width=3.5in,angle=0]{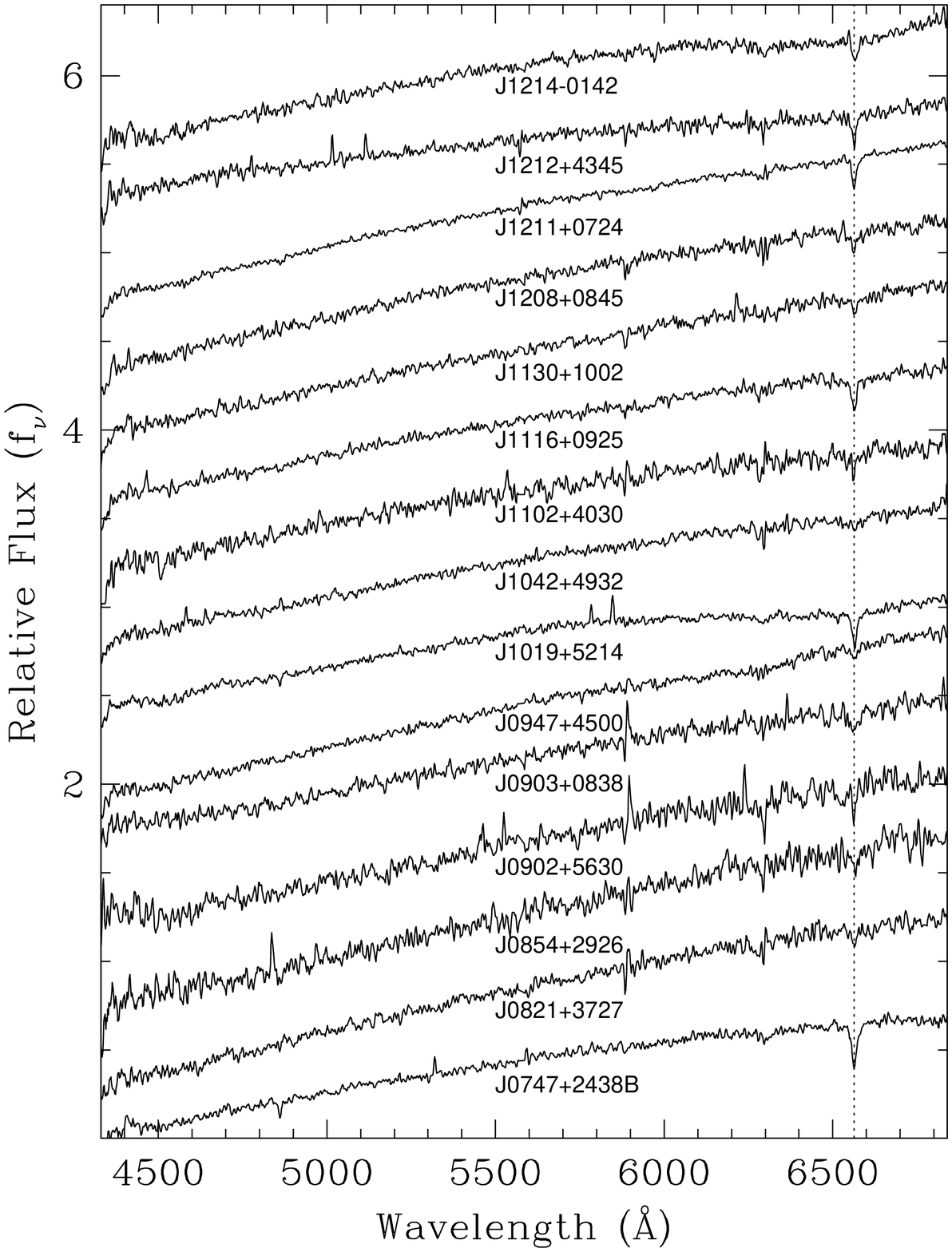}
\includegraphics[width=3.5in,angle=0]{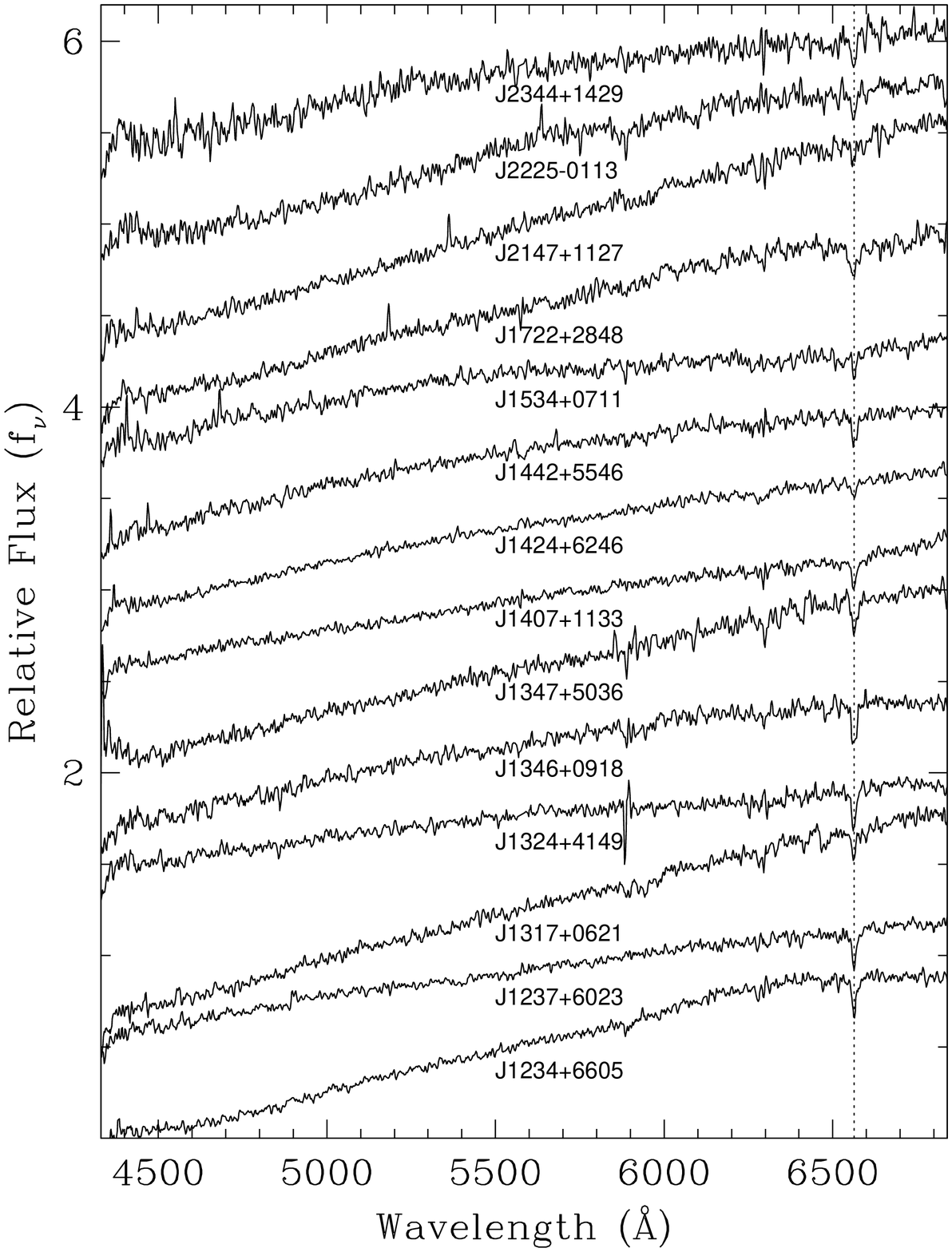}
\caption{Optical spectra of the DA WDs observed at the HET. The spectra
are normalized at 5700 \AA, and are shifted vertically from each other by 0.4 units.
The dotted line marks H$\alpha$.}
\end{figure}

\begin{figure}
\includegraphics[width=3.5in,angle=0]{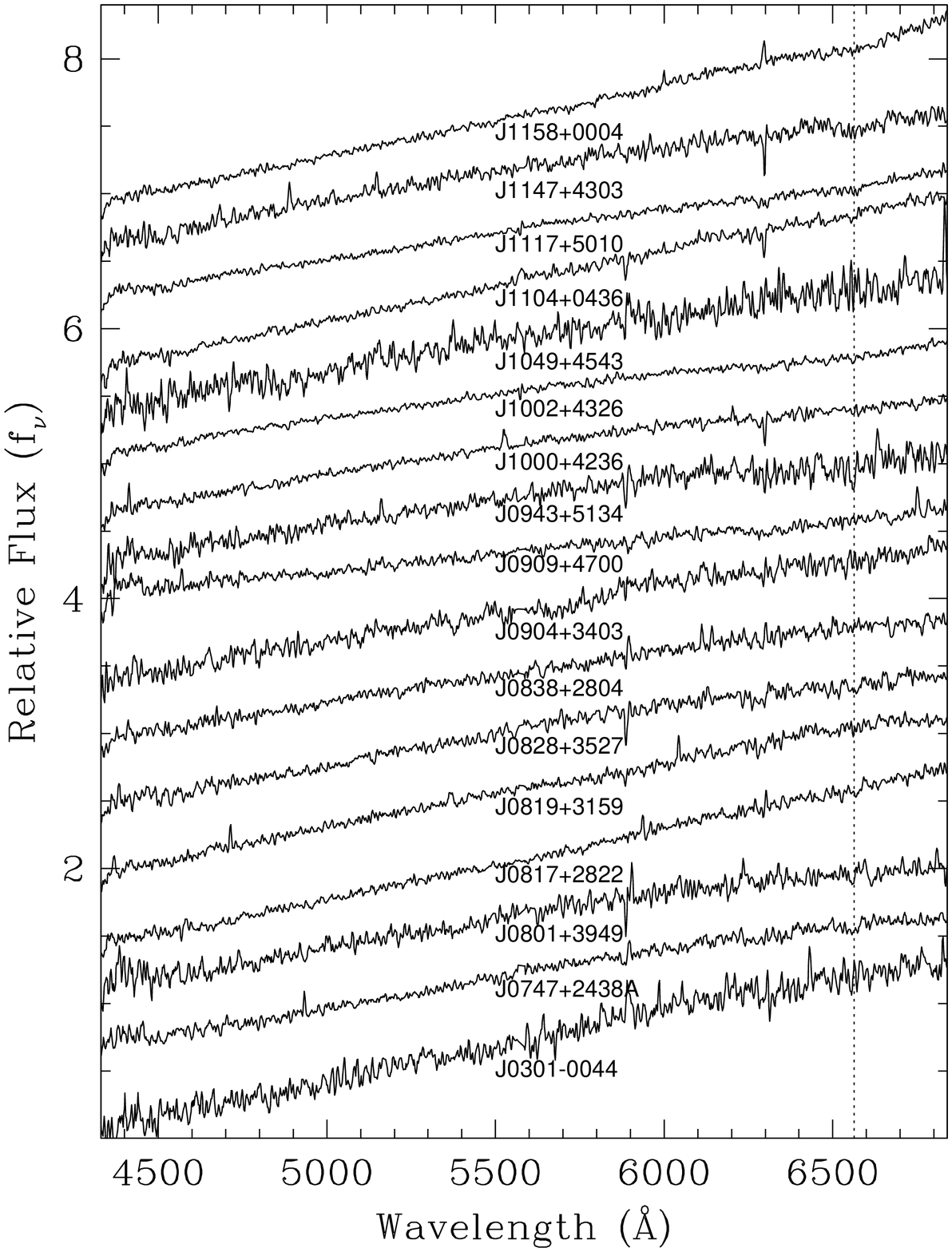}
\includegraphics[width=3.5in,angle=0]{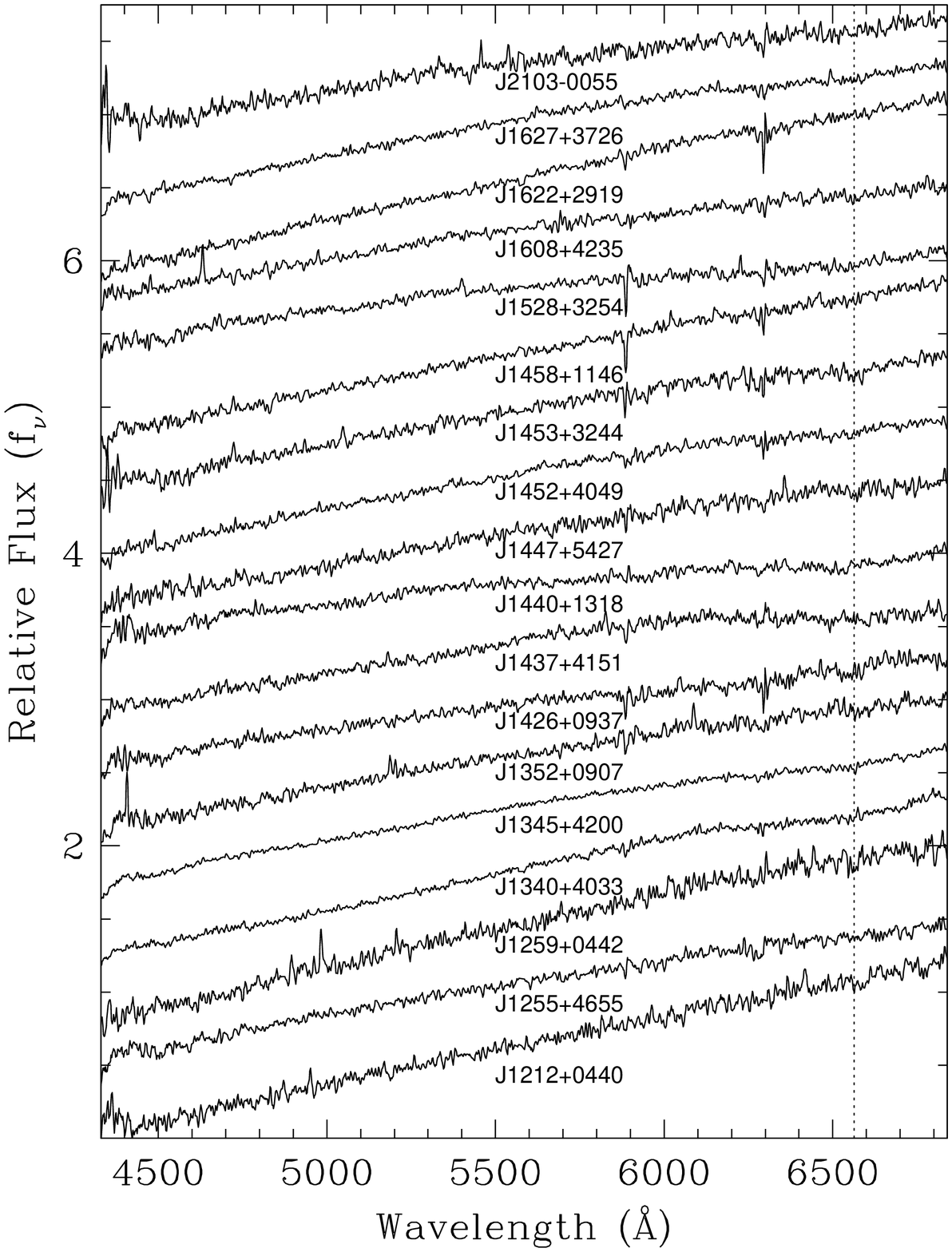}
\caption{Optical spectra of the DC WDs observed at the HET.}
\end{figure}

\begin{figure}
\plotone{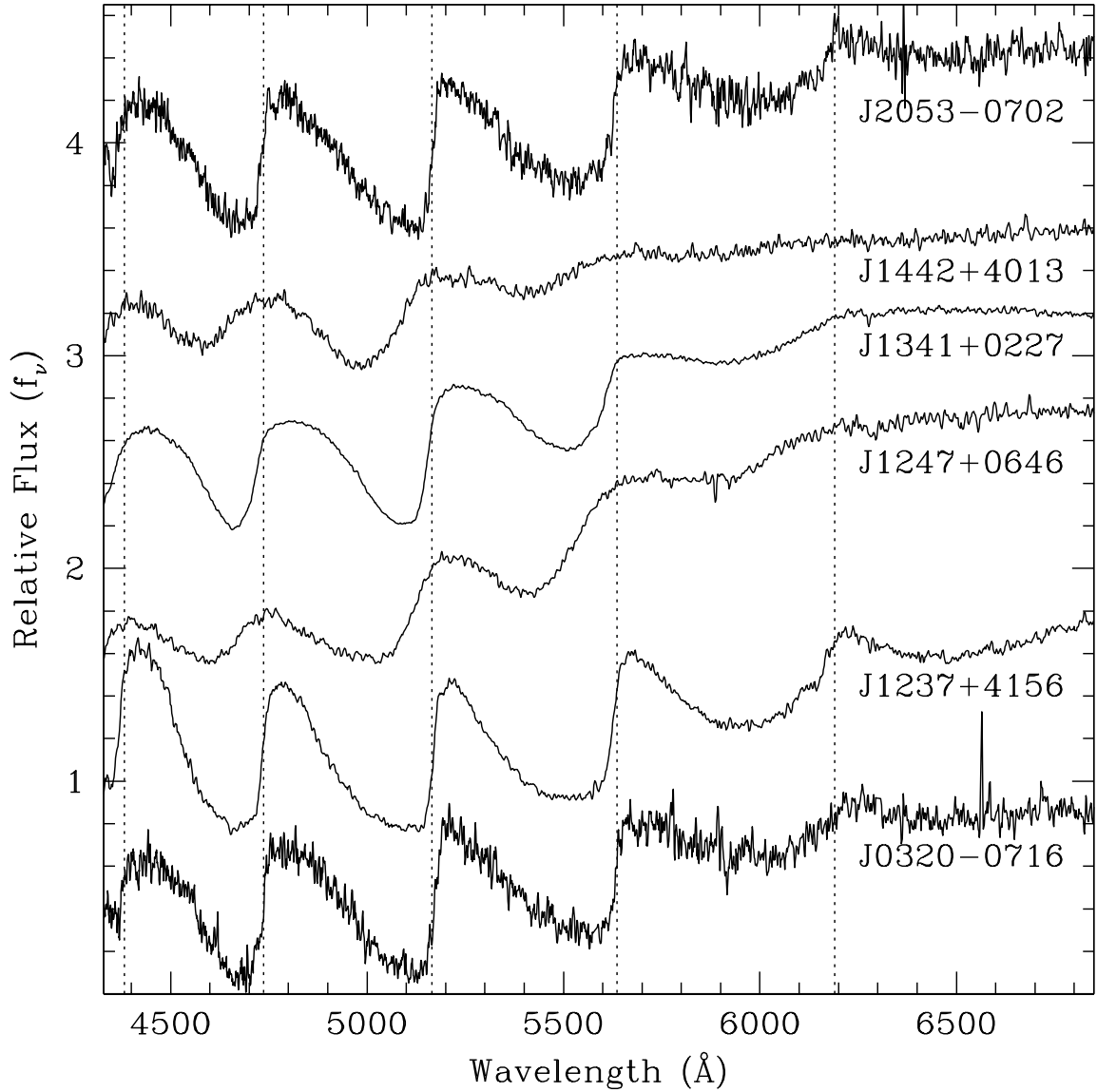}
\caption{Optical spectra of the normal and peculiar DQ WDs in our sample. Dotted lines mark the expected
locations of C$_2$ bandheads. The spectra for J0320$-$0716 and J2053$-$0702 are from the SDSS.}
\end{figure}

\begin{figure}
\plotone{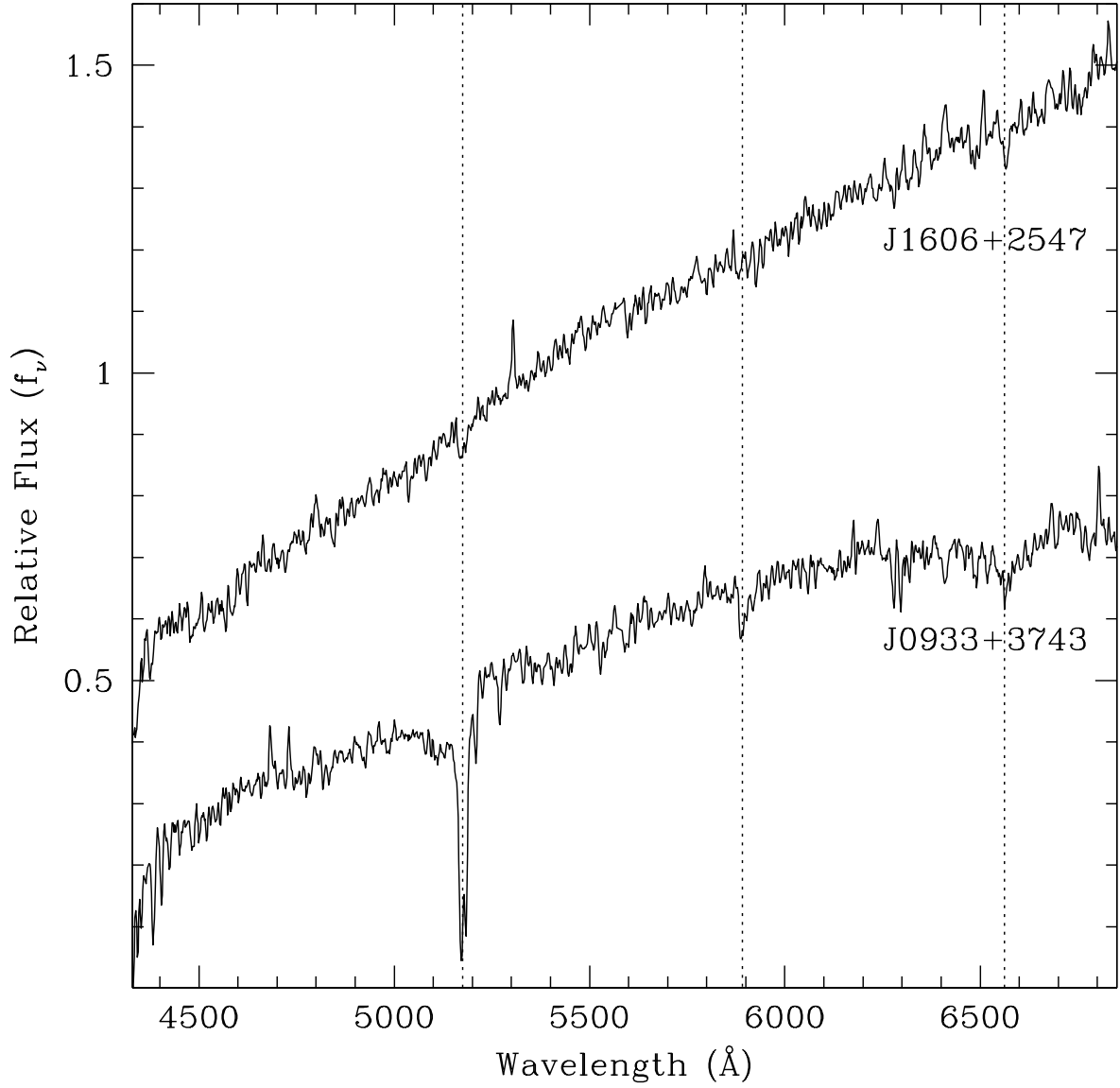}
\caption{HET spectra of the DZA WDs. The dotted lines mark the positions
of Mg, Na, and H absorption lines.}
\end{figure}

\begin{figure}
\plotone{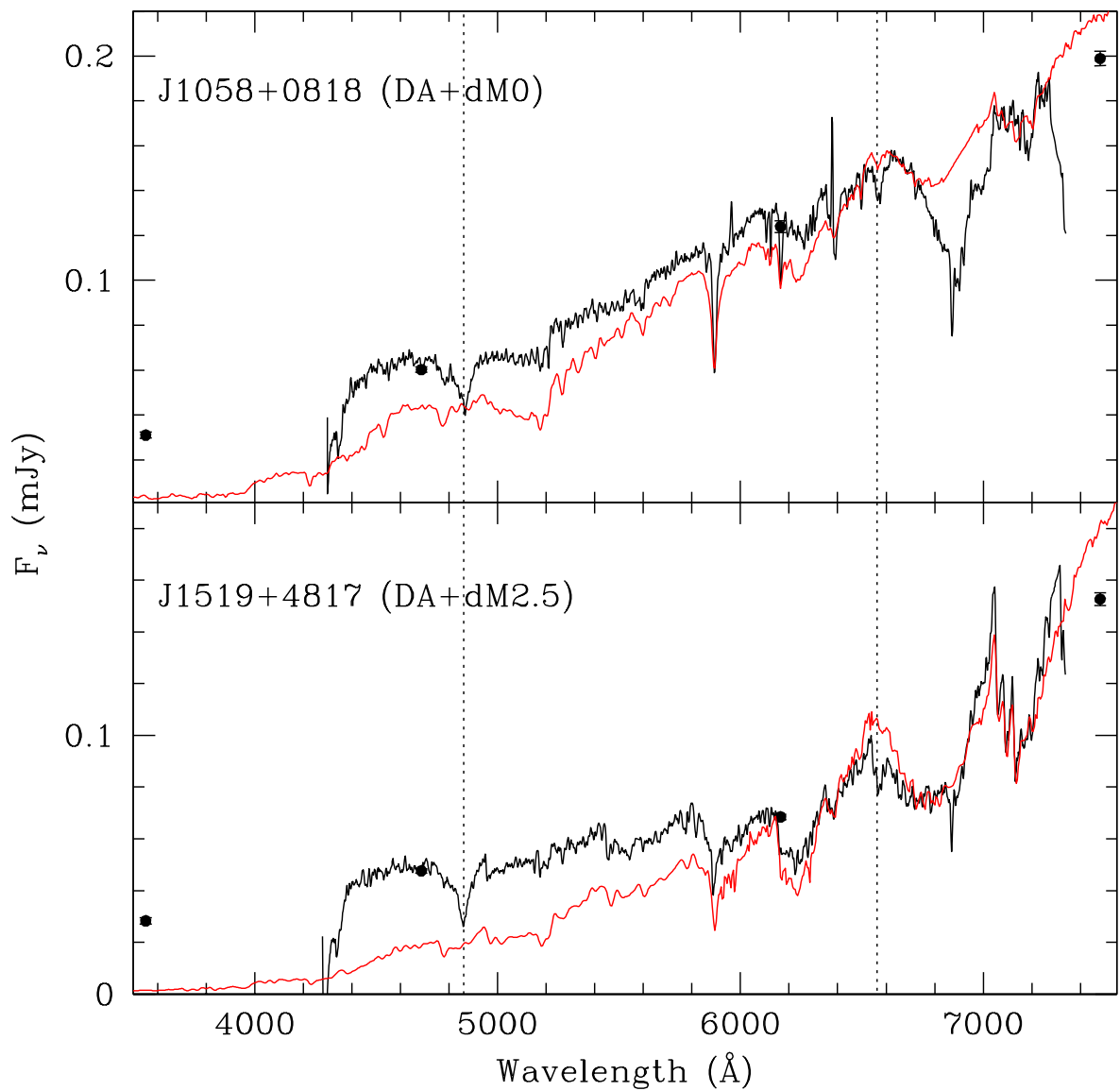}
\caption{HET spectra of the WD+dM binaries. The observed spectra are normalized
to match the SDSS $r-$band photometry. Red lines show the predicted contribution
from dM companions based on Pickles (1998) templates. The blue excess from WD companions is evident
in our HET spectra and the SDSS $u-$ and $g-$band photometry. The dotted lines mark H$\beta$ and H$\alpha$.}
\end{figure}

\begin{figure}
\plotone{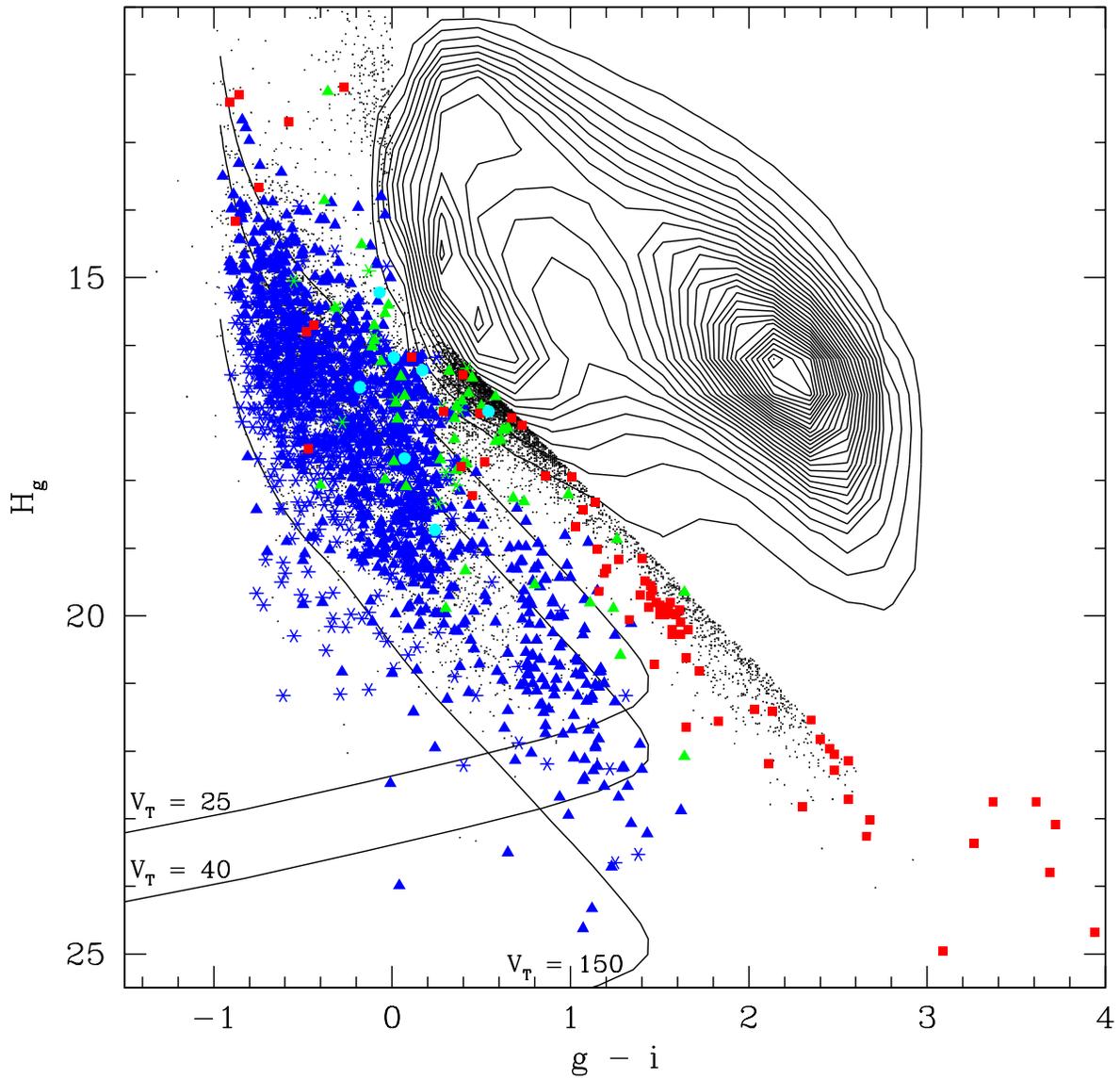}
\caption{The reduced proper motion diagram for stars in the SDSS DR3.
Individual stars are plotted only in the region of interest for white dwarfs,
the remaining stars are represented by the contours. WDs, WDs plus late type star
binaries, subdwarfs, and quasars are shown as blue triangles, green triangles,
red squares, and cyan circles, respectively.
White dwarfs that do not meet our criteria for reliable proper motions \citep[see][]{kilic06} are plotted as blue asterisks.
White dwarf cooling curves for different tangential velocities are shown as solid
lines. The $V_{\rm T}=$ 25--40 km s$^{-1}$ curves mark the expected location of disk white dwarfs,
whereas the $V_{\rm T}=$ 150 km s$^{-1}$ curve represents the halo white dwarfs.}
\end{figure}

\begin{figure}
\plottwo{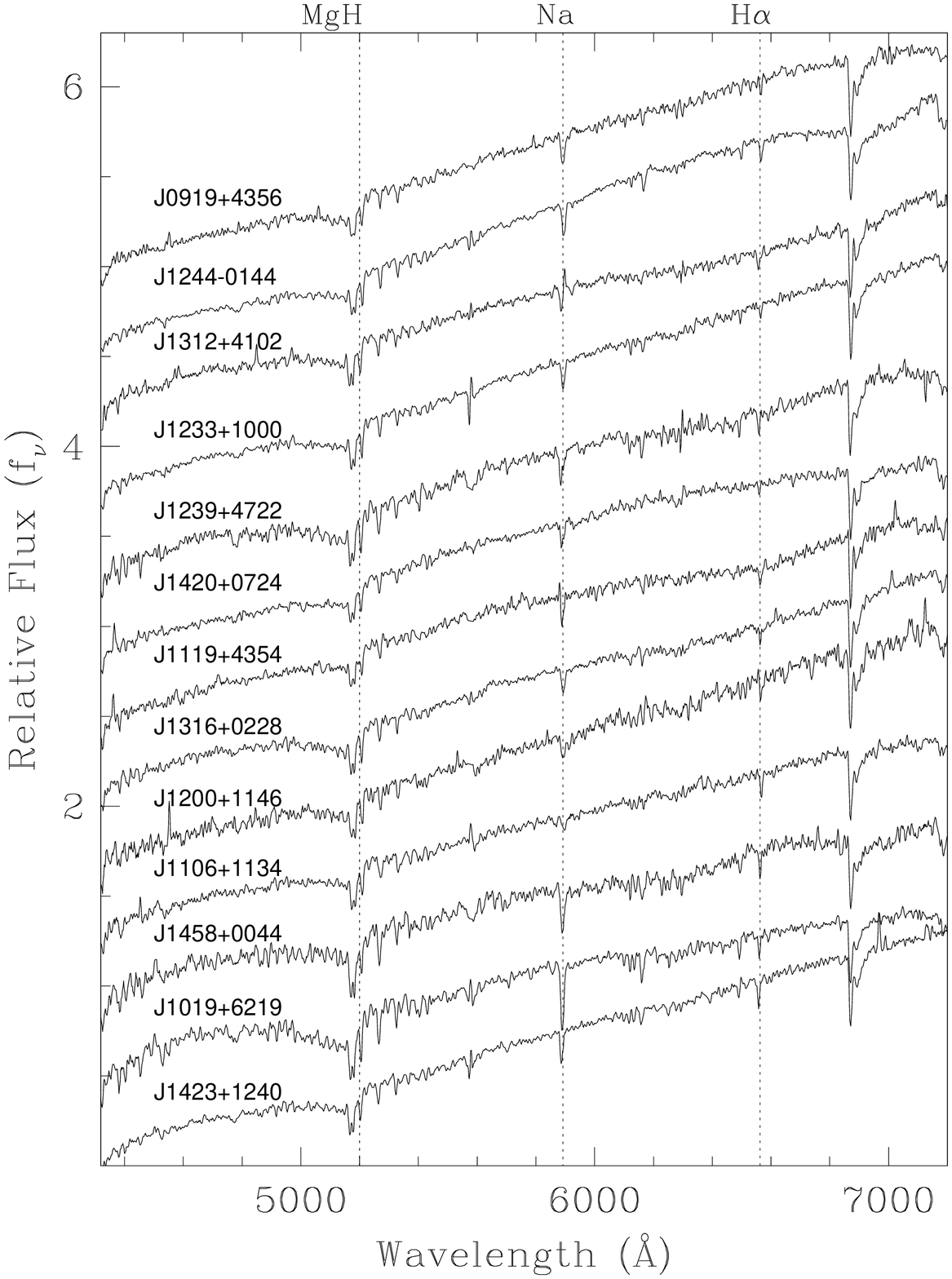}{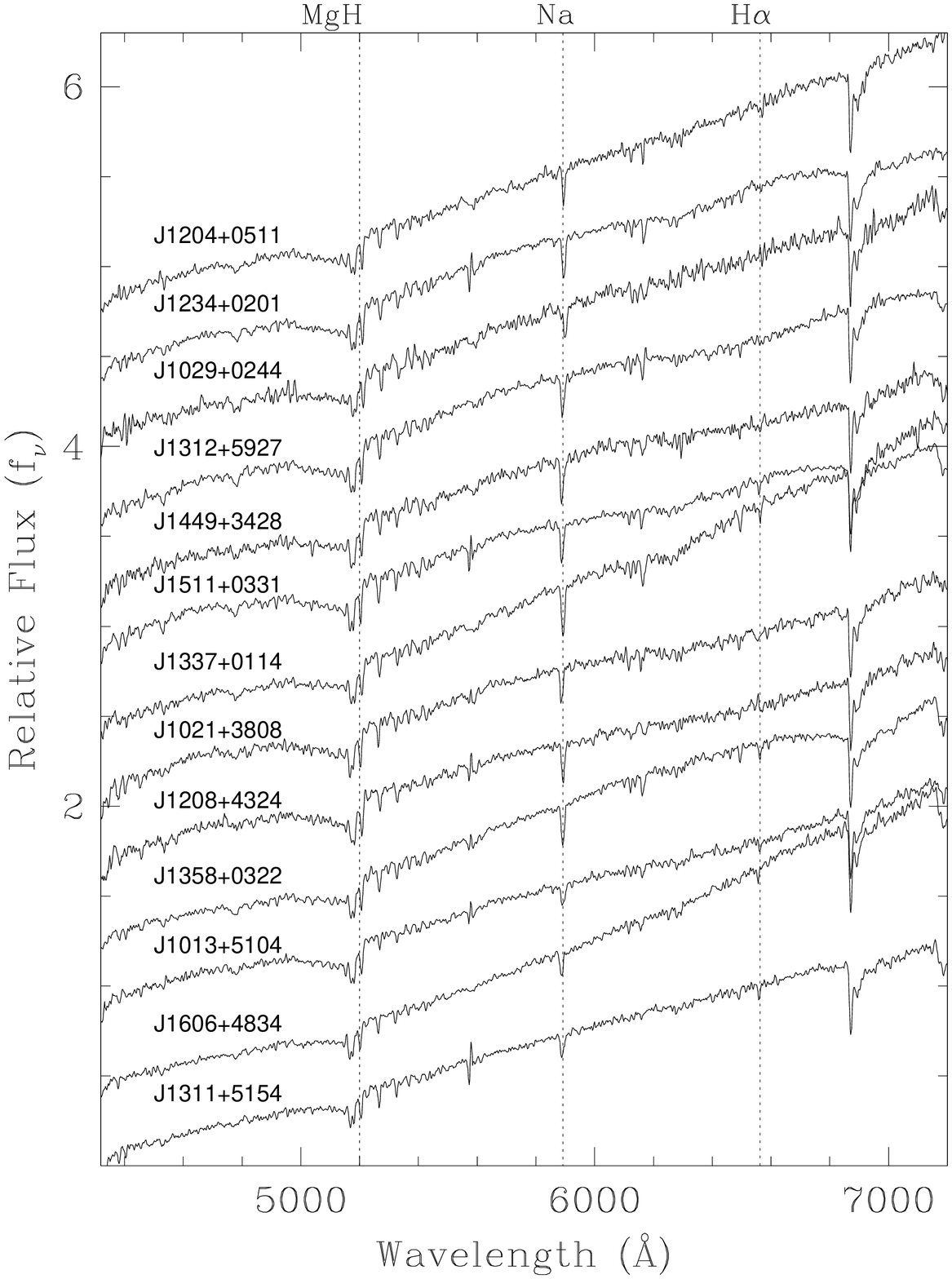}
\caption{HET spectra of the contaminants in our survey; subdwarfs. The
spectra are ordered in increasing $g-i$ color.}
\end{figure}

\begin{figure}
\hspace*{-0.5in}
\includegraphics[width=3.7in,angle=0]{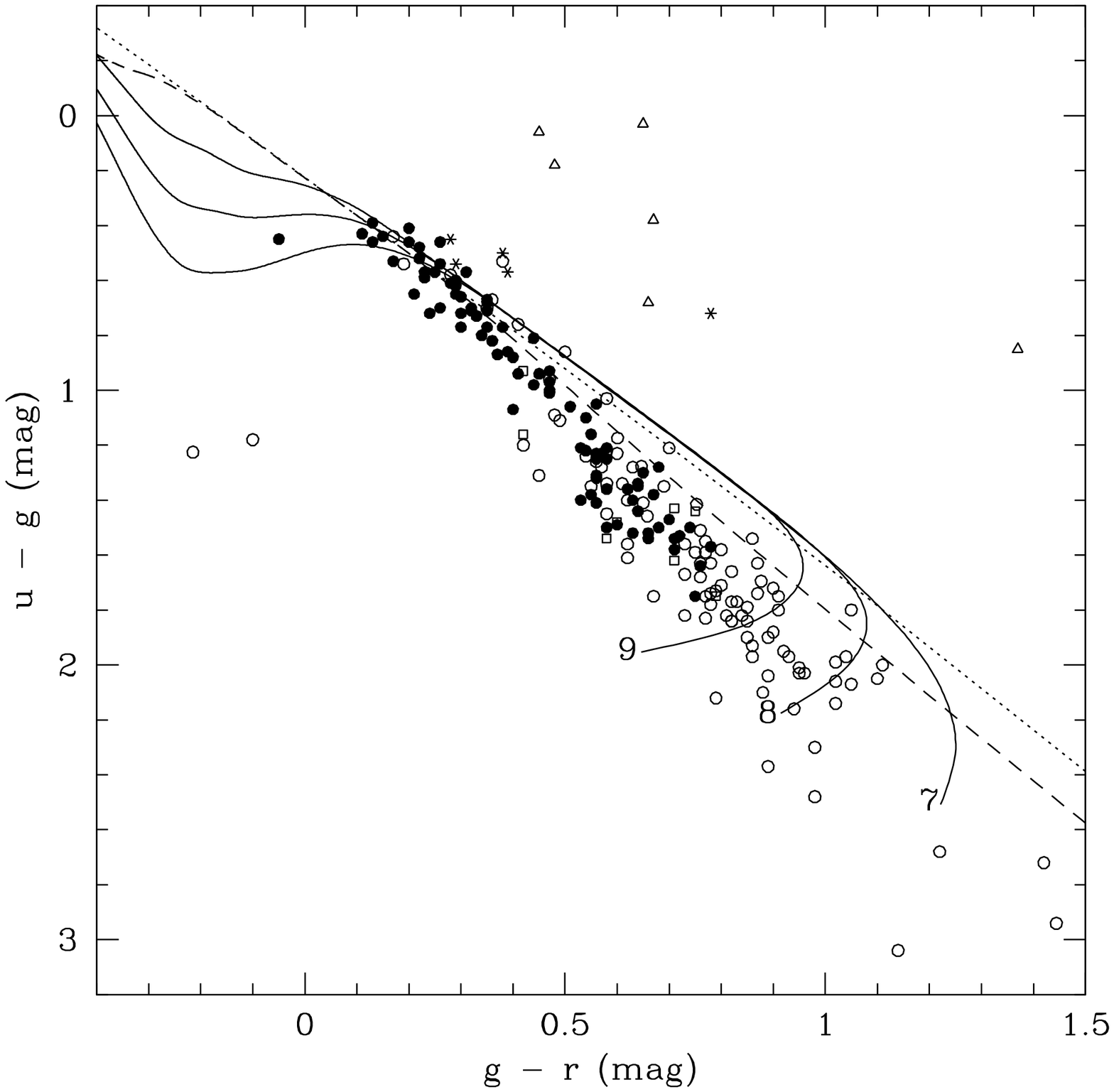}
\includegraphics[width=3.7in,angle=0]{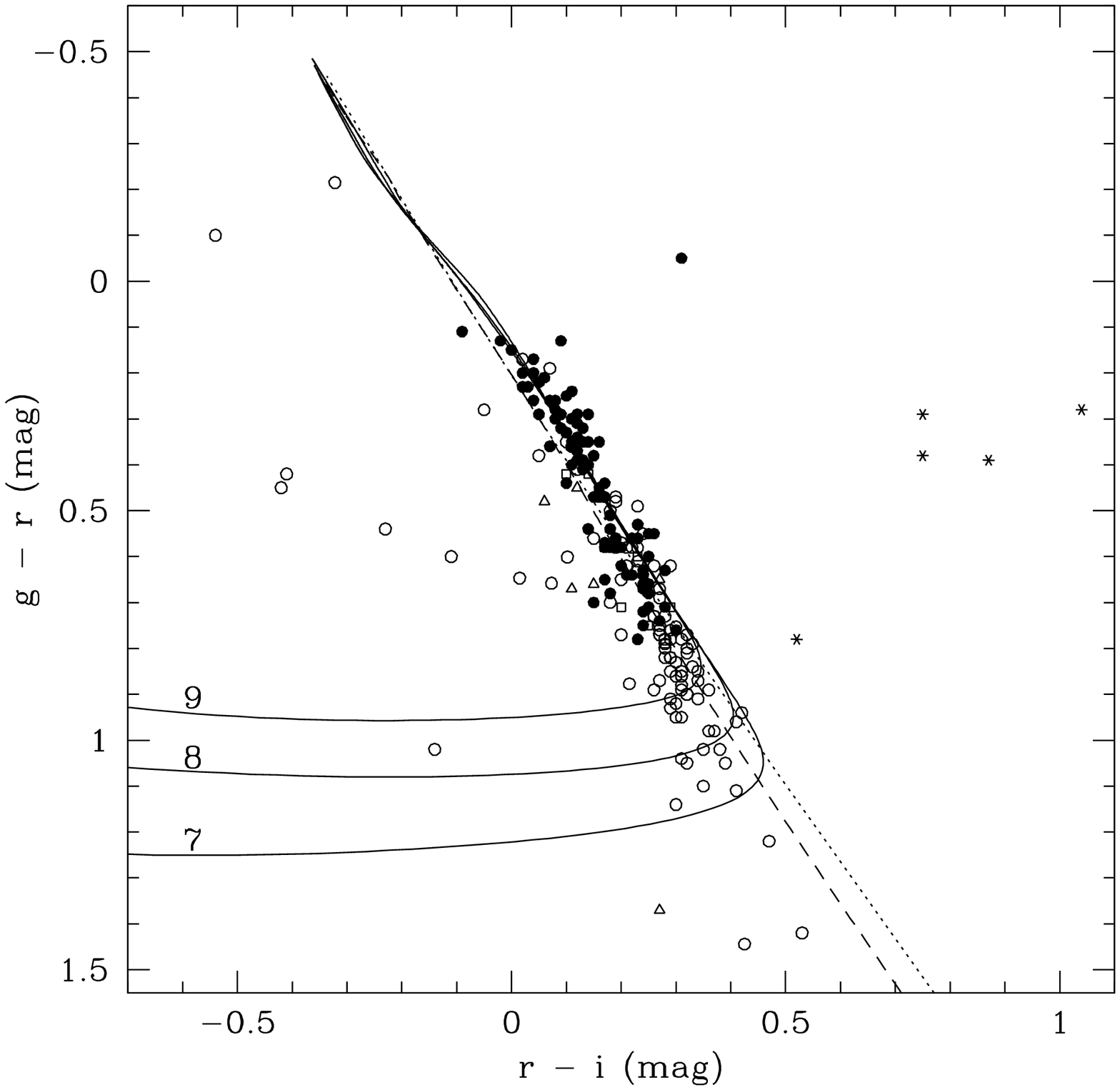}
\includegraphics[width=3.7in,angle=0]{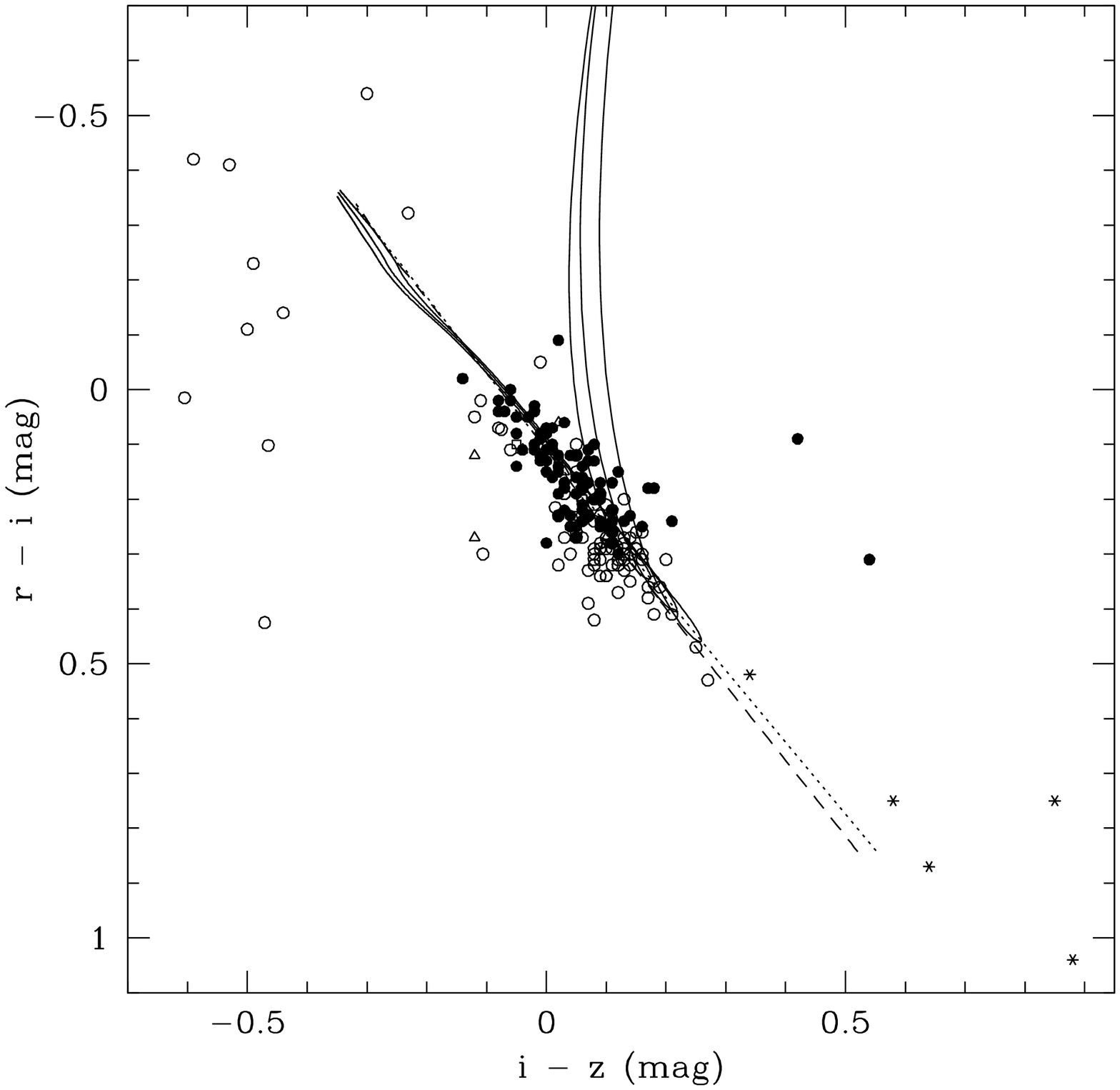}
\caption{Color-color diagrams for DA (filled circles), DC (open circles), DQ
(open triangles), DZ (open squares), WD+dM (asterisks) in the SDSS. 
The solid lines
show the predicted colors for pure hydrogen atmosphere WDs with $T_{\rm eff}= 2000-30000$ K and $\log g=$ 7, 8,
and 9. The dashed line shows a pure helium atmosphere WD sequence with
$T_{\rm eff}= 3000-30000$ K and $\log g=8$, whereas the dotted line shows
the colors for blackbody SEDs for the same temperature range.}
\end{figure}

\begin{figure}
\hspace*{-0.5in}
\includegraphics[width=3.7in,angle=0]{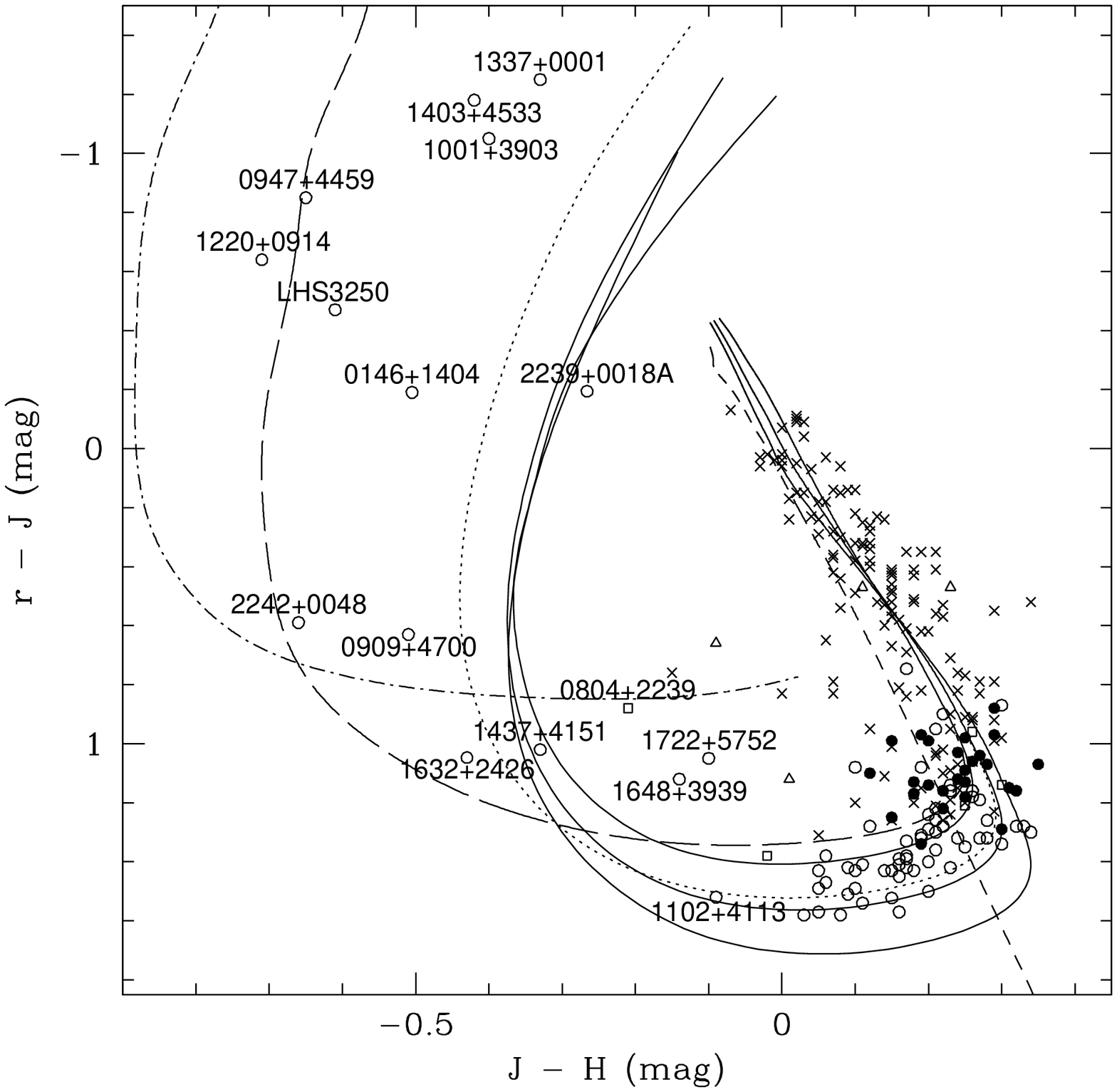}
\includegraphics[width=3.7in,angle=0]{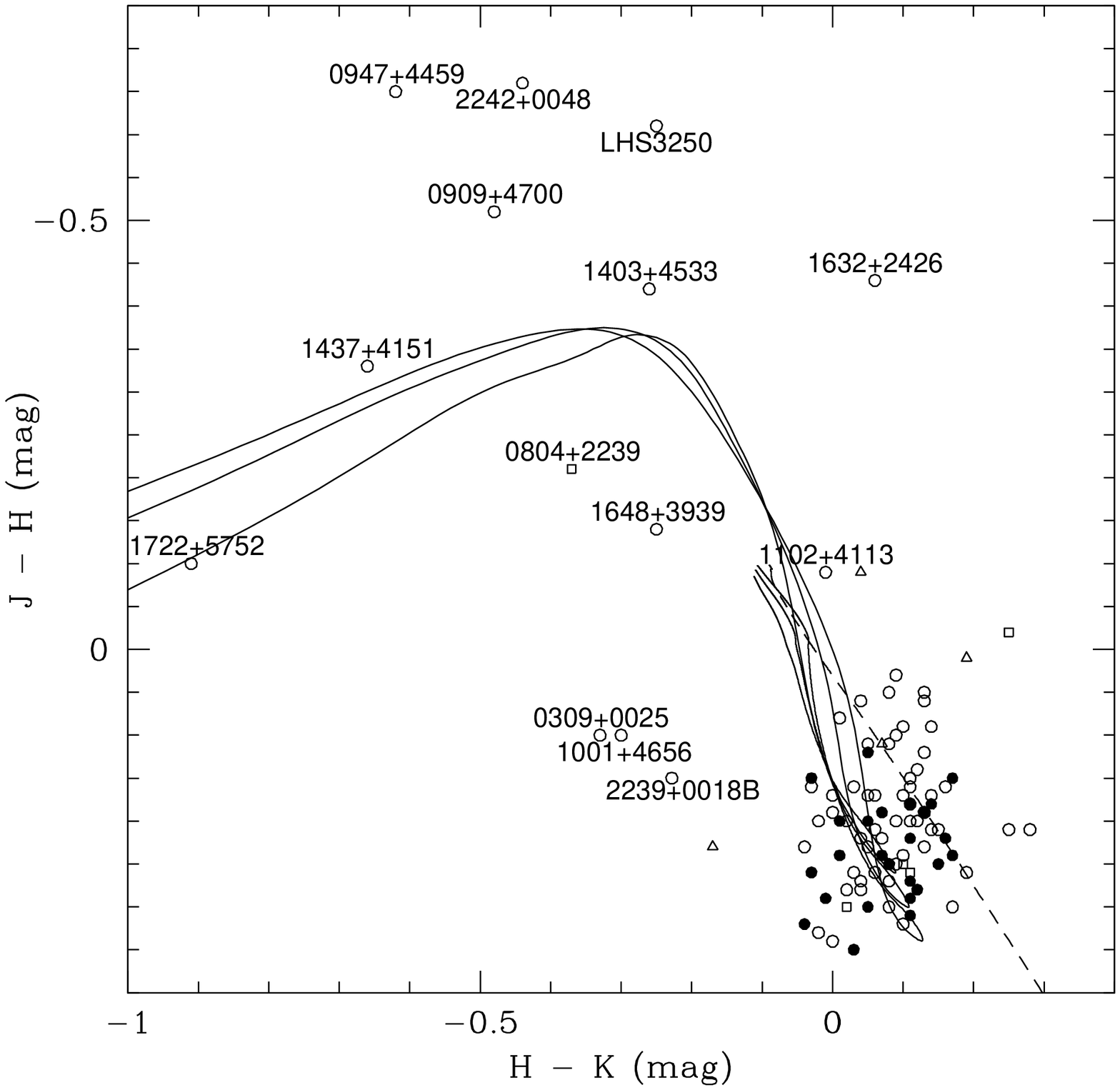}
\caption{Near-infrared color-color diagrams for WDs studied by \citet[][crosses in the left panel]{bergeron01}
and our sample (circles, triangles, and squares). The symbols are the same as in Figure 8.
WDs with significant infrared flux deficits, and the halo WD candidate J1102+4113
are labeled. In addition to the pure hydrogen and pure helium atmosphere model sequences, the dotted, long-dashed, and dashed-dotted
lines show the predicted color sequences for mixed atmosphere models with ($T_{\rm eff}= 2000-6000$ K and
$\log g=8$) H/He = 10, 1, and 0.01, respectively.}
\end{figure}

\begin{figure}
\includegraphics[width=6.0in,angle=0]{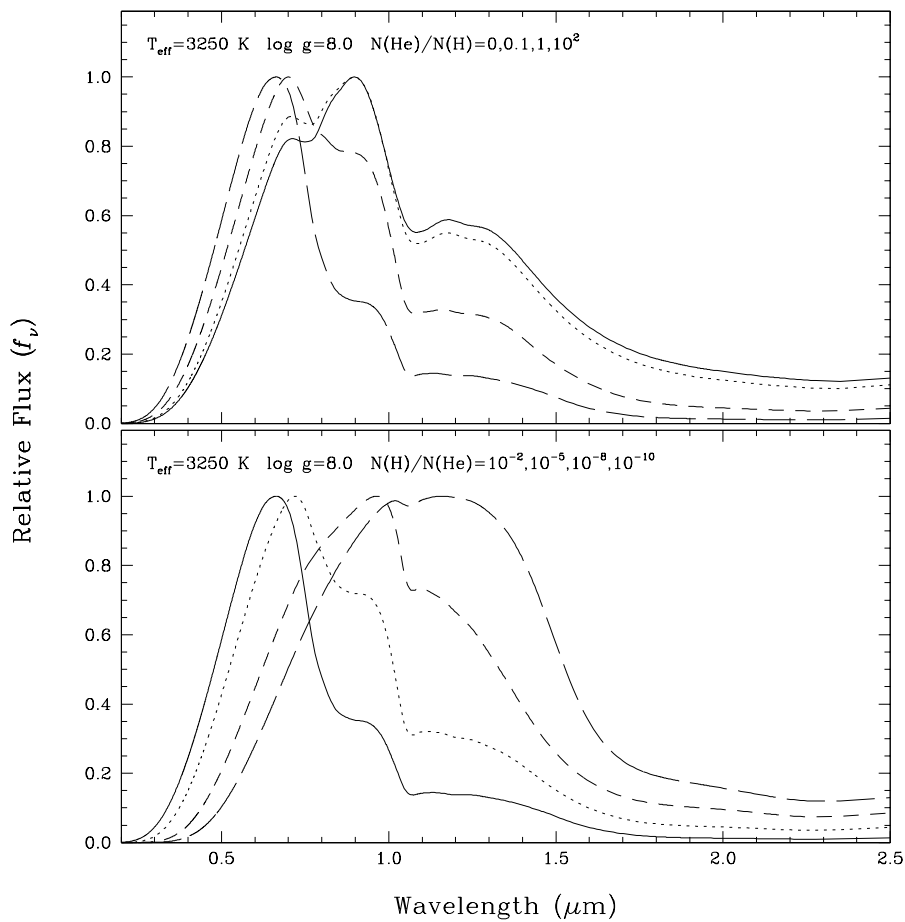}
\caption{Similar to Figure 5 of \citet{bergeron02} but with our updated
models for mixed H/He compositions. The top panel compares models at
T$_{\rm eff}$ = 3250 K and $\log g = 8.0$ from a pure hydrogen composition
({\it solid line}) to a value of H/He = 10$^{-2}$, where the infrared flux
deficiency is the strongest. In the bottom panel, the hydrogen abundance
is further decreased from a value of H/He = 10$^{-2}$ ({\it solid line})
to 10$^{-10}$.}
\end{figure}

\begin{figure}
\plotone{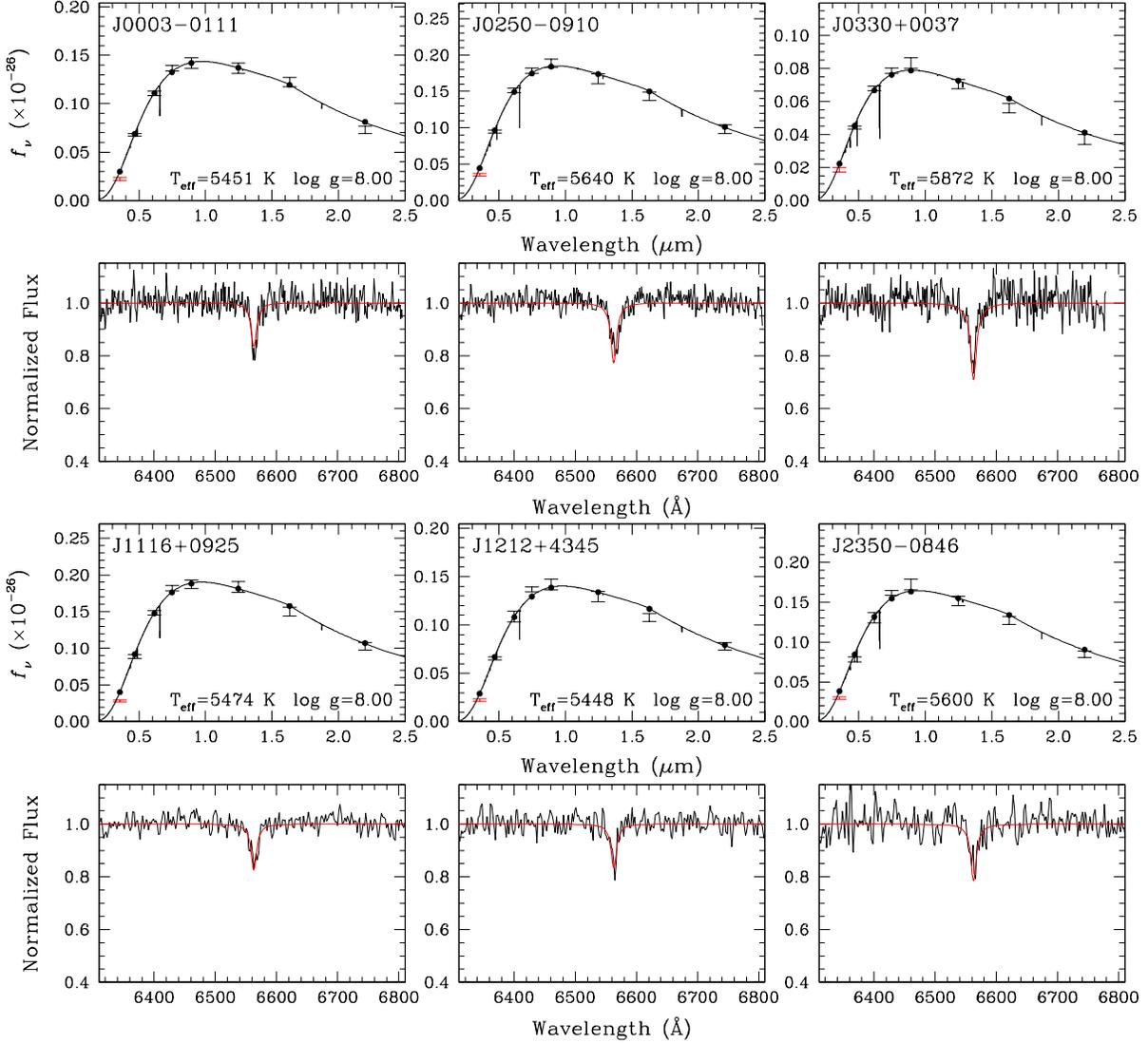}
\caption{Fits to the SEDs of 6 DA stars with pure hydrogen models. Here and in the following figures, the $ugriz$ and $JHK$
photometric observations are represented by error bars, while the model monochromatic fluxes are shown as solid lines.
The error bars shown in red indicate bandpasses that are not included explicitly in the fit.
The filled circles represent the average over the filter bandpasses. The lower panels show the
normalized spectra together with the synthetic line profiles for the parameters obtained from the SED fits.}
\end{figure}

\begin{figure}
\plotone{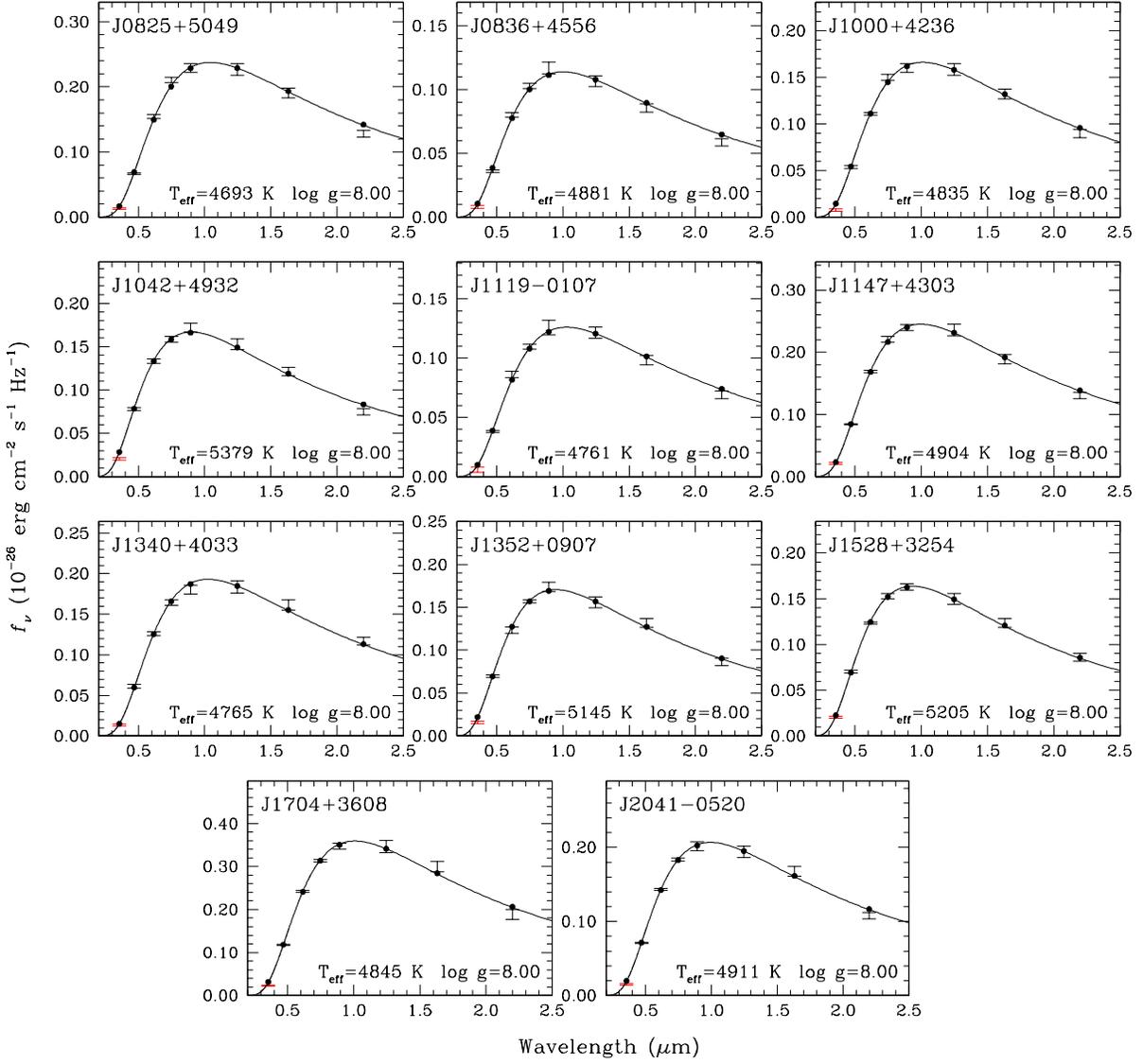}
\caption{Fits to the SEDs of 11 DC WDs with pure helium models (filled circles). All objects have featureless spectra near the
H$\alpha$ region.}
\end{figure}

\begin{figure}
\plotone{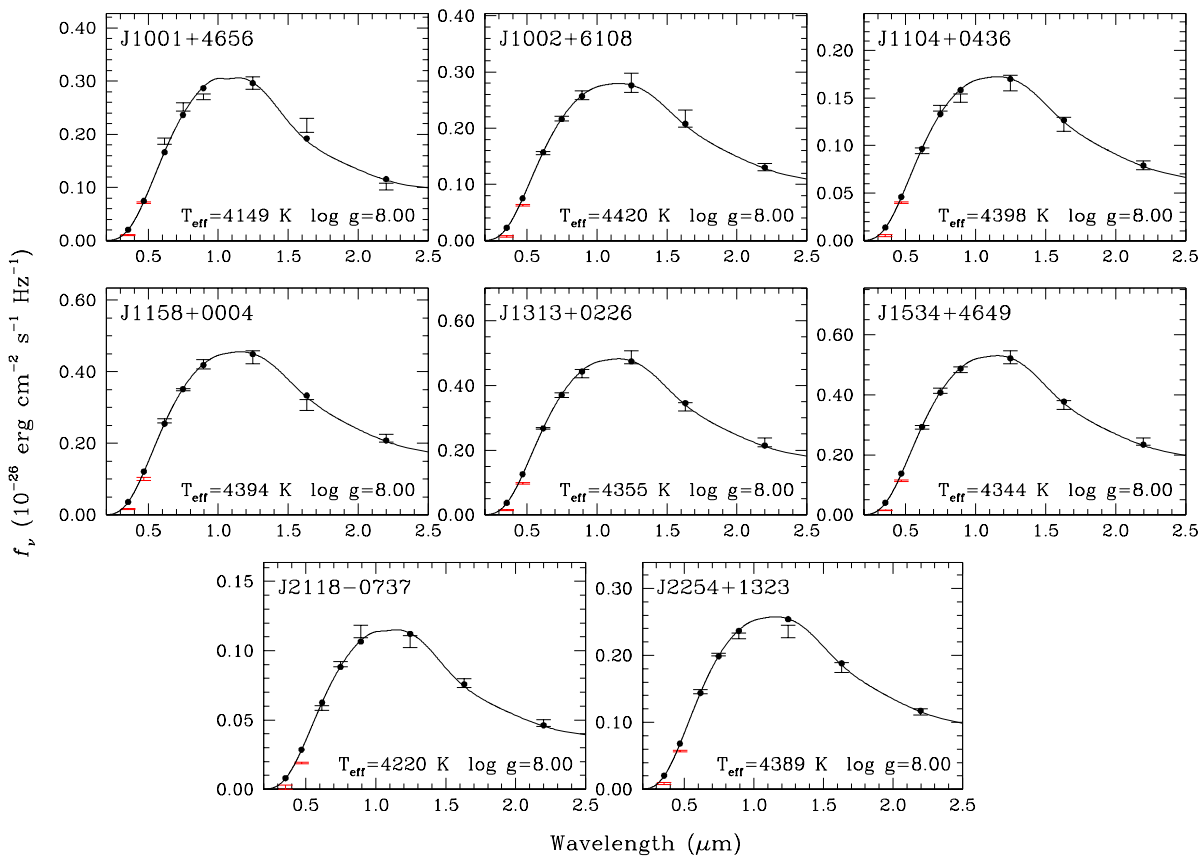}
\caption{Fits to the SEDs of the coolest 8 DC WDs in our sample, excluding the ultracool WDs. All objects have featureless spectra near the
H$\alpha$ region, and the SEDs are best explained with pure hydrogen atmosphere models.}
\end{figure}

\begin{figure}
\plotone{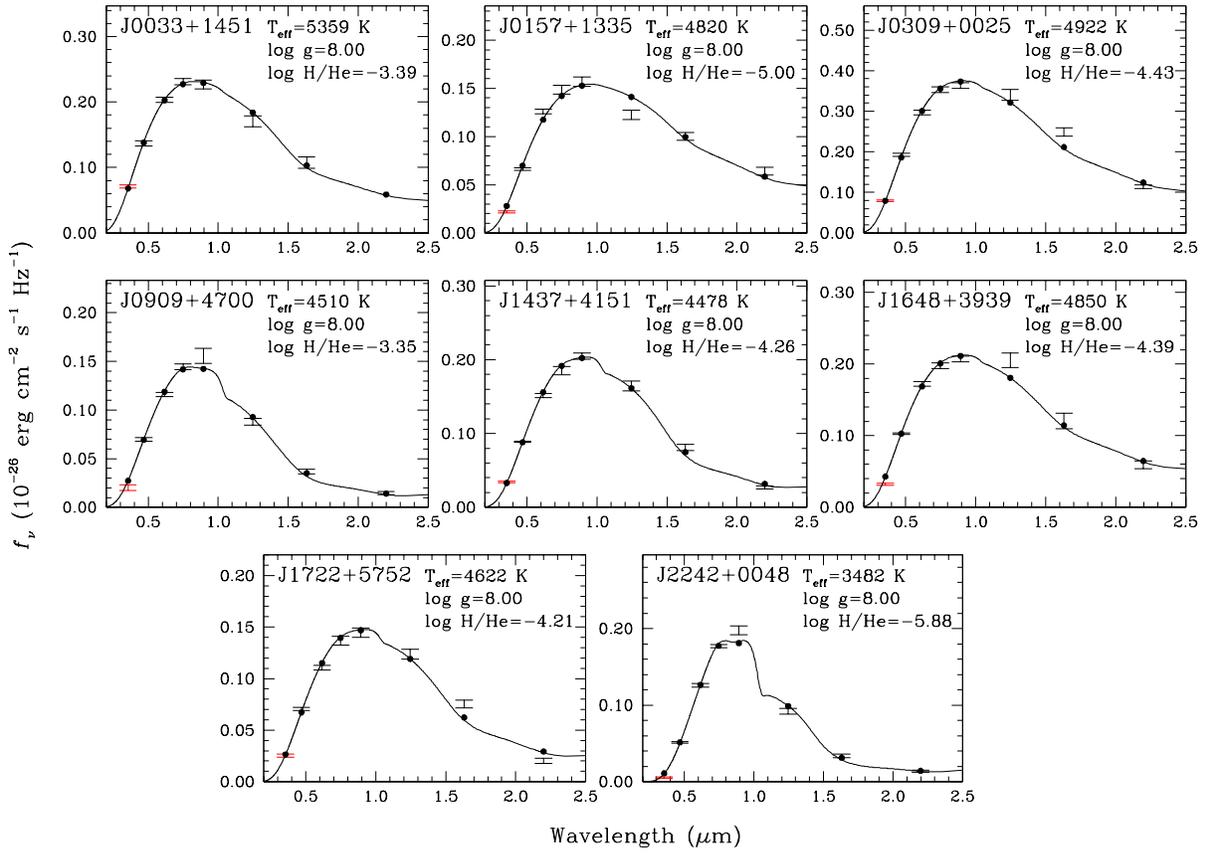}
\caption{Fits to the SEDs of 8 DC WDs with mixed H/He atmosphere models.}
\end{figure}

\begin{figure}
\plotone{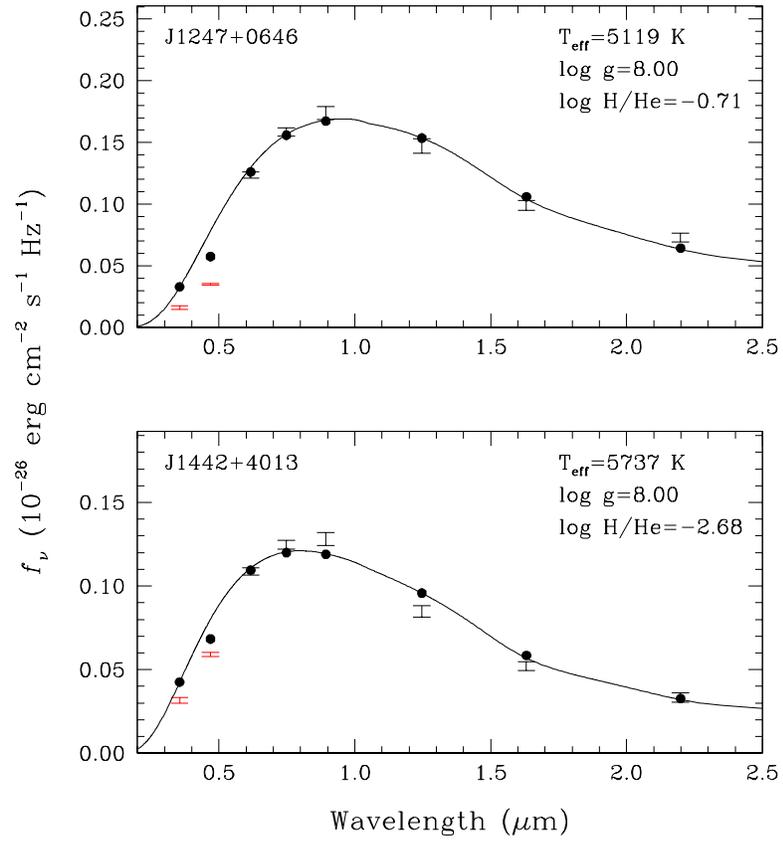}
\caption{Fits to the SEDs of 2 peculiar DQ WDs with mixed H/He atmosphere models.}
\end{figure}

\begin{figure}
\plotone{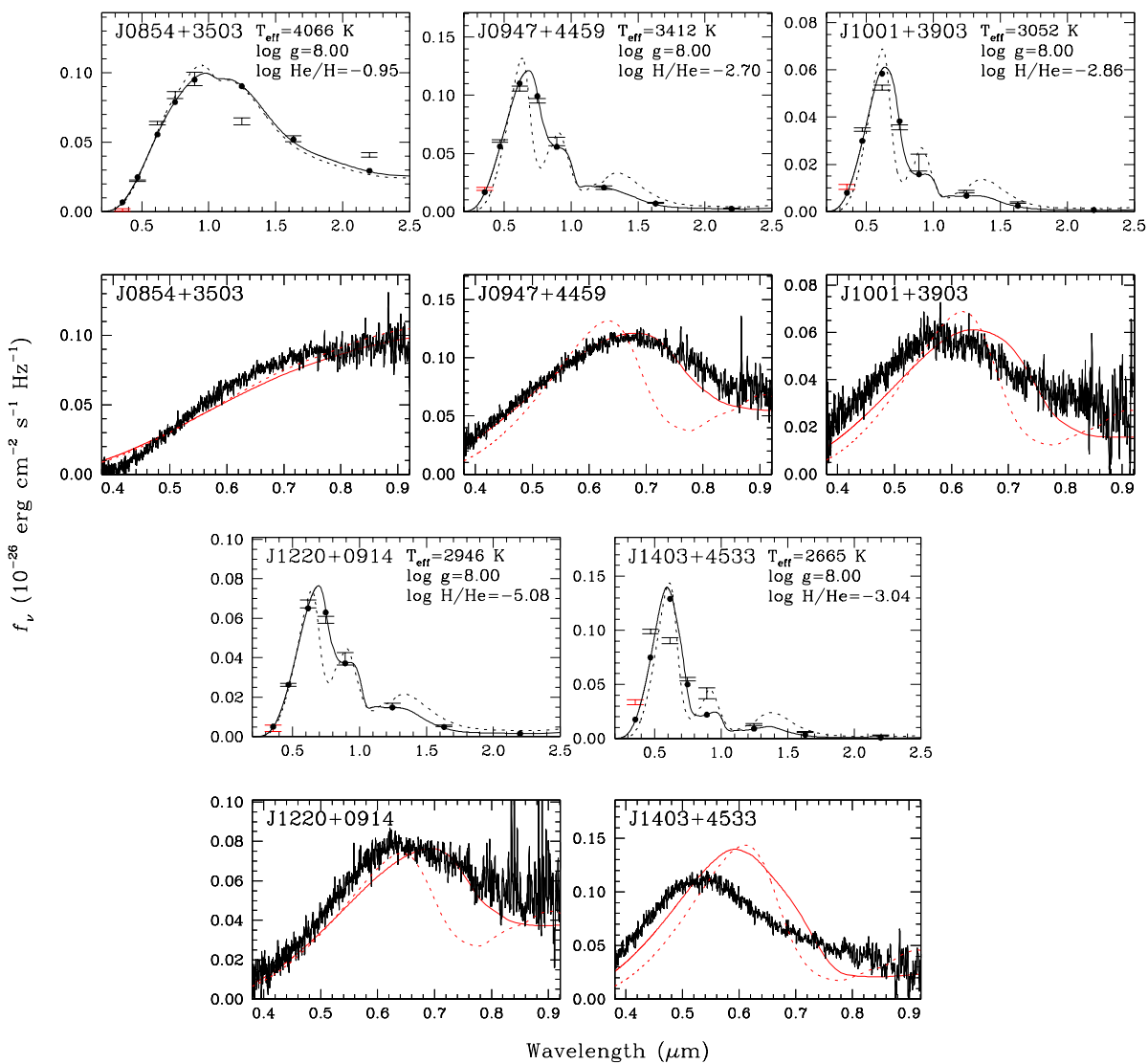}
\caption{Fits to the SEDs of the ultracool WDs discovered by Gates et al. (2004) using pure hydrogen (dotted lines)
and mixed H/He (solid lines) atmosphere models. The top panels show the optical and near-infrared photometry,
whereas the bottom panels show the SDSS spectrum of each object.}
\end{figure}

\clearpage
\begin{figure}
\plotone{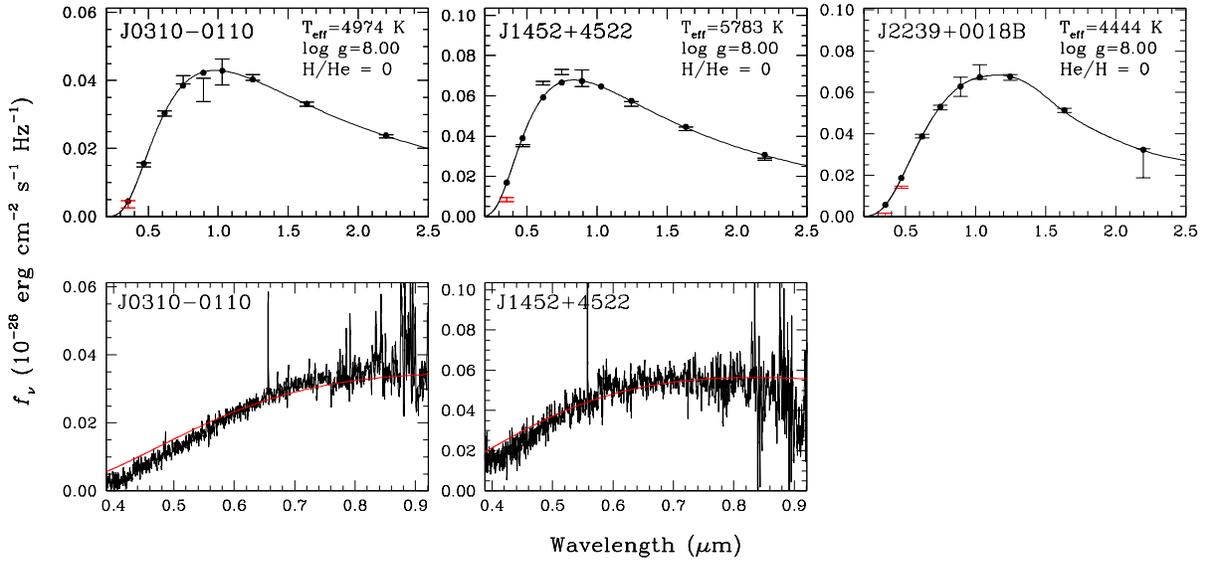}
\caption{Similar to Figure 16, but for three ultracool WD candidates from \citet{harris08}. Our model fits show that
J0310$-$0110 and J1452+4522 are best explained as pure helium atmosphere WDs and J2239+0018B as a pure hydrogen atmosphere
WD. The best-fit models have temperatures above 4000 K, implying that these three stars are not ultracool.}
\end{figure}

\begin{figure}
\plotone{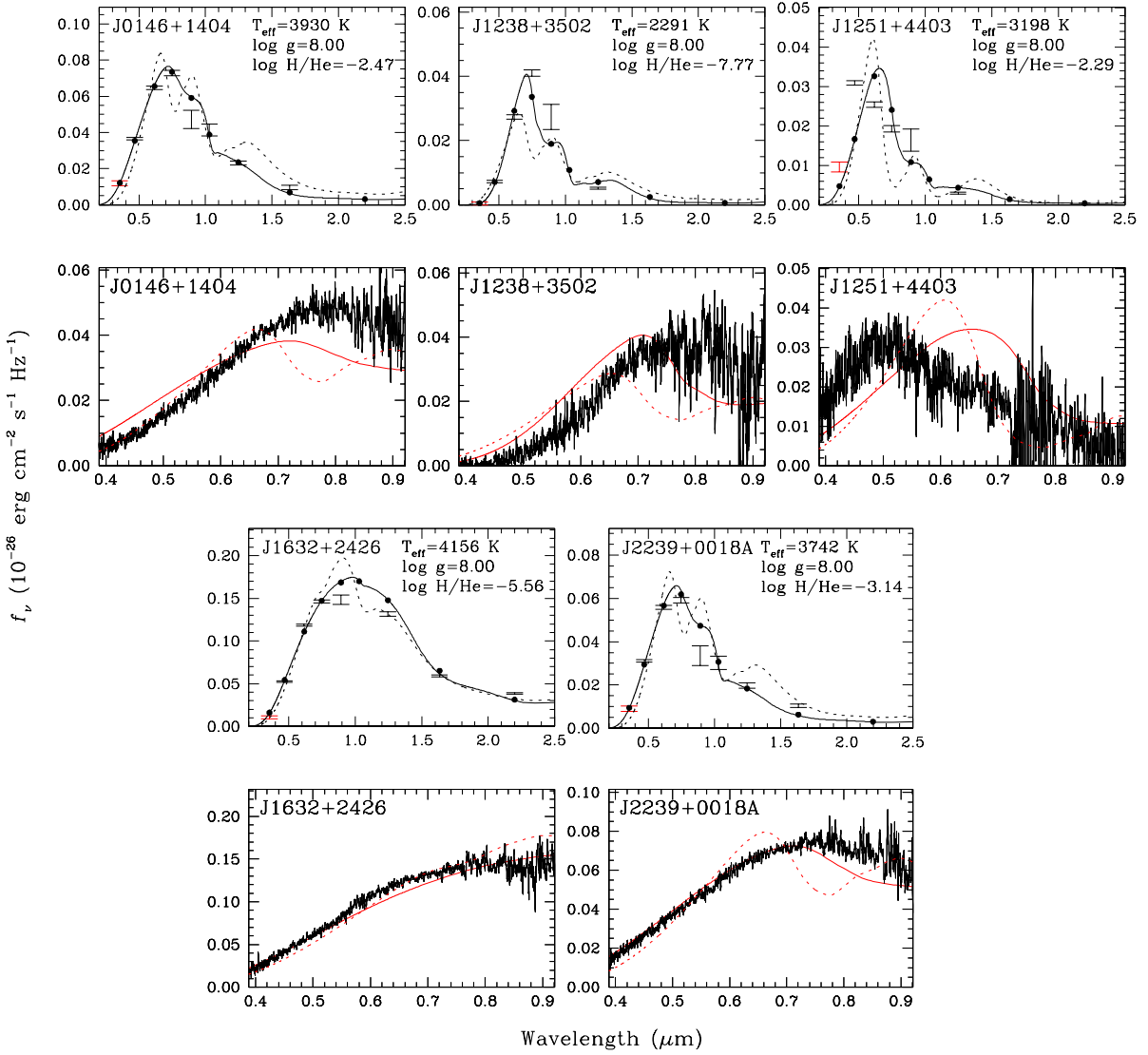}
\caption{Similar to Figure 16, but for five ultracool WD candidates from \citet{harris08}. All five stars are best-fit with mixed
H/He atmosphere models (solid lines).}
\end{figure}

\begin{figure}
\includegraphics[width=4.5in,angle=-90]{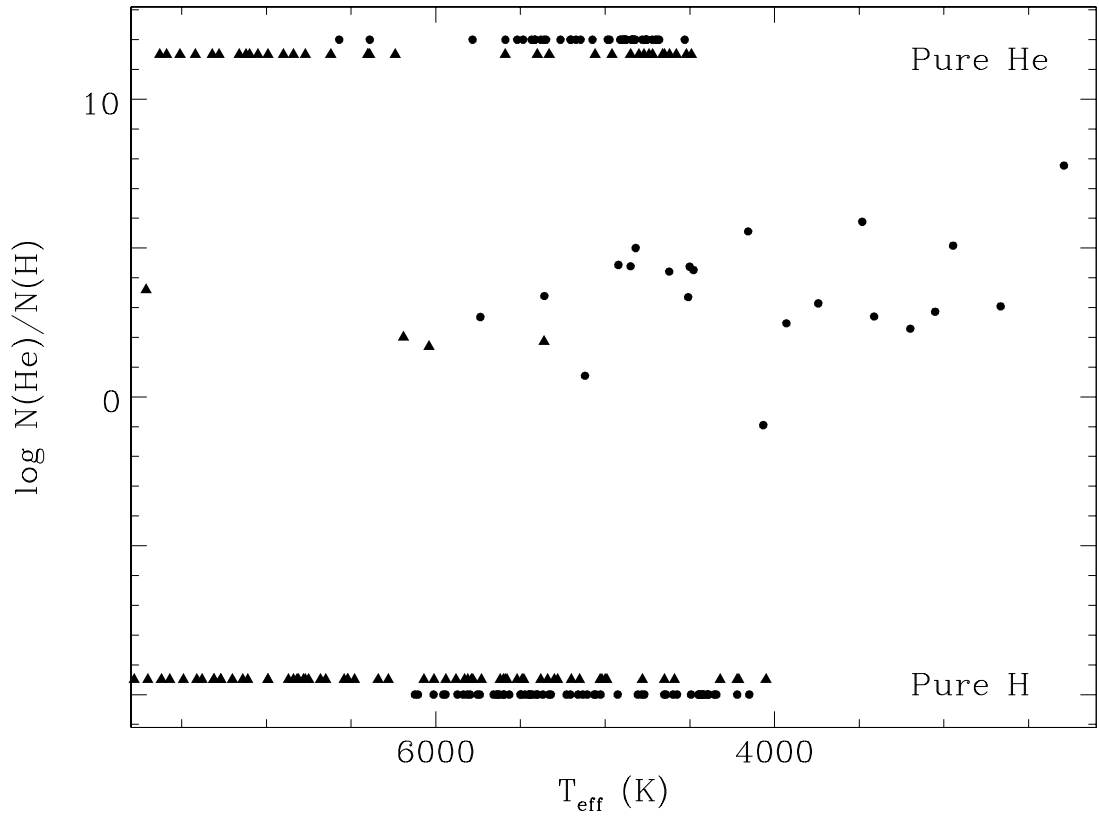}
\caption{The distribution of hydrogen and helium abundances for our sample of cool WDs (circles) and
that of Bergeron et al. (2001, triangles).}
\end{figure}
\clearpage

\begin{figure}
\includegraphics[width=4.5in,angle=-90]{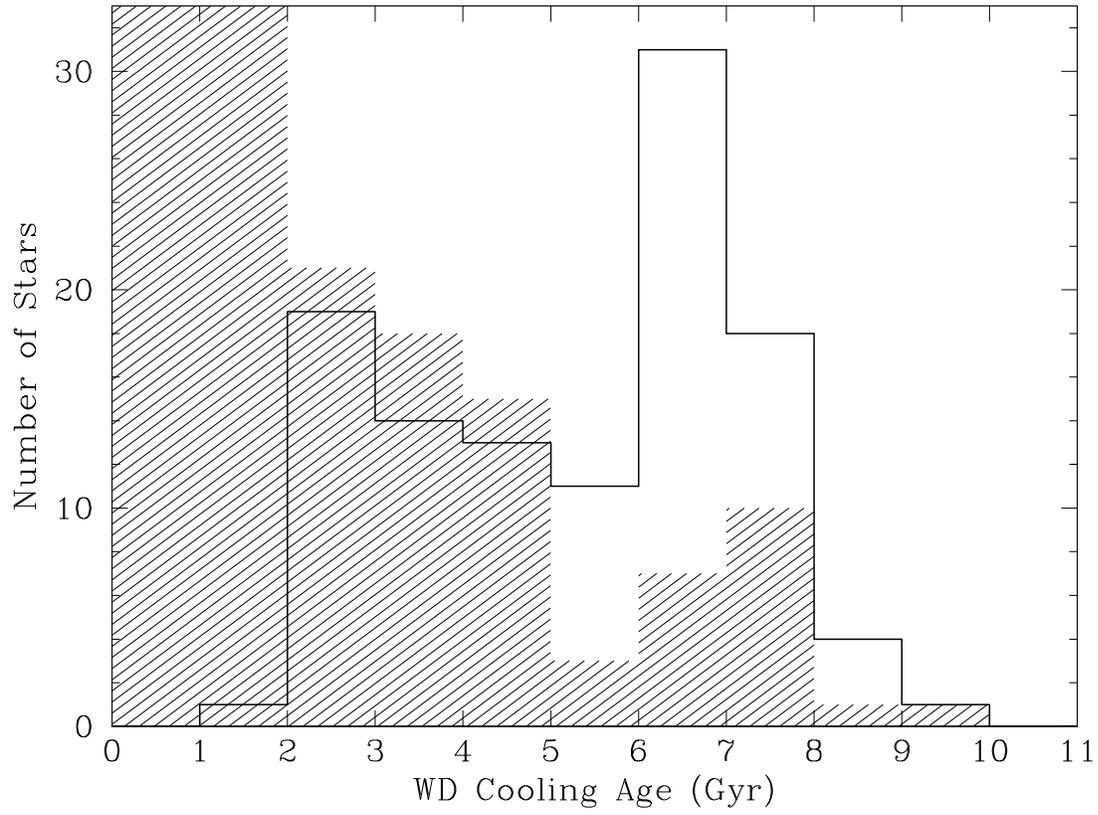}
\caption{Age distribution of our sample of WDs (solid histogram, excluding the ultracool WD candidates
due to poor model fits)
compared to that of \citet[][shaded histogram]{bergeron01}.}
\end{figure}

\begin{figure}
\plotone{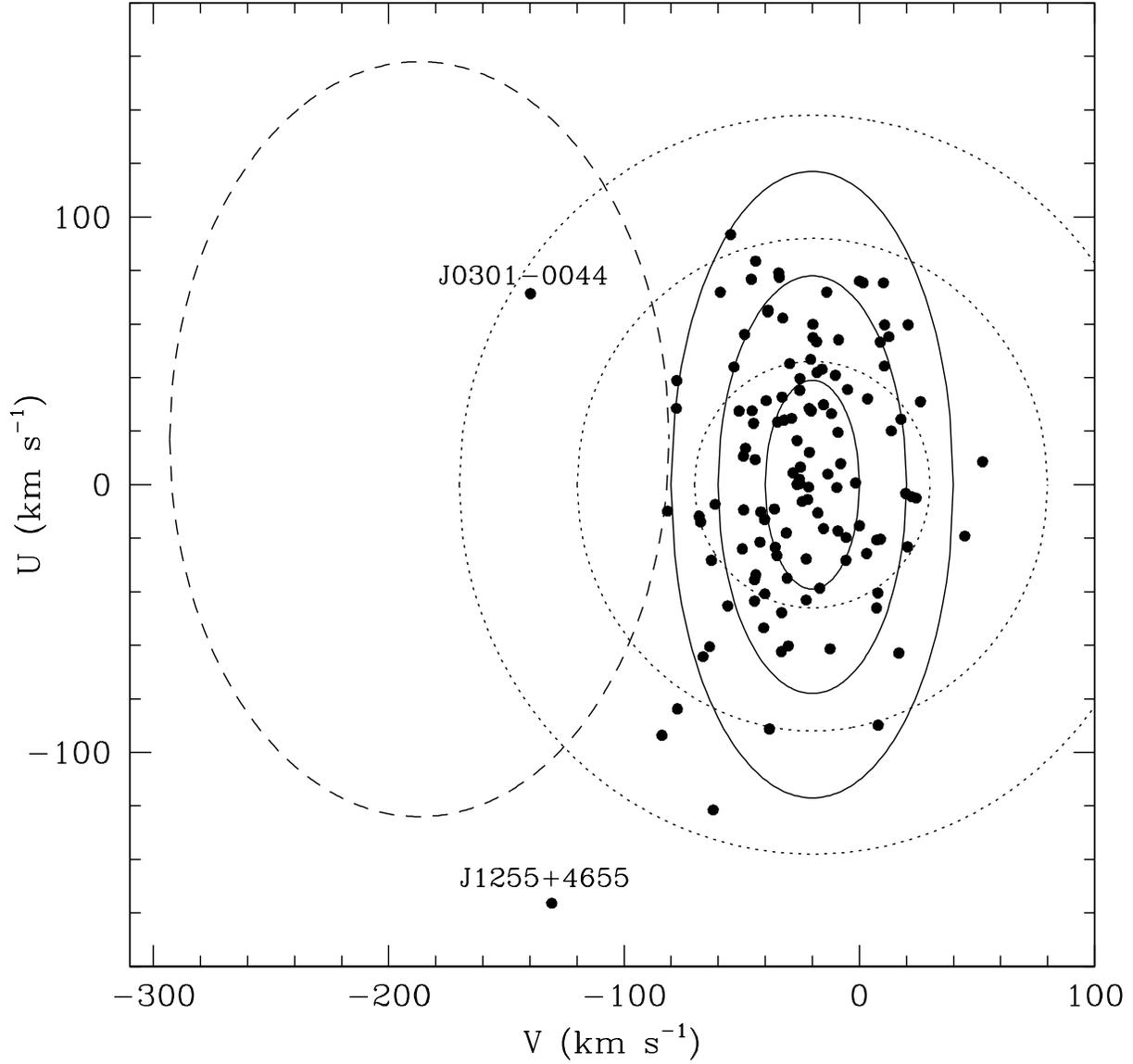}
\caption{$U$ vs. $V$ velocities for our targets assuming $\log~g=$ 8.0 and zero radial velocity.
The 1$\sigma$, 2$\sigma$, and 3$\sigma$ velocity ellipses of the thin disk (solid line) and the thick disk (dotted line), and the
1$\sigma$ ellipse of the halo (dashed line) are also shown.}
\end{figure}

\begin{figure}
\plotone{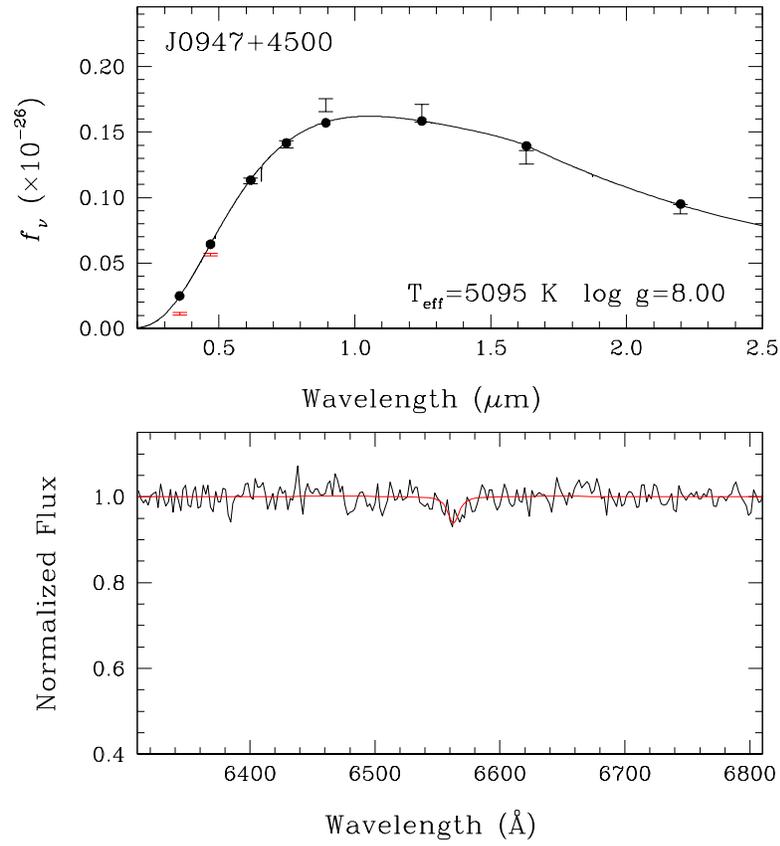}
\caption{Companion to the ultracool WD J0947+4459.}
\end{figure}

\end{document}